\documentclass[11pt,a4paper]{article}

\usepackage[T1]{fontenc}
\usepackage[utf8]{inputenc}

\usepackage{lmodern}
\usepackage{microtype}
\usepackage{comment}

\usepackage[a4paper,margin=1in]{geometry}

\usepackage{amsmath,amssymb,amsfonts,amsthm,mathtools}
\usepackage{mathrsfs}

\usepackage{enumitem}
\usepackage{booktabs}
\usepackage{graphicx}
\usepackage{xcolor}
\usepackage{hyperref}
\usepackage[nameinlink]{cleveref}
\usepackage{here}
\usepackage{multirow}
\usepackage{tikz}
\usetikzlibrary{automata,positioning}
\usepackage{changepage}

\hypersetup{
    colorlinks=true,
    linkcolor=blue,
    citecolor=blue,
    urlcolor=blue
}

\newtheorem{theorem}{Theorem}[section]
\newtheorem{lemma}[theorem]{Lemma}
\newtheorem{proposition}[theorem]{Proposition}
\newtheorem{corollary}[theorem]{Corollary}

\theoremstyle{definition}
\newtheorem{definition}[theorem]{Definition}
\newtheorem{example}[theorem]{Example}

\newtheorem{exercise}{Exercise}[section]
\newenvironment{starexercise}
  {\refstepcounter{exercise}\par\noindent
   \textbf{Exercise \theexercise\textsuperscript{\,$\star$}.}\ }
  {\par}
\newtheorem{problem}{Problem}[section]

\newcommand{\Ppos}{\mathcal{P}}
\newcommand{\Npos}{\mathcal{N}}
\newcommand{\Pk}{\mathsf{Pk}}
\newcommand{\WSsuf}{\mathtt{WordSubtraction}^\mathsf{suf}}
\newcommand{\WSpre}{\mathtt{WordSubtraction}^\mathsf{pre}}
\newcommand{\LPsuf}{L_\Ppos^\mathsf{suf}}
\newcommand{\LNsuf}{L_\Npos^\mathsf{suf}}
\newcommand{\LPpre}{L_\Ppos^\mathsf{pre}}
\newcommand{\LNpre}{L_\Npos^\mathsf{pre}}
\newcommand{\bqed}{\hfill$\blacksquare$}
\newcommand{\diaqed}{\hfill$\diamond$}

\newcommand{\versiondate}{November 27, 2026}

\title{
    \LARGE\bfseries Word Subtraction Games
}

\author{
    {\Large Yusuke Inoue}\thanks{Nagoya University, \href{mailto:y-inoue@sqlab.jp}{\texttt{y-inoue@sqlab.jp}}} \and
    {\Large Hiroki Inazu}\thanks{Hiroshima University, \href{mailto:hiroki.inazu@gmail.com}{\texttt{hiroki.inazu@gmail.com}}}
}

\date{%
    {\small
    \textbf{Version of \versiondate}\\
    \begin{flushleft}
    This is a work in progress.
    Comments, corrections, and suggestions are very welcome.
    Please feel free to contact the authors if you find
any errors or anything unclear.
    We recommend including the version date when citing this note.
    \end{flushleft}
}
}

\begin{document}

\maketitle

\begin{abstract}
    This note is a self-contained text on word subtraction games, combinatorial games in which players alternately remove a nonempty suffix belonging to a prescribed subtraction set from the current word. 
    It collects and develops basic results on these games from the viewpoints of combinatorial game theory and formal language theory.
    We focus on the relationship between subtraction sets and the resulting languages of winning and losing positions.
    The topics include finite, regular, and context-free subtraction sets, as well as several variants and further developments.
\end{abstract}

\section*{Preface}
Around the beginning of 2026, the first author began studying word subtraction games, a class of combinatorial games in which positions are finite words and moves consist of removing prescribed nonempty suffixes. They can be viewed as a natural generalization of classical subtraction games, obtained by replacing heaps with finite words.

Word subtraction games connect combinatorial game theory with automata theory and formal language theory. Both the subtraction set specifying the legal moves and the resulting sets of winning and losing positions can be viewed as languages. 
This allows us to ask how the structure of the rules is reflected in the structure of the outcomes, and to study that relationship using tools from formal language theory.

The goal of this note is to give a self-contained
introduction to word subtraction games, their basic
properties, and several variants.
We collect the results obtained so far and discuss
questions for further study.
This note is intended to be accessible to readers without prior knowledge of combinatorial game theory or formal language theory. 
The basic notions needed for the main development are introduced along the way.

The main development begins with finite subtraction sets in Section~\ref{sec:fin}, followed by regular subtraction sets in Section~\ref{sec:reg} and context-free subtraction sets in Section~\ref{sec:cf}.
Section~\ref{sec:variant} considers what happens when moves can remove prefixes as well as suffixes, or factors from within a word.
Section~\ref{sec:advanced} presents a selection of advanced topics, including misère play and algorithmic questions about word subtraction games.
The final section describes the research background and discusses broader directions for further study.

Although this note is primarily intended to be read in order, readers who have completed Section~\ref{sec:reg} can proceed to
Section~\ref{sec:cf}, Section~\ref{sec:variant}, or any of the topics in Section~\ref{sec:advanced}, according to their interests.

Several sections include exercises and open problems. 
The exercises are intended to help readers check their understanding, fill in details of selected proofs, and explore further results not developed in the main text. Some of the more challenging exercises are marked with a star. The open problems collect questions that the authors currently find interesting. Some have not yet been investigated in depth, and we hope that readers will find useful starting points for their own investigations among them.

AI tools were used to assist with drafting and editing parts of the English text and to help check and improve some mathematical arguments.
The authors carefully reviewed the AI-assisted material and take full responsibility for the content of this note.

\section*{Acknowledgments}
We are especially grateful to Yoshiaki Takata, Hiroyuki Seki, Yushin Kusajima, Shunya Oguchi, Hiyu Inoue, and Takuto Maki for their contributions to the mathematical content of this note and for many valuable discussions related to the problems considered here. Some of the results, ideas, and observations in this note grew out of these discussions. Their contributions are described in more detail in the relevant parts of the text and in \emph{Research Background and Further Directions}.

We would also like to thank Shun-ichi Kimura and the members of his research group, including Hiyu Inoue, Yuto Terakura, Haruki Wada, Kosaku Watanabe, and Takahiro Yamashita. 
The second author, as a member of this group, is especially grateful for the continuous guidance and stimulating discussions that have supported his research in combinatorial game theory. 
The first author would also like to thank the group for the many fruitful discussions and interactions in combinatorial game theory.

The first author would particularly like to thank Yuji Komatsu, who first introduced him to combinatorial game theory and continues to discuss these ideas with him.
We are also deeply grateful to Ryuya Hora, whose broad knowledge and insight have contributed greatly to our understanding of both formal language theory and combinatorial game theory, as well as the connections between these two areas.
In addition, we would like to thank 
Tomoaki Abuku, Kengo Hashimoto, Hironori Kiya, Hikaru Manabe,  Nanako Omiya, Yu Tokumoto, Takuho Yamada, and Ryo Yoshinaka for many valuable discussions on several occasions during the summer of 2026.

As mentioned above, this note was written with the help and support of many people in various ways. 
Nevertheless, any errors or inaccuracies in this note are entirely the responsibility of the authors.
\newpage
\tableofcontents

\newpage

\section{Word Subtraction Games}\label{sec:intro}
In this note, let $\mathbb{N}$ and $\mathbb{N}_+$ denote the set of all non-negative integers and the set of all positive integers, respectively.
For a set $X$, let $\#(X)$ denote the cardinality of $X$.

\subsection{Combinatorial Games}\label{subsec:intro-games}
Combinatorial game theory studies a broad class of two-player games in which the players move alternately, all information is available to both players, and no chance moves are involved. 
In this note, we mainly focus on a class of combinatorial games satisfying the following assumptions:
\begin{itemize}
    \item \emph{Impartial}: The available moves from a position depend only on the position itself and not on whose turn it is.
    \item \emph{Loop-free}: Every play eventually terminates after finitely many moves. Games that do not satisfy this assumption are called \emph{loopy}.
    \item \emph{Finite-option}: Every position has only finitely many available moves. A game that is both loop-free and finite-option is often called \emph{short}.
    \item \emph{Normal-play}: The player making the last move wins. Equivalently, the player who has no legal move loses. In contrast, under the \emph{misère} convention, the player who has no legal move wins.
\end{itemize}

A position that can be reached from a position $p$ by a single legal move is called an \emph{option} of $p$.
A position with no legal moves is called a \emph{terminal position}.
Under the normal-play convention, the player to move
at a terminal position loses.

\begin{example}[Subtraction Games]\label{ex1-subtraction}
    A classical example of a combinatorial game is the \emph{subtraction game}. Let $S\subseteq \mathbb{N}_+$ be a set of positive integers. 
    The game $\mathtt{Subtraction}(S)$ is played as follows. 
    A position consists of a single \emph{heap} (or \emph{pile}) of stones, represented by a nonnegative integer $n\in \mathbb{N}$, where $n$ denotes the number of stones in the heap. 
    Starting from an initial heap, the two players alternately remove $k$ stones from the heap, where $k\in S$ and $k\leq n$, with $n$ denoting the current number of stones.

    For example, if $S=\{1,2\}$, then each player can remove either one or two stones on each turn. 
    Starting from a heap of five stones, one possible play is\[
    5~\longrightarrow ~3~\longrightarrow ~1~\longrightarrow ~0
    \]
    where the first player removes two stones, the second player also removes two stones, and finally the first player removes the last remaining stone.
    Under the normal-play convention, the first player wins because they make the last move.
    Representing each stone by a symbol $\bullet$, the same play can be visualized as\[
{\bullet\bullet\bullet}\,\underline{\bullet\,\bullet}~\longrightarrow ~\bullet\,\underline{\bullet\,\bullet}~\longrightarrow ~\underline{\bullet}~\longrightarrow ~\varepsilon
    \]
    where $\varepsilon$ denotes the empty heap, namely the position with no remaining stones.\bqed
\end{example}

One of the most fundamental notions in combinatorial game theory is the classification of positions into \emph{$\Ppos$-positions} and \emph{$\Npos$-positions}.
An $\Npos$-position is a position from which the player to move has a winning strategy, regardless of how the opponent plays.
A $\Ppos$-position is a position from which the opponent has such a strategy.
Here, $\Npos$ stands for \emph{Next} and $\Ppos$ for \emph{Previous}: after a move has been made, the opponent of the player to move is the player who made the previous move.
These terms apply to initial positions as well.
For impartial, short, normal-play games, the following recursive characterization holds:
\begin{proposition}[Partition Theorem]
    For an impartial short normal-play game,
    \begin{itemize}
        \item a position is a $\Ppos$-position if and only if all of its options are $\Npos$-positions, and
        \item a position is an $\Npos$-position if and only if it has at least one option that is a $\Ppos$-position.
    \end{itemize}
    In particular, the set of positions is partitioned into $\Ppos$-positions and $\Npos$-positions.\bqed
\end{proposition}
\noindent
Every terminal position is a $\Ppos$-position, since the player to move has no legal move and therefore loses.
For any fixed starting position, there are only finitely many possible plays because each position has only finitely many options and no infinite play is possible.
Thus, we can classify the starting position by working backwards from the terminal positions and applying the rules above.
The game can have infinitely many positions in total, but the classification of any particular position requires only finitely many steps.

\begin{example}
Let us return to $\mathtt{Subtraction}(\{1,2\})$ in Example~\ref{ex1-subtraction}.
We first observe that position $0$ is a terminal position, and hence it is a $\Ppos$-position. 
The positions $1$ and $2$ are $\Npos$-positions because the next player can move directly to the $\Ppos$-position $0$.
Next, both options from $3$ are $\Npos$-positions, and hence $3$ is a $\Ppos$-position.
Continuing this procedure, we observe that a position $n$ is a $\Ppos$-position if and only if $n$ is a multiple of $3$.
Indeed, every legal move from a multiple of $3$ leads to a number that is not a multiple of $3$, while from any other position one can reach a multiple of $3$ by removing one or two stones.
\bqed
\end{example}

The $\Ppos$/$\Npos$ classification distinguishes only between winning and losing positions. 
We can obtain more information by assigning
a nonnegative integer, called the \emph{Grundy number},
to each position.
Grundy numbers play a central role in the theory of normal-play impartial games. 
Because they are not needed for the main development of this note, we do not explore them in detail here.
For a general introduction to combinatorial game theory, including Grundy numbers, see, for example, Albert, Nowakowski, and Wolfe~\cite{Albert2019} and Siegel~\cite{CGT}.

The games studied in this note generalize subtraction games from integers to \emph{words}.
To illustrate the basic idea, consider a row of stones of two types, labeled $a$ and $b$. 
A position is represented by the sequence of these labels. 
Suppose that, on each turn, a player can remove $a$, $b$, or $aa$ from the right end of the sequence, whenever the sequence ends with the chosen word. 
For example, one possible play is\[
aabba\underline{aa}~\longrightarrow~aabb\underline{a}~\longrightarrow~aab\underline{b}~\longrightarrow~aa\underline{b}~\longrightarrow~\underline{aa}~\longrightarrow~\varepsilon\,.
\]
Although this rule is easy to describe informally, studying such games systematically requires a mathematical framework for words and sets of words.
The next subsection introduces the basic notions of formal language theory, which provides such a framework.

\subsection{Words and Languages}
A set $\Sigma$ of symbols is called an \emph{alphabet}. 
The elements of an alphabet are called \emph{letters}. For example, $\Sigma=\{a,b\}$ is an alphabet consisting of two letters $a$ and $b$. 
Throughout this note, we assume that alphabets are finite and nonempty.
A \emph{word} (or \emph{string}) over $\Sigma$ is a finite sequence of letters from $\Sigma$. 
For example, $(a,b,b,a,a)$ is a word of length $5$ over the alphabet $\Sigma=\{a,b\}$. We usually omit commas and parentheses and write this word simply as $abbaa$.
The length of a word $w$ is denoted as $|w|\in\mathbb{N}$.
The \emph{empty word}, denoted by $\varepsilon$, is the unique word of length $0$. 
The set of all words over $\Sigma$, including the empty word $\varepsilon$, is denoted by $\Sigma^*$.
The \emph{concatenation} of two words $u$ and $v$, denoted by $uv$, is obtained by writing $v$ immediately after $u$.
For example, the concatenation of $abb$ and $aa$ is $abbaa$.
The order matters: in general, $uv\neq vu$.
For every $w\in\Sigma^*$, we have $\varepsilon w=w\varepsilon=w$.
Because concatenation is associative, $\Sigma^*$ forms a monoid with $\varepsilon$ as its identity element, and this monoid is called the \emph{free monoid} on $\Sigma$.\footnote{Some of the terminology and additional remarks concerning monoids and monoid actions are included mainly to help readers with a background in algebra. These remarks are not essential for following the main discussion, and readers who are not familiar with the terminology can skip them.}

A subset $L$ of $\Sigma^*$ is called a \emph{language} over $\Sigma$. 
In other words, a language $L\subseteq \Sigma^*$ is a set of words over $\Sigma$.
The concatenation of words naturally extends to an operation on languages.
For two languages $L_1,L_2\subseteq \Sigma^*$, their concatenation is defined by
\[
L_1L_2=\{w_1w_2\in \Sigma^*: w_1\in L_1,\,w_2\in L_2\}.
\]
We often identify a word $w\in\Sigma^*$ with the singleton language $\{w\}\subseteq \Sigma^*$. 
For example, $wL$ and $Lw$ stand for $\{w\}L$ and $L\{w\}$, respectively.
For a language $L\subseteq\Sigma^*$ and a nonnegative integer $n$, we write $L^n$ for the $n$-fold concatenation of $L$, with the convention $L^0=\{\varepsilon\}$.
Equivalently, $L^{n+1}=L^nL$ for every $n\geq 0$.
For a word $w\in\Sigma^*$, we similarly write
\[
    w^n=\underbrace{ww\cdots w}_{n\text{ copies}},
\]
with $w^0=\varepsilon$.
The \emph{Kleene star} of a language $L\subseteq \Sigma^*$ is defined by \[
L^*=\bigcup_{n\ge 0}L^n.
\] 
Thus, $L^*$ consists of all words obtained by concatenating zero or more words from $L$; in particular, it always contains $\varepsilon$.
Equivalently, $L^*$ is the smallest submonoid of $\Sigma^*$ containing $L$. 
Similarly, the \emph{positive closure} (or \emph{Kleene plus}) of $L\subseteq \Sigma^*$ is defined by\[
L^+=\bigcup_{n\ge 1}L^n.
\]
Thus, $L^+$ consists of all words obtained by concatenating one or more words from $L$.
In particular, $\Sigma^+=\Sigma^*\setminus\{\varepsilon\}$ for an alphabet $\Sigma$.
For a language $L\subseteq\Sigma^*$, its complement is taken relative to $\Sigma^*$ and is denoted by
\[
    L^c=\Sigma^*\setminus L.
\]

\begin{example}The following examples illustrate languages over $\Sigma=\{a,b\}$:
\begin{itemize}
    \item $\{\varepsilon\}$ is the language consisting only of the empty word.
It is different from the \emph{empty language} $\emptyset$.
    \item $\Sigma^*$ is the \emph{full language}, consisting of all words over $\Sigma$, including the empty word.
    \item $\{a,b,aa\}$ is the finite language consisting of three words $a$, $b$, and $aa$.
    \item $(\Sigma\Sigma)^*=\{aa,ab,ba,bb\}^*$ is the language of even length words.
    \item $a\Sigma(\Sigma\Sigma)^*\cup b(\Sigma\Sigma)^*$ is the language of words that either start with $a$ and have even length, or start with $b$ and have odd length.
    \item $a^*b^*$ is the language of all words formed by an arbitrary number of $a$'s followed by an arbitrary number of $b$'s.
    \item $\Sigma^*ba\Sigma^*$ is the language of all words containing $ba$ as a consecutive pair of letters.
This language is the complement of $a^*b^*$: a word contains no such pair if and only if all its $a$'s precede all its $b$'s.
    \item $\{a^nb^n:n\ge 0\}$ is the language of all words formed by some number of $a$'s followed by the same
number of $b$'s.\bqed
\end{itemize}
    
\end{example}

Let $w\in\Sigma^*$ be a word.
A word $p\in\Sigma^*$ is called a \emph{prefix} of $w$ if there exists a word $v\in \Sigma^*$ such that $w=pv$.
If $v\ne \varepsilon$, that is, if $p\ne w$, then $p$ is also called a \emph{proper prefix} of $w$.
For example, $\varepsilon,a,aa$, and $aab$ are prefixes of $aab\in\{a,b\}^*$. In this case, $\varepsilon$, $a$, and $aa$ are proper prefixes.
Similarly, a word $s\in\Sigma^*$ is called a \emph{suffix} of $w$ if there exists a word $u\in \Sigma^*$ such that $w=us$.
A word $f\in\Sigma^*$ is called a \emph{factor} (or an \emph{infix}) of $w$ if there exist words $u,v\in \Sigma^*$ such that $w=ufv$.
A suffix or factor of $w$ is called \emph{proper} if it is different from $w$ itself.
Note that both $\varepsilon$ and $w$ are prefixes, suffixes, and factors of $w$.

\subsection{Word Subtraction}
We are now ready to give the formal definition of a word subtraction game.
\begin{definition}[Word Subtraction Games]
Let $S\subseteq\Sigma^*$ be a language over an alphabet $\Sigma$ such that $\varepsilon\notin S$.
We call $S$ the \emph{subtraction set}.
The \emph{word subtraction game} $\mathtt{WordSubtraction}(S)$ is the normal-play game in which each player removes a suffix belonging to $S$ from the current word.
More formally, a position is a word over $\Sigma$, and a move from a word $w\in\Sigma^*$ to a word $w'\in\Sigma^*$ is legal if
\[
w = w's
\]
for some $s\in S$.\bqed
\end{definition}
The game is impartial because the legal moves depend only on the current word.
It is also short.
Indeed, every word has only finitely many suffixes, so every position has finitely many options.
Moreover, since $\varepsilon\notin S$, every move strictly decreases the length of the current word, and hence no infinite play is possible.
A word has a legal move if and only if it has a suffix belonging to $S$, or equivalently, if it belongs to $\Sigma^*S$.
Thus, the language of terminal positions is $(\Sigma^*S)^c$.
\begin{example}\label{ex1-ws_a_b_aa}
    Let $\Sigma=\{a,b\}$ and $S=\{a,b,aa\}\subseteq \Sigma^*$.
    Then, $\mathtt{WordSubtraction}(S)$ is the game introduced informally in Section~\ref{subsec:intro-games}:
    A position is represented by a word over $\Sigma=\{a,b\}$, and each player alternately removes `$a$', `$b$', or `$aa$' from the right end of the position.\footnote{\label{fn:prefix-closed}
In this example, the set of $\Ppos$-positions is
$\{ab,ba,bb,aaa,aab\}^*$.
More generally, in the case where $S\subseteq\Sigma^+$ is closed under taking nonempty prefixes, the game is a \emph{strongly Bernoulli type
truncation game} in the sense of Hetyei \cite{DBLP:journals/jct/Hetyei10}.
Theorem~2.9 of that paper gives $\bigl((S\Sigma\cup\Sigma)\setminus S\bigr)^*$ as the set of $\Ppos$-positions.
} \bqed
\end{example}
\begin{example}
    Let $\Sigma=\{a,b\}$ and $S=a^*b^*\setminus\{\varepsilon\}=a^+b^*\cup a^*b^+$.
    Then, $\mathtt{WordSubtraction}(S)$ is the game in which a player can remove any nonempty suffix consisting of an arbitrary number of $a$'s followed by an arbitrary number of $b$'s.
    One possible play is given by\[             ba\underline{abb}~\longrightarrow~b\underline{a}~\longrightarrow~\underline{b}~\longrightarrow~\varepsilon\,.
    \]
    In this example, the empty word $\varepsilon$ is the only terminal position.
More generally, $\varepsilon$ is the unique terminal position if and only if $\Sigma\subseteq S$.
Indeed, if every letter belongs to $S$, then every nonempty word has a legal move obtained by removing its last letter.
Conversely, if some letter $a\in\Sigma$ does not belong to $S$, then the one-letter word $a$ is a terminal position.
    \bqed
\end{example}
\begin{example}
    If $\Sigma$ is \emph{unary}, that is, $\Sigma$ consists of only one letter $\bullet$, then $\mathtt{WordSubtraction}(S)$ is equivalent to $\mathtt{Subtraction}(S')$ where\[
    S'=\{n\in\mathbb{N}_+:\bullet^n\in S\}.
    \]
    For example, $\mathtt{WordSubtraction}(\{\bullet,\bullet\bullet\})$ is equivalent to $\mathtt{Subtraction}(\{1,2\})$.\bqed
\end{example}

Note that the sets of $\Ppos$- and $\Npos$-positions are also languages over $\Sigma$.
We denote these languages by $L_{\Ppos}(S)$ and $L_{\Npos}(S)$, respectively.
By the Partition Theorem, these languages satisfy
\[
    L_{\Ppos}(S)=\bigl(L_{\Npos}(S)\bigr)^c.
\]
Moreover, a word is an $\Npos$-position if and only if it can be written as $ps$, where $p$ is a $\Ppos$-position and $s\in S$.
Thus,
\[
    L_{\Npos}(S)=L_{\Ppos}(S)S.
\]
The central theme of this note is to understand the relationship between the subtraction set $S$ and the resulting $\Ppos$- and $\Npos$-position languages.
In particular, we ask the following question:
\begin{quote}
\textbf{For a given class $\mathcal{S}$ of languages, what can we say about $L_\Ppos(S)$ and $L_{\Npos}(S)$ when $S\in \mathcal{S}$?}
\end{quote}

In this note, we mainly consider word subtraction games in which suffixes are removed. 
A prefix-removal version can also be defined, but it is essentially equivalent to the suffix-removal version: reversing every position and every word in the subtraction set gives a direct correspondence between the two games. 
Moreover, many standard language classes are closed under reversal. 
Thus, for our main purpose of studying how the language class of the subtraction set is related to those of the resulting $\Ppos$- and $\Npos$-position languages, there is little essential difference between prefix and suffix removal. 
Therefore, we take suffix removal as our basic convention.

\subsection*{Exercises}
\begin{exercise}
    Let $\Sigma=\{a,b\}$.
    Give a brief description of each of the following languages:
\begin{enumerate}
    \item $a^*$
    \item $(ab)^+$
    \item $(a\Sigma)^*$
    \item $(\Sigma^3)^*\Sigma$
    \item $(b^*aa)^*$\diaqed
\end{enumerate}
\end{exercise}
\begin{exercise}
    Let $\Sigma=\{a,b\}$.
    Determine whether $aabaa$ is a $\Ppos$-position or an $\Npos$-position in $\mathtt{WordSubtraction}(S)$ with $S=\{a,b,aa\}\subseteq\Sigma^+$.\diaqed
\end{exercise}
\begin{exercise}\label{ex1:a*b*}
    Let $\Sigma=\{a,b\}$ and consider $\mathtt{WordSubtraction}(S)$ with $S=a^+b^*\cup a^*b^+\subseteq\Sigma^*$.
    \begin{enumerate}
        \item Verify that $ba$ is a $\Ppos$-position.
        \item Verify that $baabba$ is a $\Ppos$-position.
        \item Prove that a word $w\in \Sigma^*$ is a $\Ppos$-position if and only if $w\in\Sigma^*ba\cup\{\varepsilon\}$. (\textbf{Hint}: Any word that does not contain $ba$ as a factor belongs to $a^*b^*$.)\diaqed
    \end{enumerate}
\end{exercise}

\begin{exercise}\label{ex1:monochromatic-subtraction}
Let $\Sigma=\{a,b\}$ and $S=a^+\cup b^+$.
Use the Partition Theorem to show that $L_{\Ppos}(S)=(a^+b\cup b^+a)^*$.\diaqed
\end{exercise}

\begin{starexercise}\label{ex1:additional-moves}
Let $L\subseteq\Sigma^*$ be
any language with $\Sigma=\{a,b\}$, and let
\[
    S'=a^+\cup b^+\cup Laa.
\]
Show that
\[
    L_{\Ppos}(S')=L_{\Ppos}(a^+\cup b^+).
\]
(\textbf{Hint}: Every position from which a suffix
in $Laa$ can be removed is already an $\Npos$-position
of $\mathtt{WordSubtraction}(a^+\cup b^+)$.
Consider how the additional moves affect the two
conditions in the Partition Theorem.)

Together with the previous exercise, this shows that
\[
    L_{\Ppos}(a^+\cup b^+\cup Laa)
    =(a^+b\cup b^+a)^*
\]
for every language $L\subseteq\{a,b\}^*$.
The $\Ppos$-position language is fixed and simple:
it is regular, in the terminology introduced in the next section.
Nevertheless, the subtraction set can be arbitrarily complex, depending on $L$.
Thus, the simplicity of a $\Ppos$-position language alone does not imply simplicity of the subtraction set.
\diaqed
\end{starexercise}

\begin{exercise}
    Let $\Sigma=\{a,b\}$.
    Find an infinite subtraction set $S\subseteq\Sigma^+$ such that there is a constant $C\ge 0$ for which every position has at most $C$ options.\diaqed
\end{exercise}
\begin{exercise}\label{ex-sufffree}
    A language $S\subseteq\Sigma^*$ is called \emph{suffix-free} if no word in $S$ is a proper suffix of another word in $S$.
    For example, $S=ba^*$ with $\Sigma=\{a,b\}$ is suffix-free.
    Prove that $S\subseteq\Sigma^+$ is suffix-free if and only if every position in $\mathtt{WordSubtraction}(S)$ has at most one option.\diaqed
\end{exercise}
\begin{starexercise}\label{ex1:density}
Let $\Sigma$ be an alphabet.
For a language $L\subseteq\Sigma^*$, the value
\[
    d_n(L)=\frac{\#(L\cap\Sigma^n)}{\#(\Sigma^n)}
\]
is the proportion of words of length $n$ that belong to $L$.
Equivalently, it is the probability that a uniformly chosen word of length $n$ belongs to $L$.
We define the \emph{density} of $L$ by
\[
    \delta(L)
    =
    \lim_{n\to\infty}
    \frac{1}{n}\sum_{k=1}^{n}d_k(L),
\]
provided that the limit exists.
Thus, the density measures the limiting average of these proportions over word lengths.
For example, the language of even-length words has density $1/2$.
Note that this limit need not exist for every language. 
For more details on the density of languages, see, e.g., Section~III.6 of \cite{DBLP:books/daglib/0067812}.

Now let $S\subseteq\Sigma^+$ satisfy $\Sigma\subseteq S$.
For each $n\geq 0$, put
\[
    p_n=d_n\bigl(L_{\Ppos}(S)\bigr).
\]
\begin{enumerate}
    \item Show that $p_{n+1}\leq 1-p_n$ for every $n\geq 0$.
    \item Show that
    \[
        \frac{1}{n}\sum_{k=1}^{n}p_k\leq\frac12
    \]
    for every $n\geq 1$.
    Hence, if $\delta(L_{\Ppos}(S))$ exists, then
    $\delta(L_{\Ppos}(S))\leq 1/2$.
\end{enumerate}
Therefore, in this density-based sense, the first player is at no disadvantage when $\Sigma\subseteq S$.\diaqed
\end{starexercise}

\newpage
\section{Finite Subtraction Sets}\label{sec:fin}
\subsection{Motivation}\label{subsec:fin-motivation}
In this section, we study word subtraction games with finite subtraction sets.
For ordinary subtraction games, it is well known that when the subtraction set is finite, the set of $\Ppos$-positions is ultimately periodic (see, e.g., Theorem~2.4 in Chapter~IV of~\cite{CGT}).
A similar phenomenon occurs for word subtraction games: for a finite subtraction set $S$, the language $L_{\Ppos}(S)$ has a certain form of ``periodicity.''
However, there is an important difference.
In word subtraction games, positions are words rather than nonnegative integers, and their outcomes can depend on the arrangement of letters, not just on their length.
What might play the role of ultimate periodicity in this setting?
Finite automata and regular languages provide a natural framework for answering this question.
Before introducing these notions formally, let us first examine the concrete example $S=\{a,b,aa\}$.

For a fixed subtraction set $S$, we introduce a useful notation called the \emph{$\Ppos/\Npos$ pattern}.
For a word $w$, we record the $\Ppos/\Npos$ status of each prefix by placing a $\Ppos$ or $\Npos$ symbol at each boundary between letters, including the beginning and the end of the word.
More precisely, for a word $w=a_1a_2\cdots a_n$ with $a_1,a_2,\ldots,a_n\in\Sigma$, its \emph{$\Ppos/\Npos$ pattern}, denoted by $\mathsf{pat}(w)$, is the sequence
\[
\mathsf{pat}(w)=\langle \theta_0,a_1,\theta_1,a_2,\ldots,a_n,\theta_n\rangle,
\]
where\[
\theta_i=\begin{cases}
    \Ppos&\text{if the prefix $a_1\cdots a_i$ is a $\Ppos$-position},\\
    \Npos&\text{if the prefix $a_1\cdots a_i$ is an $\Npos$-position}
\end{cases}
\] for each $0\le i\le n$. Note that $\theta_0=\Ppos$ because $\varepsilon$ is always a $\Ppos$-position.

Let us see how the $\Ppos/\Npos$ pattern is computed in a concrete example.
Consider the subtraction set $S=\{a,b,aa\}$ and the word $w=aabaa$.
Since the pattern is computed from left to right, let us start with $\mathsf{pat}({\color{red}\varepsilon})=\langle {\color{red}\Ppos}\rangle$.
Next, the pattern of $a=a_1$ is
\[
\mathsf{pat}({\color{red}a})=\langle {\color{blue}\Ppos},a,{\color{red}\Npos}\rangle
\]
because the only option from $a$ is $\varepsilon$, which is a
$\Ppos$-position, as indicated by the last entry of $\mathsf{pat}(\varepsilon)$. Hence $a$ is an $\Npos$-position.
Next,
\[
\mathsf{pat}({\color{blue} a}{\color{red}a})=\langle {\color{blue}\Ppos},a,{\color{blue}\Npos},a,{\color{red}\Npos}\rangle.
\]
The options from $aa$ are $\varepsilon$ and $a$, whose $\Ppos/\Npos$ statuses are $\Ppos$ and $\Npos$, respectively. 
Because a move to the $\Ppos$-position $\varepsilon$ is possible, $aa$ is an $\Npos$-position.
We then obtain
\[
\mathsf{pat}(a{\color{blue}a}{\color{red}b})=\langle \Ppos,a,\Npos,a,{\color{blue}\Npos},b,{\color{red}\Ppos}\rangle.
\]
The only option from $aab$ is $aa$, which is an $\Npos$-position.
Thus, $aab$ is a $\Ppos$-position.
On the other hand, we have
\[
\mathsf{pat}(aa{\color{blue}b}{\color{red}a})=\langle \Ppos,a,\Npos,a,\Npos,b,{\color{blue}\Ppos},a,{\color{red}\Npos}\rangle
\]
since $aaba$ has $aab$ as an option, and $aab$ is a
$\Ppos$-position.
Finally, we have
\[
\mathsf{pat}(aa{\color{blue}ba}{\color{red}a})
 =\langle \Ppos,a,\Npos,a,\Npos,b,{\color{blue}\Ppos},a,{\color{blue}\Npos},a,{\color{red}\Npos}\rangle
\]
because the options from $aabaa$ are $aab$ and $aaba$, and $aab$ is a $\Ppos$-position.

The key observation is that only a short final portion of the pattern is needed to compute the next status.
Let $w=a_1\cdots a_{n-1}$ with $n\geq 2$, and suppose that a new letter $a_n$ is appended to $w$.
Removing the final letter is always legal, and it leaves the prefix with status $\theta_{n-1}$.
The only other possible move is to remove $aa$.
This move is legal precisely when $a_{n-1}=a_n=a$, and it leaves the prefix with status $\theta_{n-2}$.
Thus, together with the new letter $a_n$, the last three entries
\[
    \langle{\color{blue}\theta_{n-2}},a_{n-1},{\color{blue}\theta_{n-1}}\rangle
\]
of $\mathsf{pat}(w)$ determine $\theta_n$.
After reading $a_n$, we retain
\[
    \langle{\color{blue}\theta_{n-1}},a_n,{\color{red}\theta_n}\rangle.
\]
Together with the next letter $a_{n+1}$, this information is sufficient to determine the status of $wa_na_{n+1}$.
The initial case is handled separately: the empty word has pattern $\langle\Ppos\rangle$, and either first letter produces an $\Npos$-position.

This observation can be illustrated as follows:
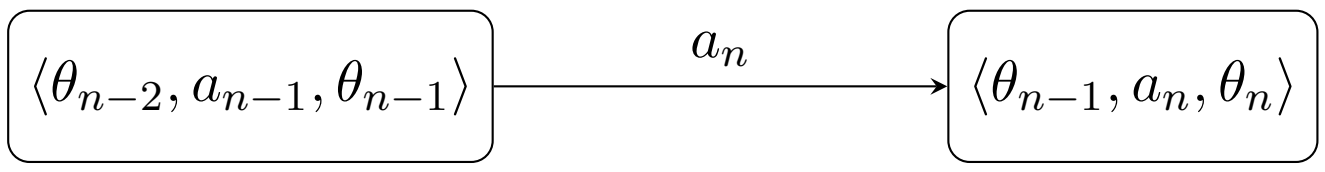
\begin{figure}[H]
\centering
\begin{tikzpicture}[
  ->,
  >=stealth,
  node distance=3cm,
  every state/.style={
    rectangle,
    rounded corners,
    draw,
    minimum height=1cm
  }
]
  \node[state] (q) {
    $\langle\theta_{n-2},a_{n-1},\theta_{n-1}\rangle$
  };
  \node[state] (qprime) [right=of q] {
    $\langle\theta_{n-1},a_n,\theta_n\rangle$
  };

  \path (q) edge node[above] {$a_n$} (qprime);
\end{tikzpicture}
\label{fig:pattern-update}
\end{figure}
\noindent
This figure shows that, when a letter $a_n$ is appended to the right of a word whose $\Ppos/\Npos$ pattern ends with $\langle \theta_{n-2},a_{n-1},\theta_{n-1}\rangle$,
the resulting word has a $\Ppos/\Npos$ pattern ending with $\langle \theta_{n-1},a_n,\theta_n\rangle$.
Notice that there are at most
\[
\#(\{\Ppos,\Npos\})
\cdot \#(\Sigma)
\cdot \#(\{\Ppos,\Npos\})
= 8
\]
possible triples of the form
\[
\langle\theta_{n-1},a_n,\theta_n\rangle.
\]
Together with the initial state $\langle\Ppos\rangle$, these triples give only finitely many possible states.
In fact, for $S=\{a,b,aa\}$, the $\Ppos$/$\Npos$ status of a word can be determined by the following finite-state diagram:
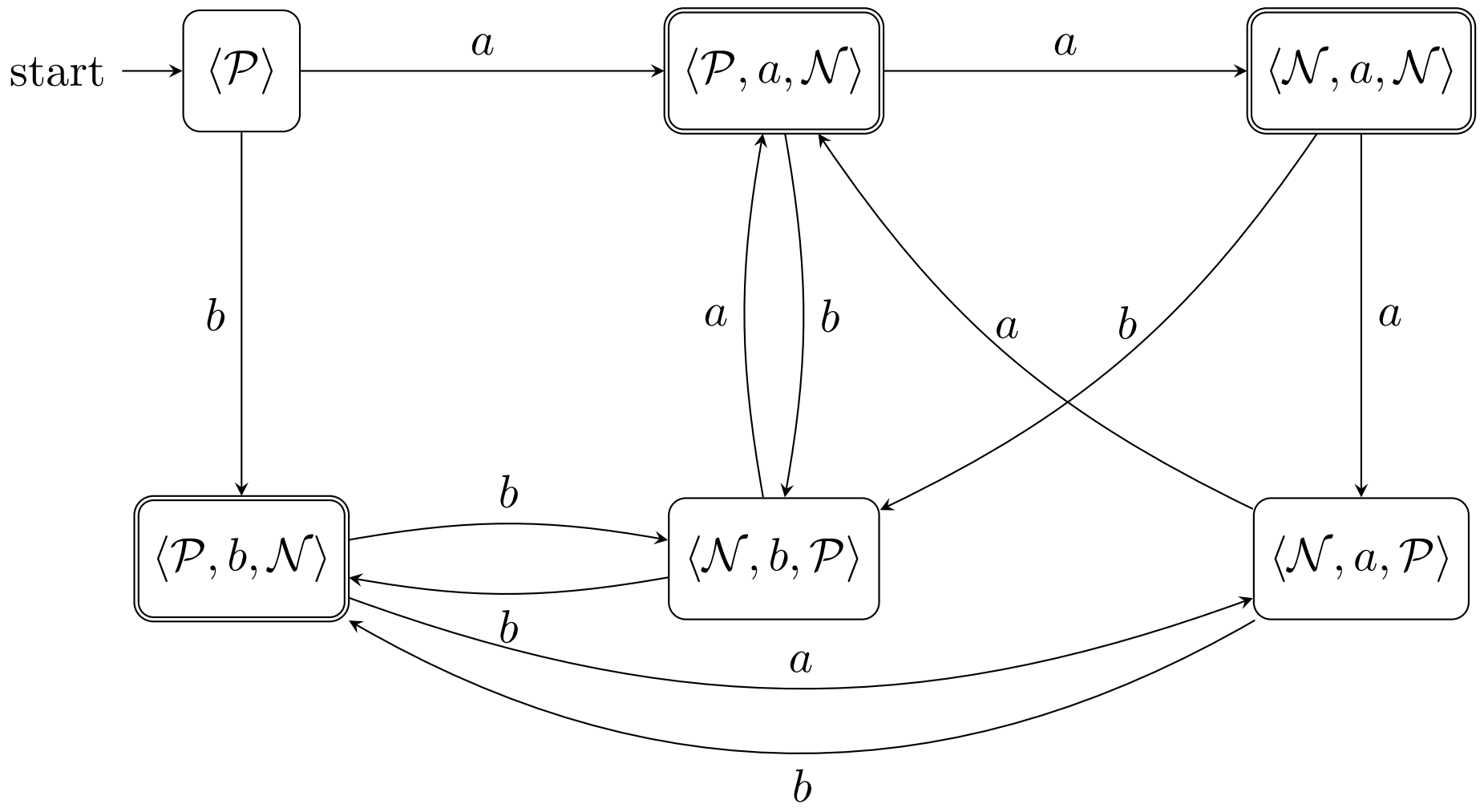
\begin{figure}[H]
\centering
\begin{tikzpicture}[
  ->,
  >=stealth,
  node distance=3cm,
  every state/.style={
    rectangle,
    rounded corners,
    draw,
    minimum height=1cm
  }
]

  \node[state, initial] (P) {$\langle\Ppos\rangle$};
\node[state, accepting] (PaN) [right=of P]
  {$\langle\Ppos,a,\Npos\rangle$};
\node[state, accepting] (PbN) [below=of P]
  {$\langle\Ppos,b,\Npos\rangle$};

\node[state, accepting] (NaN) [right=of PaN]
  {$\langle\Npos,a,\Npos\rangle$};
\node[state] (NaP) [below=of NaN]
  {$\langle\Npos,a,\Ppos\rangle$};

\node[state] (NbP) [below=of PaN]
  {$\langle\Npos,b,\Ppos\rangle$};

  \path
    (P) edge node[above] {$a$} (PaN)
        edge node[left] {$b$} (PbN)

    (PaN) edge node[above] {$a$} (NaN)
           edge[bend left=10] node[right] {$b$} (NbP)

    (PbN) edge[bend right=20] node[above] {$a$} (NaP)
           edge[bend left=10] node[above] {$b$} (NbP)

    (NaN) edge node[right] {$a$} (NaP)
           edge[bend left=15] node[above] {$b$} (NbP)

    (NaP) edge[bend left=15] node[above] {$a$} (PaN)
           edge[bend left=30] node[below] {$b$} (PbN)

    (NbP) edge[bend left=10] node[left] {$a$} (PaN)
           edge[bend left=10] node[below] {$b$} (PbN);
\end{tikzpicture}
\caption{A finite-state diagram for $\mathtt{WordSubtraction}(\{a,b,aa\})$.}
\label{fig:ws-example-dfa}
\end{figure}
\noindent
The diagram should be read as follows.
Let $w=a_1a_2\cdots a_n$ be the word under consideration.
Starting from the state $\langle \Ppos\rangle$ in the upper left, follow the edges labeled
$a_1,a_2,\ldots,a_n$ in order.
If the final state is marked with a double border, then $w$ is an $\Npos$-position; otherwise, $w$ is a $\Ppos$-position.
For example, when the input is $aabaa$, we start at
$\langle\Ppos\rangle$ and successively visit
\[
\langle\Ppos,a,\Npos\rangle,\quad
\langle\Npos,a,\Npos\rangle,\quad
\langle\Npos,b,\Ppos\rangle,\quad
\langle\Ppos,a,\Npos\rangle,\quad
\langle\Npos,a,\Npos\rangle,
\]
and the corresponding \emph{run} is\[
\langle\Ppos\rangle\xrightarrow{a}
\langle\Ppos,a,\Npos\rangle\xrightarrow{a}
\langle\Npos,a,\Npos\rangle\xrightarrow{b}
\langle\Npos,b,\Ppos\rangle\xrightarrow{a}
\langle\Ppos,a,\Npos\rangle\xrightarrow{a}
\langle\Npos,a,\Npos\rangle.
\]
Because the final state $\langle\Npos,a,\Npos\rangle$ is marked with a double border, we conclude that $aabaa$ is an $\Npos$-position.

The finite-state diagram we have just constructed is, in fact, a \emph{finite automaton}.
Although we have considered only the case $S=\{a,b,aa\}$, the same idea works for every nonempty finite subtraction set $S$.
Let $\ell$ be the maximum length of a word in $S$.
To determine the status after appending a new letter, it suffices to remember the last $2\ell-1$ entries of the current $\Ppos/\Npos$ pattern, or the entire pattern if it is shorter.
Indeed, this information identifies every removable suffix of the new word and the status of the position left after its removal.
Because the alphabet is finite, only finitely many such patterns are possible.
We can therefore construct a finite-state diagram that determines whether a given word is a $\Ppos$- or an $\Npos$-position.
This observation will be formalized in Theorem~\ref{th2-fin}.

We will now introduce finite automata more formally and examine some of their basic properties.
Also, we will explain how regular languages generalize the periodicity found in ordinary subtraction games.

\subsection{Finite Automata}
We have seen that, for a finite subtraction set, the $\Ppos/\Npos$ status of a word can be determined by reading it from left to right and updating a record with only finitely many possible states.
This naturally leads us to the notion of a \emph{finite
automaton}.
A finite automaton is a machine that classifies words using finitely many \emph{states}. 
As illustrated in Figure~\ref{fig:ws-example-dfa}, it can be represented by a directed graph whose edges are labeled by letters
from an alphabet. 
One state is distinguished as the \emph{initial state},
and zero or more states are designated as \emph{accepting states}.
Starting from the initial state, the automaton reads the input word from left to right, following the edge labeled by each successive letter. 
After the entire word has been read, the word is \emph{accepted} if the automaton is in an accepting state, and \emph{rejected} otherwise.
The formal definition is as follows:
\begin{definition}[Deterministic Finite Automata]
A \emph{deterministic finite automaton} (\emph{DFA} for short) is a $5$-tuple
\[
\mathbf{A}=(Q,\Sigma,\delta,q_0,F),
\]
where\begin{itemize}
    \item $Q$ is a finite set of \emph{states},
    \item $\Sigma$ is an alphabet,
    \item $\delta:Q\times \Sigma\to Q$ is the \emph{transition function},
    \item $q_0\in Q$ is the \emph{initial state}, and
    \item $F\subseteq Q$ is the set of \emph{accepting states} (or \emph{final states}).
\end{itemize}
For a word $w=a_1a_2\cdots a_n\in\Sigma^*$, its \emph{run} on
$\mathbf{A}$ is the sequence of states
\[
q_0,q_1,\ldots,q_n,
\]
where $q_0$ is the initial state and
\[
q_i=\delta(q_{i-1},a_i)
\]
for each $1\le i\le n$.
The run can also be written as
\[
q_0\xrightarrow{a_1}q_1\xrightarrow{a_2}\cdots
\xrightarrow{a_n}q_n.
\]
We say that $\mathbf{A}$ \emph{accepts} $w$ if $q_n\in F$, and
\emph{rejects} $w$ otherwise. 
In particular, $\varepsilon$ is accepted if and only if $q_0\in F$.
The language\[
L(\mathbf{A})=\{w\in\Sigma^*: \mathbf{A}\text{ accepts }w\}
\] is called the \emph{language recognized (or accepted) by $\mathbf{A}$}.\bqed
\end{definition}
The word ``deterministic'' means that, for each input word, there is a unique run on the automaton.
Indeed, from any state $q_{i-1}\in Q$ and for each input letter $a_i\in \Sigma$, there is exactly one possible next state $\delta(q_{i-1},a_i)\in Q$. 
Thus, there is exactly one run for every input word.

Note that the transition function $\delta$ naturally extends from
letters to words. Define
\[
\delta^*:Q\times\Sigma^*\to Q
\]
recursively by
\begin{eqnarray*}
    \delta^*(q,\varepsilon)&=&q,\\
    \delta^*(q,wa)&=&\delta(\delta^*(q,w),a)
\end{eqnarray*}
for each $q\in Q$, $w\in\Sigma^*$, and $a\in\Sigma$.
By definition,
\[
w\in L(\mathbf A)
\quad\Longleftrightarrow\quad
\delta^*(q_0,w)\in F
\]
holds.
Moreover,
\[
    \delta^*(q,uv)=\delta^*(\delta^*(q,u),v)
\]
for all $q\in Q$ and $u,v\in\Sigma^*$.
In other words, reading $uv$ is the same as first reading $u$ and then reading $v$.
Together with $\delta^*(q,\varepsilon)=q$, this identity says that $\delta^*$ defines a right action of the free monoid $\Sigma^*$ on $Q$.

As illustrated in Figure~\ref{fig:ws-example-dfa}, a finite automaton can be visualized as a directed graph:
the states are vertices, and each transition $\delta(q,a)=q'$ is represented by an edge from $q$ to $q'$ labeled by $a$. 
The initial state is usually indicated by an arrow coming from nowhere, while accepting states are distinguished by double circles.

\begin{example}\label{ex-DFAs}
The following examples are DFAs with $\Sigma=\{a,b\}$:
\begin{itemize}
\item The finite language
\[
L_1=\{a,b,aa\}
\]is recognized by the DFA $\mathbf{A}_1=(\{q_0,q_{a},q_{b},q_{aa},q_d\},\Sigma,\delta_1,q_0,\{q_{a},q_{b},q_{aa}\})$ defined as follows:
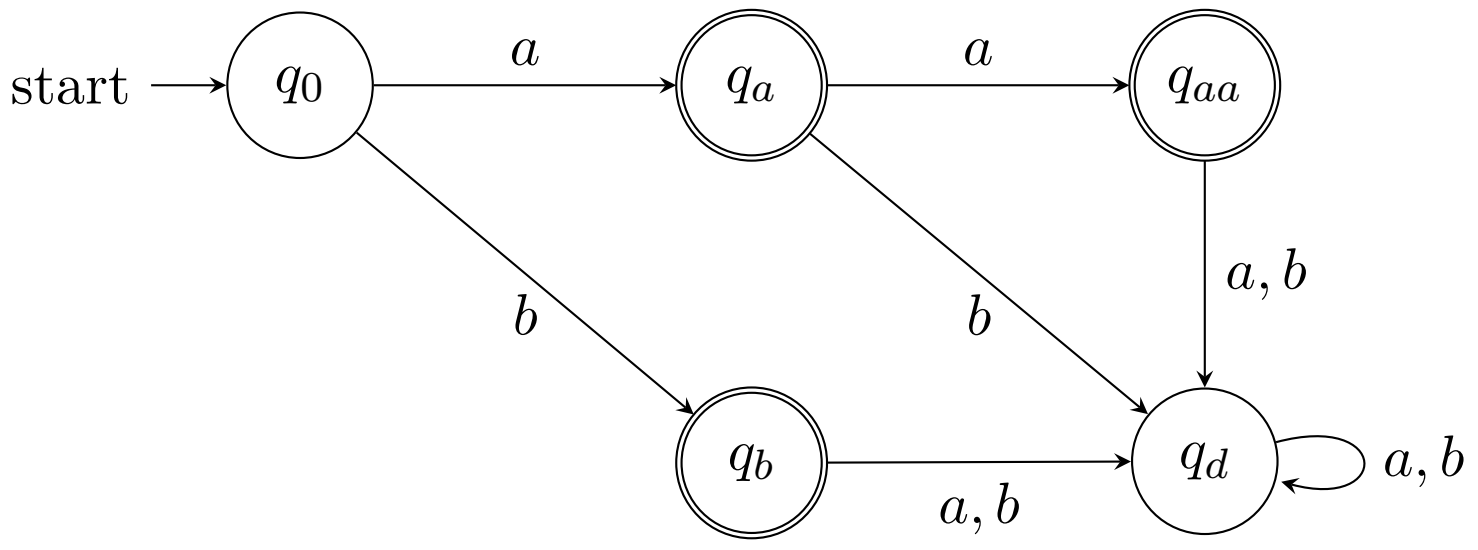
\begin{figure}[H]
\centering
\begin{tikzpicture}[
  ->,
  >=stealth,
  node distance=2cm
]
  \node[state, initial] (q0) {$q_0$};
  \node[state, accepting] (q1) [right=of q0] {$q_a$};
  \node[state, accepting] (q2) [right=of q1] {$q_{aa}$};
  \node[state, accepting] (q3) [below=1.5cm of q1] {$q_b$};
  \node[state] (qd) [below=1.5cm of q2] {$q_d$};

  \path
    (q0) edge node[above] {$a$} (q1)
         edge node[below] {$b$} (q3)
    (q1) edge node[above] {$a$} (q2)
         edge node[below] {$b$} (qd)
    (q2) edge node[right] {$a,b$} (qd)
    (q3) edge node[below] {$a,b$} (qd)
    (qd) edge[loop right] node {$a,b$} ();
\end{tikzpicture}
\end{figure}
More generally, every finite language is recognized by some DFA.
\item The language
\[
L_2=(\Sigma\Sigma)^*
\]
of even length words is recognized by the DFA $\mathbf{A}_2=(\{q_{\mathrm{even}},q_{\mathrm{odd}}\},\Sigma,\delta_2,q_{\mathrm{even}},\{q_{\mathrm{even}}\})$ defined as follows:
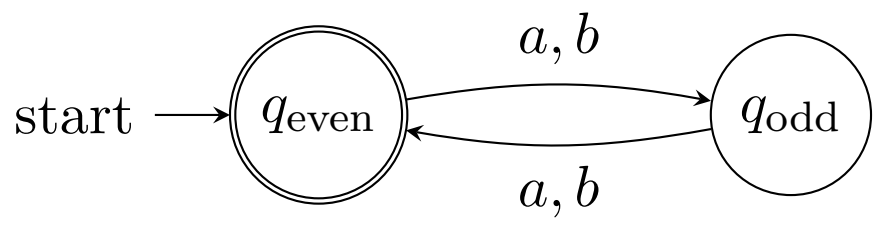
\begin{figure}[H]
\centering
\begin{tikzpicture}[
  ->,
  >=stealth,
  node distance=2cm
]
  \node[state, initial, accepting] (q0) {$q_{\mathrm{even}}$};
  \node[state] (q1) [right=of q0] {$q_{\mathrm{odd}}$};

  \path
    (q0) edge[bend left=10] node[above] {$a,b$} (q1)
    (q1) edge[bend left=10] node[below] {$a,b$} (q0);
\end{tikzpicture}
\end{figure}
More generally, for every integer $k\geq 1$, the language $(\Sigma^k)^*$ is recognized by a DFA that keeps track of the input length modulo $k$.

\item The language
\[
L_3=a\Sigma(\Sigma\Sigma)^*\cup b(\Sigma\Sigma)^*
\]
is recognized by the DFA $\mathbf{A}_3=(\{q_0,q_1,q_2\},\Sigma,\delta_3,q_0,\{q_{2}\})$ defined as follows:
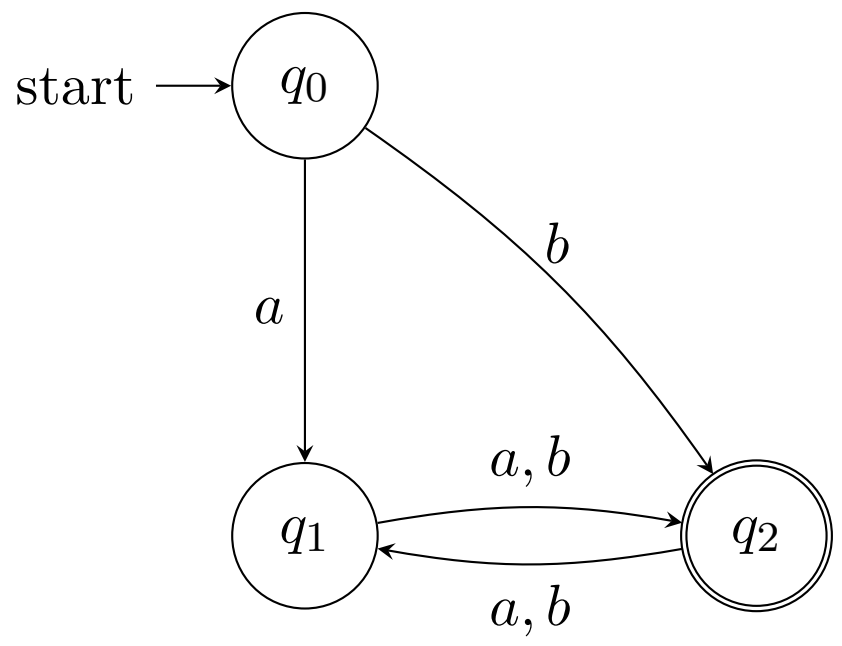
\begin{figure}[H]
\centering
\begin{tikzpicture}[
  ->,
  >=stealth,
  node distance=2cm
]
  \node[state, initial] (q0) {$q_0$};
  \node[state] (q1) [below=of q0] {$q_1$};
  \node[state,accepting] (q2) [right=of q1] {$q_2$};

  \path
    (q0) edge node[left] {$a$} (q1)
    (q0) edge[bend left=10] node[above] {$b$} (q2)
    (q1) edge[bend left=10] node[above] {$a,b$} (q2)
    (q2) edge[bend left=10] node[below] {$a,b$} (q1);
\end{tikzpicture}
\end{figure}

\item The language
\[
L_4=a^*b^*
\]
is recognized by the DFA $\mathbf{A}_4=(\{q_0,q_1,q_d\},\Sigma,\delta_4,q_0,\{q_0,q_1\})$ defined as follows:
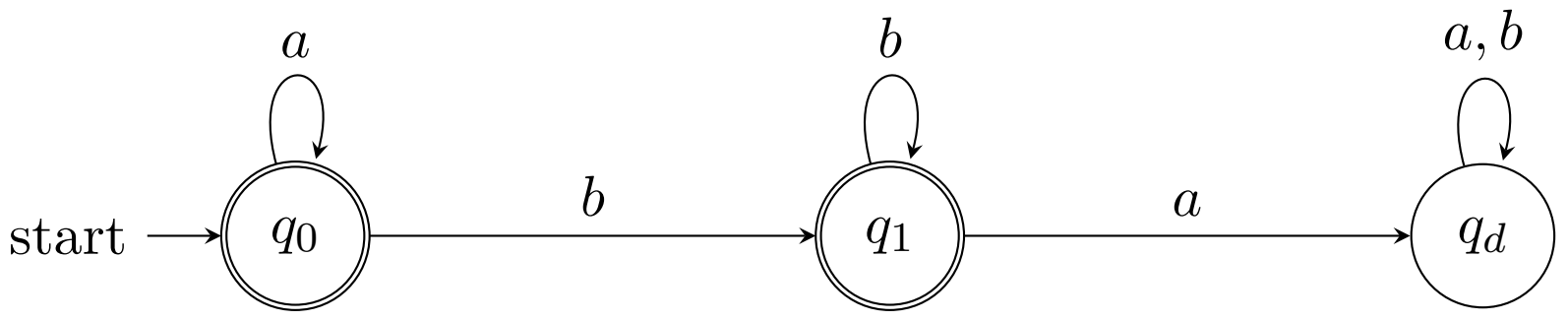
\begin{figure}[H]
\centering
\begin{tikzpicture}[
  ->,
  >=stealth,
  node distance=3cm
]
  \node[state, initial, accepting] (q0) {$q_0$};
  \node[state, accepting] (q1) [right=of q0] {$q_1$};
  \node[state] (qd) [right=of q1] {$q_d$};

  \path
    (q0) edge[loop above] node {$a$} ()
         edge node[above] {$b$} (q1)
    (q1) edge[loop above] node {$b$} ()
         edge node[above] {$a$} (qd)
    (qd) edge[loop above] node {$a,b$} ();
\end{tikzpicture}
\end{figure}
On the other hand, the complement of $L_4=a^*b^*$,
\[
(L_4)^c=\Sigma^*ba\Sigma^*,
\]
is recognized by the DFA $\mathbf{A}_4'=(\{q_0,q_1,q_d\},\Sigma,\delta_4,q_0,\{q_d\})$ defined as follows:
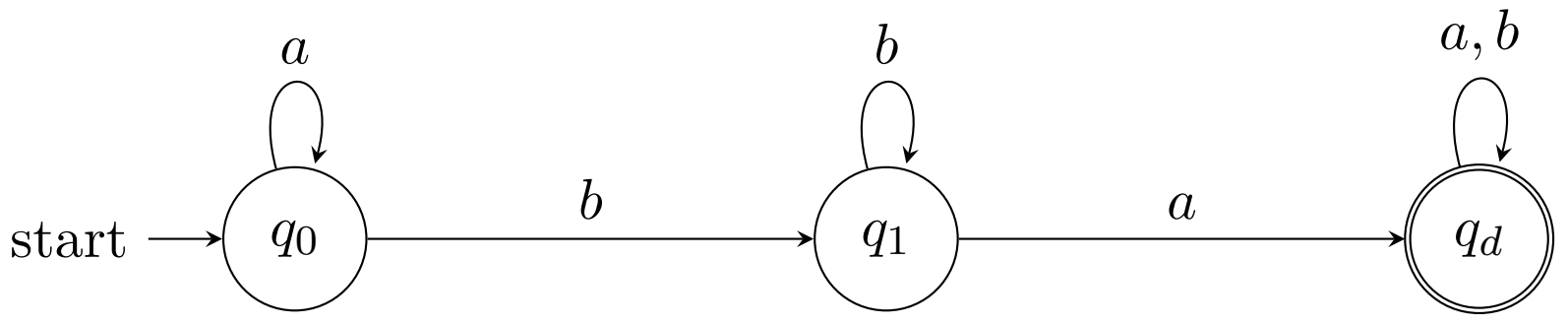
\begin{figure}[H]
\centering
\begin{tikzpicture}[
  ->,
  >=stealth,
  node distance=3cm
]
  \node[state, initial] (q0) {$q_0$};
  \node[state] (q1) [right=of q0] {$q_1$};
  \node[state, accepting] (qd) [right=of q1] {$q_d$};

  \path
    (q0) edge[loop above] node {$a$} ()
         edge node[above] {$b$} (q1)
    (q1) edge[loop above] node {$b$} ()
         edge node[above] {$a$} (qd)
    (qd) edge[loop above] node {$a,b$} ();
\end{tikzpicture}
\end{figure}
More generally, suppose that a language $L\subseteq \Sigma^*$ is recognized by a DFA
\[
\mathbf{A}=(Q,\Sigma,\delta,q_0,F).
\]
Then, its complement $L^c\subseteq \Sigma^*$ is recognized by the DFA
\[
\mathbf{A}'=(Q,\Sigma,\delta,q_0,Q\setminus F).
\]
In other words, we can obtain a DFA for the complement by switching the accepting and rejecting states.\bqed
\end{itemize}
\end{example}
As these examples illustrate, DFAs can recognize languages of many different forms. On the other hand, as we will see later, there exist languages that cannot be recognized by any DFA. This motivates the
following definition.
\begin{definition}[Regular Languages]
A language $L$ is called \emph{regular} if there exists a deterministic
finite automaton $\mathbf{A}$ such that $
L=L(\mathbf{A})$.\bqed
\end{definition}
For example, the languages $L_1$, $L_2$, $L_3$, $L_4$, and $(L_4)^c$ considered above are all regular. 
Moreover, as we saw earlier, the language $L_\Npos(\{a,b,aa\})$ of $\Npos$-positions of $\mathtt{WordSubtraction}(\{a,b,aa\})$ can be described by a finite automaton, and hence is regular as well.

Regular languages also have a number of useful closure properties.
For example, as observed at the end of Example~\ref{ex-DFAs}, the complement of a regular language is again regular. 
(Therefore, $L_\Ppos(\{a,b,aa\})$ is also regular.) 
More generally, regular languages are closed under union, intersection, concatenation, and Kleene star, as well as under images and inverse images of homomorphisms between free monoids.
These properties make regular languages particularly convenient to work with, and we will return to some of them when they become relevant.

\subsection{Periodic Structure of Regular Languages}
We have seen that the set of $\Npos$-positions of
$\mathtt{WordSubtraction}(\{a,b,aa\})$ is regular.
What does this tell us about these positions?
In particular, how is regularity related to the
periodicity found in ordinary subtraction games?

In this section, we look at two remarkable phenomena of regular languages. 
The first is the so-called \emph{pumping lemma}, which shows that sufficiently long words in a regular language contain a factor that can be repeated any number of times.
The second is \emph{semilinearity},
which appears when we forget the order of the letters in a word and record only the number of occurrences of each letter.

Let us first look at the pumping lemma. Suppose that $L\subseteq\Sigma^*$ is regular, and hence is recognized by some DFA
\[
\mathbf{A}=(Q,\Sigma,\delta,q_0,F).
\]
Let $N=\#(Q)$.
Consider an accepting run of a word $w\in L$:
\[
q_0 \overbrace{\rightarrow \cdots \rightarrow}^{w} q_n=q_f.
\]
By the assumption that $w\in L$, the final state $q_f$ is accepting.

Now suppose that, during this run, $\mathbf{A}$ reads a factor
$w'$ of $w$ whose length is at least $N$. In other words, the run has the form
\[
q_0
\overbrace{\rightarrow \cdots \rightarrow}^{u}
{\color{red} q_j
\overbrace{\rightarrow \cdots \rightarrow}^{w'}
q_{j'}}
\overbrace{\rightarrow \cdots \rightarrow}^{v} q_f
\]
with some $u,w',v\in \Sigma^*$ where $|w'|\ge N$.
By the assumption $|w'|\ge N$, the corresponding segment of the run contains at least $N+1$ state occurrences:
\[
q_j,q_{j+1},\ldots,q_{j'}.
\]
On the other hand, there are only $N$ states in total. Thus, by the
\emph{pigeonhole principle}, two of these states must be equal. That is, there
exist indices $j\le k<k'\le j'$ such that $q_k=q_{k'}$.

Let $y$ be the factor of $w'$ read between these two occurrences.
Then the run contains a loop of the form
\[
{\color{red}q_k\overbrace{\rightarrow \cdots \rightarrow}^{y}q_k}.
\]
Thus, the accepting run can be written as
\[
q_0
\overbrace{\rightarrow \cdots \rightarrow}^{u}
q_j
\underbrace{\overbrace{\rightarrow \cdots \rightarrow}^{x}
{\color{red} q_k
\overbrace{\rightarrow \cdots \rightarrow}^{y}
q_k}
\overbrace{\rightarrow \cdots \rightarrow}^{z}}_{w'} q_{j'}
\overbrace{\rightarrow \cdots \rightarrow}^{v} q_f
\]
with some $x,z\in\Sigma^*$.

The important point is that this loop can be removed, or repeated any number of times, without changing the state reached after reading $w'$. Indeed, both
\[
q_0
\overbrace{\rightarrow \cdots \rightarrow}^{u}
q_j
\overbrace{\rightarrow \cdots \rightarrow}^{x}
{\color{red} q_k}
\overbrace{\rightarrow \cdots \rightarrow}^{z} q_{j'}
\overbrace{\rightarrow \cdots \rightarrow}^{v} q_f
\]
and
\[
q_0
\overbrace{\rightarrow \cdots \rightarrow}^{u}
q_j
\overbrace{\rightarrow \cdots \rightarrow}^{x}
{\color{red} q_k\overbrace{\rightarrow \cdots \rightarrow}^{y}
q_k\overbrace{\rightarrow \cdots \rightarrow}^{y}
q_k\rightarrow \cdots \rightarrow q_k\overbrace{\rightarrow \cdots \rightarrow}^{y}
q_k}
\overbrace{\rightarrow \cdots \rightarrow}^{z} q_{j'}
\overbrace{\rightarrow \cdots \rightarrow}^{v} q_f
\]
are valid runs.
In every case, the run still starts at $q_0$ and ends at the accepting
state $q_f$. 
Therefore, for every $i\ge0$,
\[
uxy^izv\in L(\mathbf{A})
\]
holds.
This is precisely the conclusion of the pumping lemma.
\begin{theorem}[Pumping Lemma for Regular Languages]\label{th2-pump_reg}
    Let $L\subseteq \Sigma^*$ be a regular language. Then there exists a positive integer $N\ge 1$ such that, whenever
$uw'v\in L$ for some $u,w',v\in\Sigma^*$ with $|w'|\ge N$, there exist
$x,y,z\in\Sigma^*$ satisfying
\begin{itemize}
  \item $w'=xyz$,
  \item $|y|>0$, and
  \item $uxy^izv\in L$ for every $i\ge0$.\bqed
\end{itemize}
\end{theorem}
It is worth noting what the pumping lemma tells us intuitively.
If $L$ is finite, choose a positive integer $N$ larger than the length of every word in $L$.
Then the condition in the pumping lemma is automatically satisfied because there is no word $uw'v\in L$ with $|w'|\geq N$.

On the other hand, if $L$ is infinite, the pumping lemma says that $L$ must contain a language of the form
\[
\{uxy^izv:i\ge0\}
\]
for some fixed words $u,x,y,z,v$ with $|y|>0$. 
Roughly speaking, every infinite regular language contains an analogue of an infinite arithmetic progression.
For example, we have the following:
\begin{proposition}
The language
\[
L_{\mathrm{prime}}=\{a^p : p\text{ is a prime number}\}
\]
over $\{a\}$ is not regular.
\end{proposition}
\begin{proof}
Suppose, for contradiction, that $L_{\mathrm{prime}}$ is regular. Then there exists a constant $N$ as given by the
pumping lemma.

Let $p_N$ be any prime number greater than $N$, and consider the word
\[
a^{p_N}=uw'v,
\quad
u=\varepsilon,\quad w'=a^{p_N},\quad v=\varepsilon.
\]
Because $uw'v\in L_{\mathrm{prime}}$ and $|w'|\ge N$ , the pumping lemma gives a factorization
\[
w'=xyz
\]
with $|y|>0$ such that
\[
uxy^izv\in L_{\mathrm{prime}}
\]
for every $i\ge0$. Since all words are over the unary alphabet,
there exist nonnegative integers $k_x,k_y,k_z$ such that
\[
x=a^{k_x},\qquad y=a^{k_y},\qquad z=a^{k_z},
\]
where
\[
k_x+k_y+k_z=p_N
\]
and $k_y>0$. Hence
\[
\{uxy^izv:i\ge0\}
=
\left\{
a^{(k_x+k_z)+ i\cdot k_y}:i\ge0
\right\}
\subseteq L_{\mathrm{prime}}.
\]
However, taking $i=p_N+1$ gives
\[
    (k_x+k_z)+(p_N+1)k_y
    =p_N(1+k_y),
\]
which is a composite number because $p_N>1$ and $k_y>0$.
This contradicts the conclusion that every word $uxy^izv$ belongs to $L_{\mathrm{prime}}$.
Therefore, $L_{\mathrm{prime}}$ is not regular.
\end{proof}
\noindent
Similarly, the languages $\{a^{n^2}:n\ge0\}$, $\{a^{n!}:n\ge0\}$, and
$\{a^{\lfloor n\cdot\varphi\rfloor}:n\ge0, \varphi=\frac{1+\sqrt{5}}{2}\}$ over the unary alphabet $\{a\}$ are not regular.
In fact, under the identification of $a^n$ with $n$, the regular languages over $\{a\}$ are exactly the ultimately periodic subsets of $\mathbb{N}$.
These observations can be described more elegantly using the notions of Parikh images and semilinear sets. 
Before turning to these notions, let us take a moment to see another use of the pumping lemma.

We have focused on the linear periodicity inherent in regular languages. 
But there is another important limitation:
a DFA has only finitely many states, so it cannot remember arbitrarily large counts.
A standard example is the language
\[
\{a^n b a^n:n\ge 0\},
\]
which is not regular.
To see the intuition, consider feeding a word of the form
\[
a^m b a^n
\]
into a DFA from left to right.
In order to accept exactly those words for which $m=n$,
the automaton must remember the length of the first block of $a$'s.
After reading the separator $b$, this information must still be available so that the automaton can determine whether the final block contains exactly the same number of $a$'s.
However, the length of the first block can be $100$, $1000$, or arbitrarily large.
A DFA has only finitely many states, and therefore cannot keep track of all these possibilities.
The pumping lemma makes this intuition precise:
\begin{proposition}\label{prop2-pump_a^nba^n}
The language
\[
\{a^n b a^n:n\ge 0\}
\]
over $\Sigma=\{a,b\}$ is not regular.
\end{proposition}
\begin{proof}
    Suppose, for contradiction, that $L=\{a^nb a^n:n\ge0\}$ is regular. 
    Then there exists a constant $N$ as given by the pumping lemma.
    Let $u=\varepsilon$, $w'=a^N$, and $v=ba^N$.
    Because $uw'v=a^Nba^N\in L$ and $|w'|\ge N$ , the pumping lemma gives a factorization $w'=xyz$ with $|y|>0$ such that
\[
uxy^izv\in L
\]
for every $i\ge0$. 
Taking $i=2$, the pumping lemma gives
\[
    uxy^2zv=a^{N+|y|}ba^N\in L.
\]
However, by $|y|>0$, the two blocks of $a$'s have different lengths.
Thus, this word does not belong to $L$, a contradiction.
Therefore, $L$ is not regular.
\end{proof}

Similarly, the language of nonempty \emph{palindromes}
\[
\mathsf{PAL}=\{w\in\{a,b\}^+: w=w^{\mathsf{rev}}\}
\]
is also not regular.
Here, $w^{\mathsf{rev}}$ denotes the word obtained by writing the letters of $w$ in reverse order.
For example, $(aab)^\mathsf{rev}=baa$.
\begin{proposition}\label{prop2-pump_pal}
The language $\mathsf{PAL}$ is not regular.
\end{proposition}
\begin{proof}
    Suppose, for contradiction, that $\mathsf{PAL}$ is regular. 
    Then there exists a constant $N$ as given by the pumping lemma.
    Let $u=\varepsilon$, $w'=a^N$, and $v=ba^N$.
    Note that $uw'v\in \mathsf{PAL}$ because $uw'v=a^Nba^N$ is a palindrome.
    Thus, for the same reason as in Proposition~\ref{prop2-pump_a^nba^n}, the pumping lemma implies
that
\[
a^{N+k_y}ba^N\in\mathsf{PAL}
\]
with some $k_y>0$. However, this word is not a palindrome, which is a contradiction.
\end{proof}

Finally, let us briefly discuss the semilinearity of regular
languages. 
We begin with the \emph{Parikh mapping}
\[
\Pk:\Sigma^*\to\mathbb{N}^{\Sigma}\]
defined as\[
\Pk(w)=(|w|_a)_{a\in\Sigma}
\]
for each $w\in\Sigma^*$,
where $|w|_a$ denotes the number of occurrences of the letter $a\in\Sigma$
in $w$. Thus, the Parikh mapping forgets the order of the letters in
a word and records only how many times each letter occurs. For
example,
\[
\Pk(aab)=\Pk(aba)=\Pk(baa)=(2,1)
\]
with $\Sigma=\{a,b\}$.
As with any mapping, we can consider the image of a language $L$ under
the Parikh mapping,
\[
\Pk(L)=\{\Pk(w):w\in L\}\subseteq\mathbb{N}^{\Sigma},
\]
which is called the \emph{Parikh image} of $L$.
For example,\[
\Pk(\{a,b,aa\})=\{(1,0),(0,1),(2,0)\},
\]
and\[
\Pk(\Sigma^*ba\Sigma^*)=\{(i,j):i,j\ge 1\}.
\]

Now, it turns out that the Parikh image of a regular language has a particularly simple form. 
A set $X\subseteq\mathbb{N}^n$ is called
\emph{linear} if there exist vectors
$\mathbf{b},\mathbf{p}_1,\ldots,\mathbf{p}_k\in\mathbb{N}^n$ with $k\ge0$ such that
\[
    X=
    \{\mathbf{b}+i_1\mathbf{p}_1+\cdots+i_k\mathbf{p}_k
      : i_1,\ldots,i_k\in\mathbb{N}\}.
\]
For example,\[
\{(i,j):i,j\ge 1\}=\{(1,1)+i_1(1,0)+i_2(0,1):i_1,i_2\ge 0\}
\]
is a linear set.
A finite union of linear sets is called a \emph{semilinear set}.
For example, $\{(1,0),(0,1),(2,0)\}$ is a semilinear set because it is the union of linear sets $\{(1,0)\},\{(0,1)\}$, and $\{(2,0)\}$.
Then, we have the following fundamental result.
\begin{theorem}[Semilinearity of Regular Languages]
The Parikh image of every regular language is a semilinear set.\bqed
\end{theorem}
We omit the proof here.
The theorem gives another perspective on the periodic structure of regular languages: after forgetting the order of letters, the resulting set of vectors is a finite union of linear sets.

It is worth noting that neither the pumping lemma nor
semilinearity gives a sufficient condition for regularity. They only
provide necessary conditions. For example, the Parikh image of
\[
\{a^n ba^n:n\ge0\}
\]
is
\[
\{(2 i,1):i\ge0\}=\{(0,1)+i (2,0):i\ge 0\},
\]
which is a semilinear set. Nevertheless,
as we saw above, $\{a^n ba^n:n\ge0\}$ is not regular.

\subsection{Back to Word Subtraction}\label{subsec:B2WS_fin}
We have introduced the basic notions of DFAs and regular languages. 
Let us return to word subtraction games and summarize what we have learned in this section.

As we briefly saw in Section~\ref{subsec:fin-motivation}, when the subtraction set $S$ is finite, the set of $\Npos$-positions of $\mathtt{WordSubtraction}(S)$ can be recognized by a DFA, and hence is a regular language. 
The basic idea is exactly the same as in the example from
Section~\ref{subsec:fin-motivation}. 
When $S$ is finite, the $\Ppos$/$\Npos$ pattern of a word can be updated by keeping track of only finitely much information.
More precisely, for nonempty $S$, let
\[
    \ell=\max\{|s|:s\in S\}.
\]
Together with the next input letter, the last $2\ell-1$ entries of the pattern, or the entire pattern if it is shorter, determine the next status.
Thus, only finitely many states are needed:
\begin{theorem}[Regularity for Finite Subtraction Sets]\label{th2-fin}
    Let $\Sigma$ be an alphabet and $S\subseteq \Sigma^+$ be a finite language.
    Then, the $\Npos$-position language $L_{\Npos}(S)$ of $\mathtt{WordSubtraction}(S)$ is regular.
\end{theorem}
\begin{proof}
If $S=\emptyset$, then $L_{\Npos}(S)=\emptyset$, which is regular.
Assume that $S\neq\emptyset$, and let
\[
    \ell=\max\{|s|:s\in S\}.
\]
We construct a DFA
\[
    \mathbf{A}_\Npos=(Q,\Sigma,\delta,\langle\Ppos\rangle,F)
\]
as follows.
Let
\[
    Q=
    \bigl\{
        \langle\theta_0,a_1,\theta_1,\ldots,a_n,\theta_n\rangle
        :
        0\leq n<\ell,\ 
        \theta_0,\ldots,\theta_n\in\{\Ppos,\Npos\},\
        a_1,\ldots,a_n\in\Sigma
    \bigr\}.
\]
For a state
$q=\langle\theta_0,a_1,\theta_1,\ldots,a_n,\theta_n\rangle$
and an input letter $a_{n+1}\in\Sigma$, we define $\delta(q,a_{n+1})\in Q$ in the following way.
First, let
\[
    \theta_{n+1}=
    \begin{cases}
        \Npos
        & \text{if there exists }i\in\{1,\ldots,n+1\}
          \text{ such that }a_i\cdots a_{n+1}\in S
          \text{ and }\theta_{i-1}=\Ppos,\\
        \Ppos
        & \text{otherwise}.
    \end{cases}
\]
Then, we append $a_{n+1}$ and $\theta_{n+1}$ to $q$.
If the resulting sequence has more than $2\ell-1$ entries, we delete its first two entries.
The resulting state is $\delta(q,a_{n+1})$.
Finally, let
\[
    F=\{q\in Q:q\text{ ends with }\Npos\}.
\]

We claim that, after reading a word $w$, the state of $\mathbf{A}_\Npos$
is the final $2\ell-1$ entries of $\mathsf{pat}(w)$, or the entire pattern if it is shorter.
We prove this by induction on $|w|$.

For $w=\varepsilon$, the claim holds because the initial state
is $\langle\Ppos\rangle=\mathsf{pat}(\varepsilon)$.
Now suppose that the claim holds for $w$, and let $a\in\Sigma$.
Every removable suffix of $wa$ has length at most $\ell$.
Hence, the stored pattern, together with the new letter $a$, contains both the letters of every possible removable suffix and the status of the prefix left after its removal.
By the Partition Theorem, the rule defining the new status gives exactly the $\Ppos/\Npos$ status of $wa$.
Appending the new letter and its status, and deleting the first two entries if necessary, gives precisely the required final portion of $\mathsf{pat}(wa)$.
This proves the claim.

Consequently, the state reached after reading $w$ belongs to $F$
if and only if $w$ is an $\Npos$-position.
Thus, $L(\mathbf{A}_\Npos)=L_{\Npos}(S)$, and $L_{\Npos}(S)$ is regular.
\end{proof}
Because the complement of a regular language is again regular,
we immediately obtain the following corollary.
\begin{corollary}
    Let $\Sigma$ be an alphabet and $S\subseteq \Sigma^+$ be a finite language.
    Then, the $\Ppos$-position language $L_{\Ppos}(S)$ of $\mathtt{WordSubtraction}(S)$ is regular.\bqed
\end{corollary}
These results generalize the familiar fact that, in an ordinary subtraction game with a finite
subtraction set, the set of $\Ppos$-positions is ultimately periodic.
Indeed, ordinary subtraction games can be viewed as word subtraction games over a unary alphabet.
Over a unary alphabet, a word is completely determined by its length.
Then, the Parikh image of a regular language is a semilinear subset of $\mathbb{N}$, and every such set is ultimately periodic.
Thus, the result above recovers the familiar ultimate periodicity of $\Ppos$-positions in ordinary subtraction games with finite subtraction sets.

We have restricted our discussion in this section to
the binary $\Ppos$/$\Npos$ classification. 
Let us briefly note that the same idea extends to Grundy numbers as well. 
When the subtraction set is finite, every position has at most $\#(S)$ options.
Consequently, its Grundy number is at most $\#(S)$, so only finitely many distinct Grundy numbers can occur. 
Thus, if we replace each $\Ppos/\Npos$ entry in the pattern by the corresponding Grundy number, while retaining the letters, the same finite-state construction applies.
Therefore, we can use a finite automaton with a nonnegative integer attached to each state as its output.
Such an automaton is often called a \emph{Moore machine}~\cite{Moore1956}.
In particular, for every Grundy number
$g\ge 0$, the language
\[
\{w\in\Sigma^*:\text{the Grundy number of $w$ is $g$}\}
\]
is again a regular language.

\subsection*{Exercises}
\begin{exercise}
    Let $\Sigma=\{a,b\}$.
    Construct a DFA that recognizes each of the following languages:
    \begin{enumerate}
        \item $\Sigma^*ba$
        \item $(ab)^+$
        \item $\Sigma^*abab\Sigma^*$
        \item $\{w\in\Sigma^*:|w|_a\text{ is even}\}$
        \item $\{\varepsilon\}$\diaqed
    \end{enumerate}
\end{exercise}
\begin{exercise}
    Let $\Sigma=\{a,b\}$.
    Use the pumping lemma to prove that\[
    \{w\in\Sigma^*:|w|_a=|w|_b\}
    \]
    is not regular.\diaqed
\end{exercise}
\begin{starexercise}
    Let $\Sigma=\{a,b\}$.
    Use the pumping lemma to prove that\[
    \{a^n b^m : n,m\geq 0,\ \text{$n$ is even or $n=m$}\}
\]
    is not regular.
    Note that there are several variants of the pumping lemma for regular languages. The version used in this note is not the most standard one, but it can be more convenient in practice. If you know another version, try using it to solve this exercise and compare the resulting proofs.\diaqed
\end{starexercise}
\begin{exercise}
    Let $\Sigma=\{a,b\}$.
    Compute the Parikh images of the following regular languages:
    \begin{enumerate}
        \item $a^*b^*$
        \item $(ab)^+$
        \item $(b^*aa)^*$
        \item $(a\Sigma)^*$\diaqed
    \end{enumerate}
\end{exercise}
\begin{exercise}
    Consider the DFA in Figure~\ref{fig:ws-example-dfa}. Verify that, for every state $q$, the state $\delta^*(q,baa)$ has $\Npos$ as its last entry.
    What is the game-theoretic interpretation of this property?\diaqed
\end{exercise}
\begin{exercise}
    Choose a nonempty finite language $S\subseteq \Sigma^+$. Then construct a DFA recognizing $L_\Npos(S)$ by following the construction in Theorem~\ref{th2-fin}.\diaqed
\end{exercise}
\begin{starexercise}
    If you are familiar with Grundy numbers, construct an automaton that computes the Grundy number of each word in $\mathtt{WordSubtraction}(S)$ for a finite subtraction set $S\subseteq\Sigma^+$. The basic idea was mentioned at the end of Section~\ref{subsec:B2WS_fin}.\diaqed
\end{starexercise}

\subsection*{Open Problems}
\begin{problem}[Can we construct a polynomial-size DFA for $L_\Npos(S)$?]
Fix an alphabet $\Sigma$.
For a nonempty finite subtraction set $S\subseteq\Sigma^+$, let
\[
    \ell=\max\{|s|:s\in S\}.
\]
The construction in Theorem~\ref{th2-fin} uses
\[
    \sum_{n=0}^{\ell-1}2^{n+1}\#(\Sigma)^n
\]
states, which grows exponentially with $\ell$.
How much can this bound be improved?
In particular, is there a polynomial $p$, independent of $S$, such that every such language $L_{\Npos}(S)$ can be recognized by a DFA with at most $p(\ell)$ states?
A related question is whether a polynomial bound is possible in terms of the total length
\[
    \sum_{s\in S}|s|
\]
of the subtraction words.

The construction can contain both unreachable states and states that can be merged without changing the recognized language.
For example, when $S=\{a,b,aa\}$, states such as $\langle \Ppos,a,\Ppos\rangle$ are not needed, as shown in Figure~\ref{fig:ws-example-dfa}. 
Moreover, even the DFA in Figure~\ref{fig:ws-example-dfa} contains redundant information and can be reduced to the equivalent DFA with only three states:
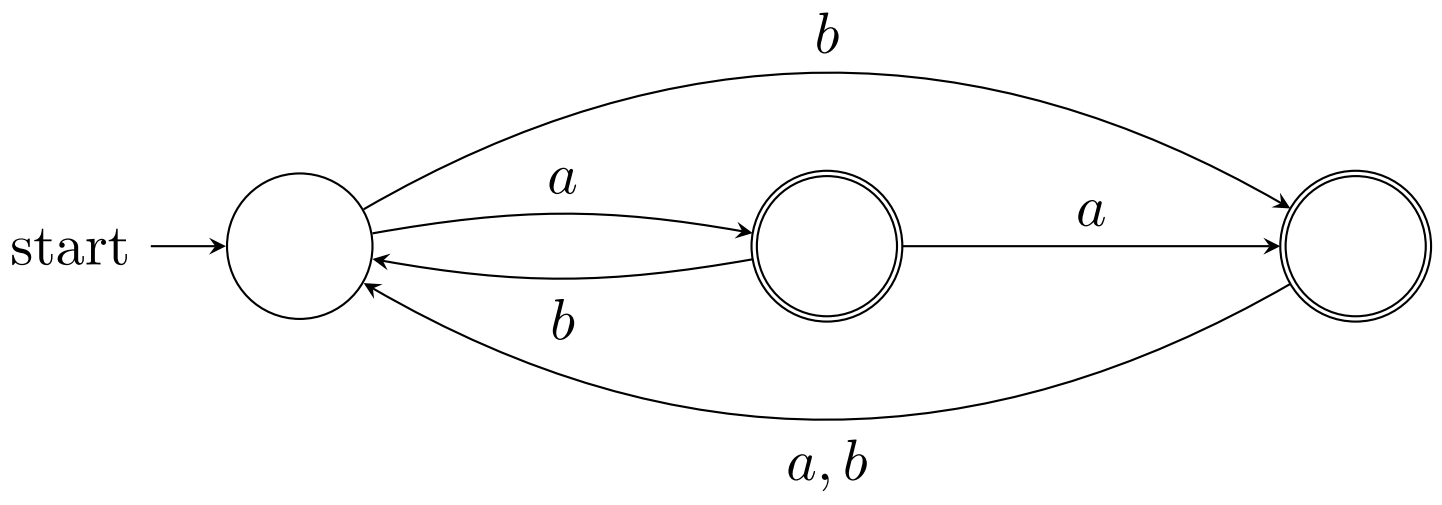
\begin{figure}[H]
\centering
\begin{tikzpicture}[
  ->,
  >=stealth,
  node distance=2.5cm
]
  \node[state, initial] (q0) {};
  \node[state, accepting] (q1) [right=of q0] {};
  \node[state, accepting] (q2) [right=of q1] {};

  \path
    (q0) edge[bend left=30] node[above] {$b$} (q2)
    (q0) edge[bend left=10] node[above] {$a$} (q1)
    (q1) edge node[above] {$a$} (q2)
    (q1) edge[bend left=10] node[below] {$b$} (q0)
    (q2) edge[bend left=30] node[below] {$a,b$} (q0);
\end{tikzpicture}
\end{figure}\noindent
The unary case, corresponding to ordinary subtraction games, is a natural starting point.
We have not yet investigated whether existing results on subtraction games settle this question.
\diaqed
\end{problem}

\newpage
\section{Regular Subtraction Sets}\label{sec:reg}
\subsection{Motivation}\label{subsec:reg-motivation}
In the previous section, we have seen that, when the subtraction set $S$ is finite, the $\Ppos$- and $\Npos$-position languages of $\mathtt{WordSubtraction}(S)$ are regular.
Recall that every finite language is itself regular. In other words, the subtraction set $S$ can be described by a finite automaton. This naturally leads to the next question: what happens when $S$ is an arbitrary regular language?

The answer is perhaps surprising: even when $S$ is an arbitrary regular language, the $\Npos$-position
language $L_{\Npos}(S)$ is still regular.
However, the construction in the previous section does not directly apply.
That construction remembers a final portion of the $\Ppos/\Npos$ pattern whose length depends on the maximum length of a word in $S$.
When $S$ is infinite, there is no such maximum, so we need a different way to record the relevant information.
The key idea is to keep track of several possible states of an automaton recognizing $S$.
This leads us to the notion of \emph{nondeterminism}.
The approach developed in this section was originally suggested by Professor Yoshiaki Takata of Kochi University of Technology.

Before going into the general construction, let us first develop some intuition through a concrete example. 
Consider the subtraction set
\[
    S=a\Sigma(\Sigma\Sigma)^*\cup b(\Sigma\Sigma)^*
\]
with $\Sigma=\{a,b\}$,
that is, the language of words that either start with $a$ and have even length, or start with $b$ and have odd length.

As we saw in Example~\ref{ex-DFAs}, this language is recognized by the following DFA:\begin{figure}[H]
\centering
\begin{tikzpicture}[
  ->,
  >=stealth,
  node distance=2cm
]
  \node[state, initial] (q0) {$q_0$};
  \node[state] (q1) [below=of q0] {$q_1$};
  \node[state,accepting] (q2) [right=of q1] {$q_2$};

  \path
    (q0) edge node[left] {$a$} (q1)
    (q0) edge[bend left=10] node[above] {$b$} (q2)
    (q1) edge[bend left=10] node[above] {$a,b$} (q2)
    (q2) edge[bend left=10] node[below] {$a,b$} (q1);
\end{tikzpicture}
\caption{A DFA recognizing $S=a\Sigma(\Sigma\Sigma)^*\cup b(\Sigma\Sigma)^*$.}
\label{fig3:DFA-A3}
\end{figure}
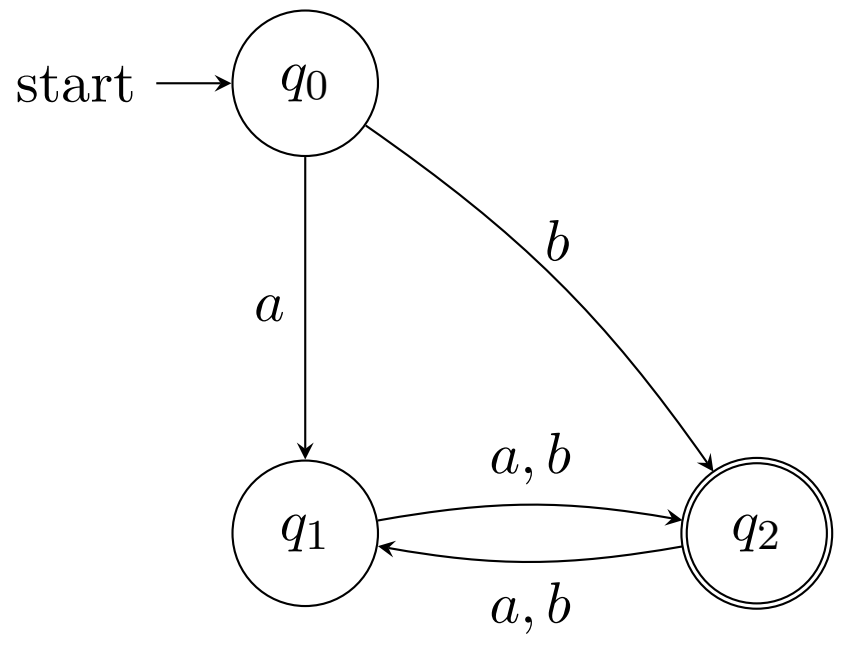
Let us think about what the $\Ppos$- and $\Npos$-position languages look like in this case. 
First, every word in $S$ is an $\Npos$-position because the player to move can remove the entire word and leave $\varepsilon$.
Thus, every word accepted by the original DFA is an $\Npos$-position.

There is another source of $\Npos$-positions. 
Suppose that $w$ is a $\Ppos$-position and $s\in S$. 
Then the concatenation $ws$ is necessarily an $\Npos$-position because we can remove the suffix $s$ and reach the $\Ppos$-position $w$.
How can we reproduce this behavior using an automaton? 
Consider, for example, $w=ba$.
Neither of its nonempty suffixes, $a$ and $ba$, belongs to $S$, so $ba$ is a terminal position and hence a $\Ppos$-position.
In the original DFA, $w$ reaches the rejecting state $q_1$ with the run\[
q_0\xrightarrow{b}q_2\xrightarrow{a}q_1.
\] 
To detect words obtained by appending an element of $S$ to $ba$, we would like to start reading again from $q_0$ after reaching this $\Ppos$-position.
As a first attempt, let us replace the transition $q_2\xrightarrow{a}q_1$ by $q_2\xrightarrow{a}q_0$.
The modified automaton then returns to $q_0$ after reading $ba$.
We call this return a \emph{reset}:
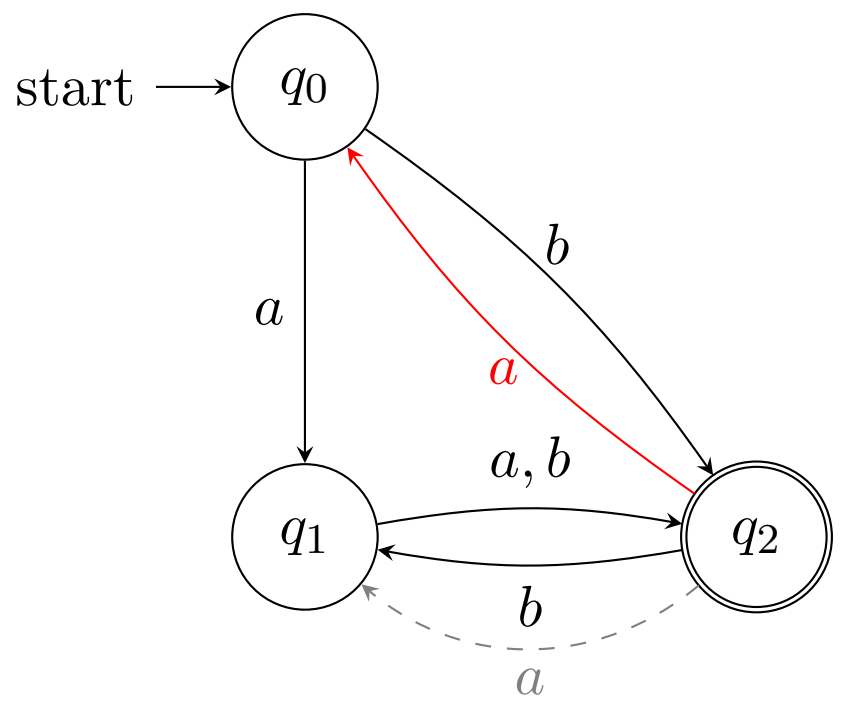
\begin{figure}[H]
\centering
\begin{tikzpicture}[
  ->,
  >=stealth,
  node distance=2cm
]
  \node[state, initial] (q0) {$q_0$};
  \node[state] (q1) [below=of q0] {$q_1$};
  \node[state,accepting] (q2) [right=of q1] {$q_2$};

  \path
    (q0) edge node[left] {$a$} (q1)
    (q0) edge[bend left=10] node[above] {$b$} (q2)
    (q1) edge[bend left=10] node[above] {$a,b$} (q2)
    (q2) edge[bend left=10, red] node[below] {$a$} (q0)
    (q2) edge[bend left=10] node[below] {$b$} (q1)
    (q2) edge[bend left=40, gray, dashed] node[below] {$a$} (q1);
\end{tikzpicture}
\end{figure}
\noindent
For example, on input $baaa$, the modified automaton follows the run
\[
    q_0\xrightarrow{b}q_2{\,\,\color{red}\xrightarrow{a}\,\,}q_0
    \xrightarrow{a}q_1\xrightarrow{a}q_2
\]
and accepts $baaa$.
This agrees with the game: removing the suffix $aa\in S$ leaves the $\Ppos$-position $ba$.
Moreover, $baaa\notin S$, so this modification detects an $\Npos$-position that the original DFA does not accept.

However, replacing the transition also loses information.
The word $baa$ belongs to $S$ and is therefore an $\Npos$-position, but the modified automaton follows
\[
    q_0\xrightarrow{b}q_2{\,\,\color{red}\xrightarrow{a}\,\,}q_0\xrightarrow{a}q_1
\]
and rejects it.
Thus, we want to retain both the original transition $q_2\xrightarrow{a}q_1$ and the reset transition $q_2{\,\,\color{red}\xrightarrow{a}\,\,}q_0$.
The original transition allows the automaton to continue checking the word being read, while the reset transition starts a new check from the $\Ppos$-prefix $ba$.
This suggests allowing more than one possible run for the same input word, and accepting the word if at least one run ends in an accepting state.
This is the basic idea of a \emph{nondeterministic finite automaton}.

One difficulty remains: a reset should be allowed only after a $\Ppos$-position.
So far, we have identified $ba$ as such a position by analyzing the game directly.
In general, how can the automaton determine when a reset is appropriate?
We will return to this question after introducing nondeterministic finite automata and the idea of keeping track of all possible current states.

\subsection{Nondeterminism}
We now introduce the \emph{nondeterministic finite automaton} (NFA)
as a variant of the DFA.
In a DFA, the current state and the next input letter uniquely determine
the next state.
In an NFA, there can be several possible next states, or none at all.
We therefore use a transition function
\[
    \delta:Q\times\Sigma\to 2^Q,
\]
where $2^Q$ denotes the set of all subsets of $Q$.
Here, $\delta(q,a)$ is the set of possible next states after reading
$a\in\Sigma$ in state $q\in Q$.
As a result, an NFA can have more than one run on the same input word.
\begin{definition}[Nondeterministic Finite Automata]
A \emph{nondeterministic finite automaton} (\emph{NFA} for short) is a $5$-tuple
\[
\mathbf{N}=(Q,\Sigma,\delta,q_0,F),
\]
where\begin{itemize}
    \item $Q$ is a finite set of states,
    \item $\Sigma$ is an alphabet,
    \item $\delta:Q\times \Sigma\to 2^Q$ is the \emph{transition function},
    \item $q_0\in Q$ is the initial state, and
    \item $F\subseteq Q$ is the set of accepting states.
\end{itemize}
A sequence of states
\[
q_0,q_1,\ldots,q_n,
\]
is a \emph{run} of a word $w=a_1a_2\cdots a_n\in\Sigma^*$ if
\[
q_i\in \delta(q_{i-1},a_i)
\]
for each $1\le i\le n$.
We say that $\mathbf{N}$ \emph{accepts} $w$ if there exists a run of $w$ such that $q_n\in F$. The language\[
L(\mathbf{N})=\{w\in\Sigma^*: \mathbf{N}\text{ accepts }w\}
\] is called the \emph{language recognized by $\mathbf{N}$}.\bqed
\end{definition}
In other words, an NFA can have several possible next states for a given state and input letter. 
If there are several possible runs of an input word, the word is accepted as long as at least one of them ends in an accepting state.
Note that $\delta(q,a)$ can also be empty, so there can be no possible transition for a given state and input letter.
A run must read the entire input word; reaching an accepting state before the input ends does not by itself imply acceptance.

\begin{example}\label{ex-NFA}
    Let $\Sigma=\{a,b\}$. 
    The following NFA $\mathbf{N}=(\{q_0,q_1,q_2\},\Sigma,\delta,q_0,\{q_{2}\})$  recognizes the language $L=\Sigma^*ba$:
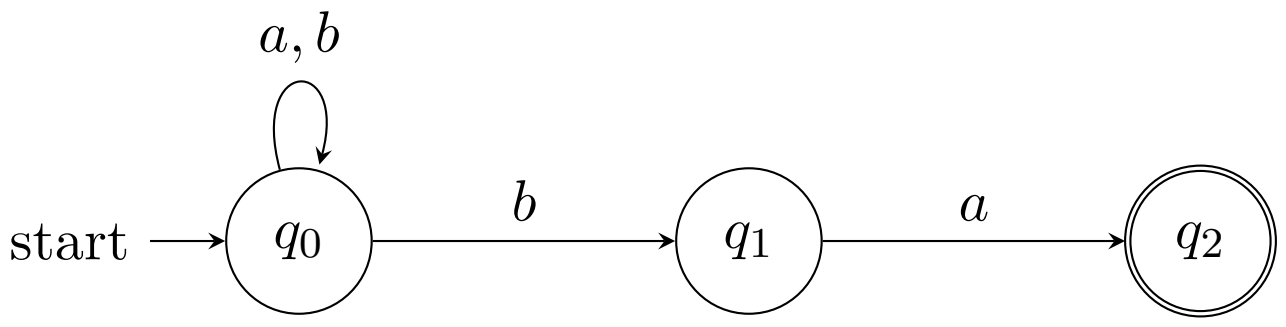
\begin{figure}[H]
\centering
\begin{tikzpicture}[
  ->,
  >=stealth,
  node distance=2cm
]
  \node[state, initial] (q0) {$q_0$};
  \node[state] (q1) [right=of q0] {$q_1$};
  \node[state,  accepting] (q2) [right=of q1] {$q_2$};

  \path
    (q0) edge[loop above] node {$a,b$} ()
         edge node[above] {$b$} (q1)
    (q1) edge node[above] {$a$} (q2);
\end{tikzpicture}
\end{figure}
For example, consider the word $aba\in L$.
As in a DFA, the automaton starts in the initial state $q_0$. There are two valid runs of $aba$:\[
q_0\xrightarrow{a}q_0\xrightarrow{b}q_0\xrightarrow{a}q_0,\]
and\[
q_0\xrightarrow{a}q_0\xrightarrow{b}q_1\xrightarrow{a}q_2.
\]
The second run ends in the accepting state $q_2$, so $aba$ is accepted.

On the other hand, consider the word $aab\notin L$.
There are two valid runs of $aab$:
\[
q_0\xrightarrow{a}q_0\xrightarrow{a}q_0\xrightarrow{b}q_0,\]
and
\[
q_0\xrightarrow{a}q_0\xrightarrow{a}q_0\xrightarrow{b}q_1.\]
Neither of them ends in $q_2$, and hence $aab$ is rejected.

Finally, consider the word $abaa$.
After reading $aba$, the possible states are $q_0$ and $q_2$, as shown above.
On the final $a$, the run ending at $q_0$ continues to $q_0$,
but the run ending at $q_2$ cannot continue because $\delta(q_2,a)=\emptyset$.
Thus, the only run of the entire word is
\[
    q_0\xrightarrow{a}q_0\xrightarrow{b}q_0
    \xrightarrow{a}q_0\xrightarrow{a}q_0,
\]
so $abaa$ is rejected.\bqed
\end{example}

An NFA is a generalization of a DFA because every DFA can be regarded as a special kind of NFA in which there is exactly one possible transition for every state and input letter.
At first sight, this may suggest that NFAs are more powerful than DFAs.
However, this is not the case.
NFAs have exactly the same expressive power as DFAs: they still recognize only regular languages.

\begin{theorem}
A language $L\subseteq \Sigma^*$ is regular if and only if it is recognized by some NFA.
\end{theorem}
\begin{proof} 
Recall that a language is regular if it is recognized by some DFA. 
Because every DFA is a special kind of NFA, the only if direction is immediate. 

For the converse, let us start with an arbitrary NFA and construct a DFA that recognizes the same language. 
This construction is well known as the \emph{subset construction}. 
The basic idea is to remember all possible current states at the same time. 
For example, consider the NFA in Example~\ref{ex-NFA} and suppose that it is currently in state $q_0$ and reads the input letter $b$. 
There are then two possible next states, $q_0$ and $q_1$. Thus, we simply remember the set \[ \{q_0,q_1\}\subseteq Q \] as a single state of the new automaton. 
More generally, each state of the new DFA represents a subset of $Q$. 
Because $Q$ is finite, there are only $2^{\#(Q)}$ possible subsets, so the new automaton still has only finitely many states. 
A subset $R\subseteq Q$ should be an accepting state
of the new automaton whenever $R$ contains an accepting
state of the original NFA.
Thus, a subset $R\subseteq Q$ is accepting if and only if $R\cap F\neq\emptyset$.

We now give the formal construction.
Let $\mathbf{N}=(Q,\Sigma,\delta,q_0,F)$ be a given NFA.
Define the DFA
\[
    \mathbf{A}=(2^Q,\Sigma,\delta',\{q_0\},F')
\]
by
\[
    \delta'(R,a)=\bigcup_{q\in R}\delta(q,a)
\]
for each $R\subseteq Q$ and $a\in\Sigma$, and
\[
    F'=\{R\subseteq Q:R\cap F\neq\emptyset\}.
\]
Initially, the DFA records the singleton set $\{q_0\}$.
If it records all states reachable in the NFA after reading a word $w$,
then the union above records exactly all states reachable after reading $wa$.
Thus, by induction on the input length, the DFA always records exactly the set of states reachable in the NFA.
By the definition of $F'$, the two automata accept the same words. \end{proof}
\begin{example}
    Applying the subset construction to the NFA in Example~\ref{ex-NFA} yields the following DFA $\mathbf{A}=(2^{\{q_0,q_1,q_2\}},\Sigma,\delta',\{q_0\},\{\{q_2\},\{q_0,q_2\},\{q_1,q_2\},\{q_0,q_1,q_2\}\})$:
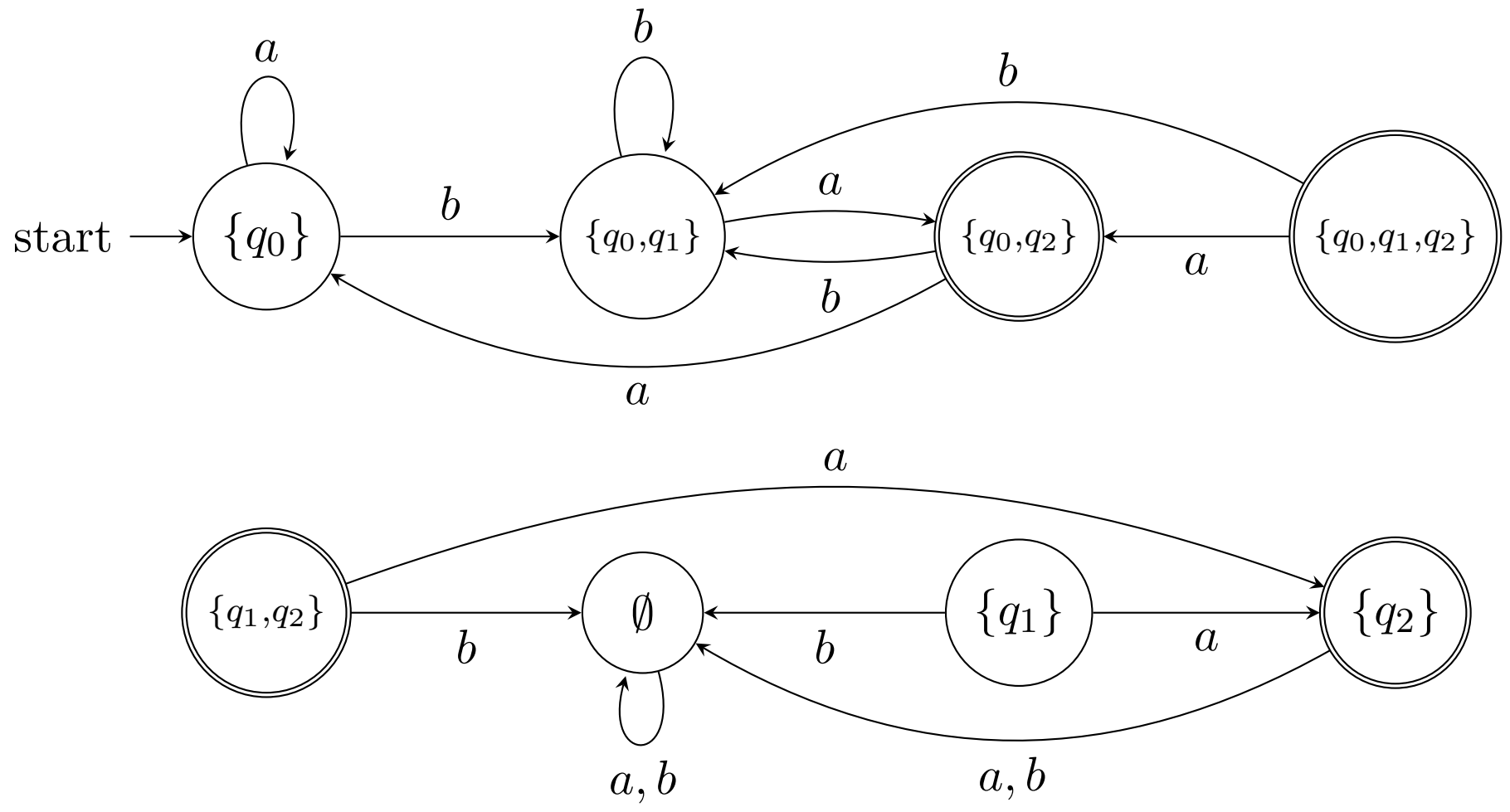
\begin{figure}[H]
\centering
\begin{tikzpicture}[
  ->,
  >=stealth,
  node distance=2cm
]
  \node[state, initial] (0) at (0,0) {$\{q_0\}$};
  \node[state] (01) at (3,0) {$\scriptstyle\{q_0,q_1\}$};
  \node[state, accepting] (02) at (6,0) {$\scriptstyle\{q_0,q_2\}$};
  \node[state, accepting] (12) at (0,-3) {$\scriptstyle\{q_1,q_2\}$};
  \node[state] (e) at (3,-3) {$\emptyset$};
  \node[state] (1) at (6,-3) {$\{q_1\}$};
  \node[state,accepting] (2) at (9,-3) {$\{q_2\}$};
  \node[state, accepting] (012) at (9,0) {$\scriptstyle \{q_0,q_1,q_2\}$};
  \path
    (0) edge[loop above] node {$a$} ()
         edge node[above] {$b$} (01)
     (1) edge node[below] {$a$} (2)
     (1) edge node[below] {$b$} (e)
    (01) edge[loop above] node {$b$} ()
    (01) edge[bend left=10] node[above] {$a$} (02)
    (02) edge[bend left=10] node[below] {$b$} (01)
    (02) edge[bend left=30] node[below] {$a$} (0)
    (12) edge[bend left=20] node[above] {$a$} (2)
    (12) edge node[below] {$b$} (e)
    (012) edge node[below] {$a$} (02) 
    (012) edge[bend right=30] node[above] {$b$} (01)
    (e) edge[loop below] node {$a,b$} ()
    (2) edge[bend left=30] node[below] {$a,b$} (e);
\end{tikzpicture}
\end{figure}
\noindent
Note that, apart from the three states
$\{q_0\}$, $\{q_0,q_1\}$, and $\{q_0,q_2\}$ in the upper left,
all other states are unreachable from the initial state and can therefore be omitted.\bqed
\end{example}
Let us briefly return to the connection with word subtraction games.
The acceptance condition of an NFA is closely related to the notion of
an $\Npos$-position.
Recall that a position is an $\Npos$-position if at least one of its options is a $\Ppos$-position.
This is analogous to an NFA accepting a word when at least one of its possible runs ends in an accepting state: in both cases, it is enough for one possible choice to succeed.

In fact, we will not use NFAs directly in our analysis of word subtraction games.
What matters is the idea behind the subset construction: instead of choosing one possibility, we can keep track of all possible states at the same time.
This idea will be the key to our construction for regular subtraction sets.

\subsection{Universal and Alternating Automata}
There is a natural variant of an NFA in which the existential acceptance condition is replaced by a universal one.
A \emph{universal finite automaton} (UFA) has the same structure as an NFA, but a word is accepted if and only if every run on that word ends in an accepting state.
Thus, while an NFA requires at least one accepting run, a UFA requires every run to be accepting.

UFAs also recognize exactly the regular languages.
Indeed, the subset construction works here as well.
After reading a word $w$, the constructed DFA records the set $R\subseteq Q$ of all states that the UFA can reach by reading $w$.
For a UFA with accepting states $F$, we simply declare a subset $R$ to be accepting when
\[
R\subseteq F.
\]
In other words, all possible current states must be accepting.
Thus, the same basic idea that turns an NFA into a DFA also shows that UFAs recognize only regular languages.

This gives a useful logical perspective on these automata.
The acceptance condition of an NFA has an existential flavor:
\[
\exists\text{ a run that is accepting}.
\]
The acceptance condition of a UFA has a universal
flavor:
\[
\forall\text{ runs are accepting}.
\]

This leads to more general models known as
\emph{Boolean automata}
\cite{DBLP:journals/tcs/BrzozowskiL80}
and \emph{alternating finite automata}
\cite{DBLP:journals/jacm/ChandraKS81}.
Informally, these models combine existential and universal choices during a computation: at some steps, the remaining input must be accepted from at least one of the possible next states; 
at others, it must be accepted from all of them. Despite this additional flexibility, alternating finite automata still recognize exactly the regular languages. We will not discuss these models further here.

The same distinction between existential and universal conditions appears in word subtraction games.
A position is an $\Npos$-position if at least one of its
options is a $\Ppos$-position, whereas it is a $\Ppos$-position if all of its options are $\Npos$-positions.
These conditions are applied recursively, in a way similar to alternating computation.
These viewpoints help to explain the construction in the next subsection, where we build a DFA for the position languages of a word subtraction game with a regular subtraction set.

\subsection{Back to Word Subtraction}\label{subsec:B2WS_reg}
We now apply these ideas to word subtraction games with a regular
subtraction set $S$.
Recall the two complementary ways to determine the outcome of a position:
it is an $\Npos$-position if some option is a $\Ppos$-position, and a $\Ppos$-position if all options are $\Npos$-positions.
These are the existential and universal viewpoints discussed above.
Because every move removes a nonempty suffix, all options are shorter prefixes, so we can determine their outcomes before determining the outcome of the current word.

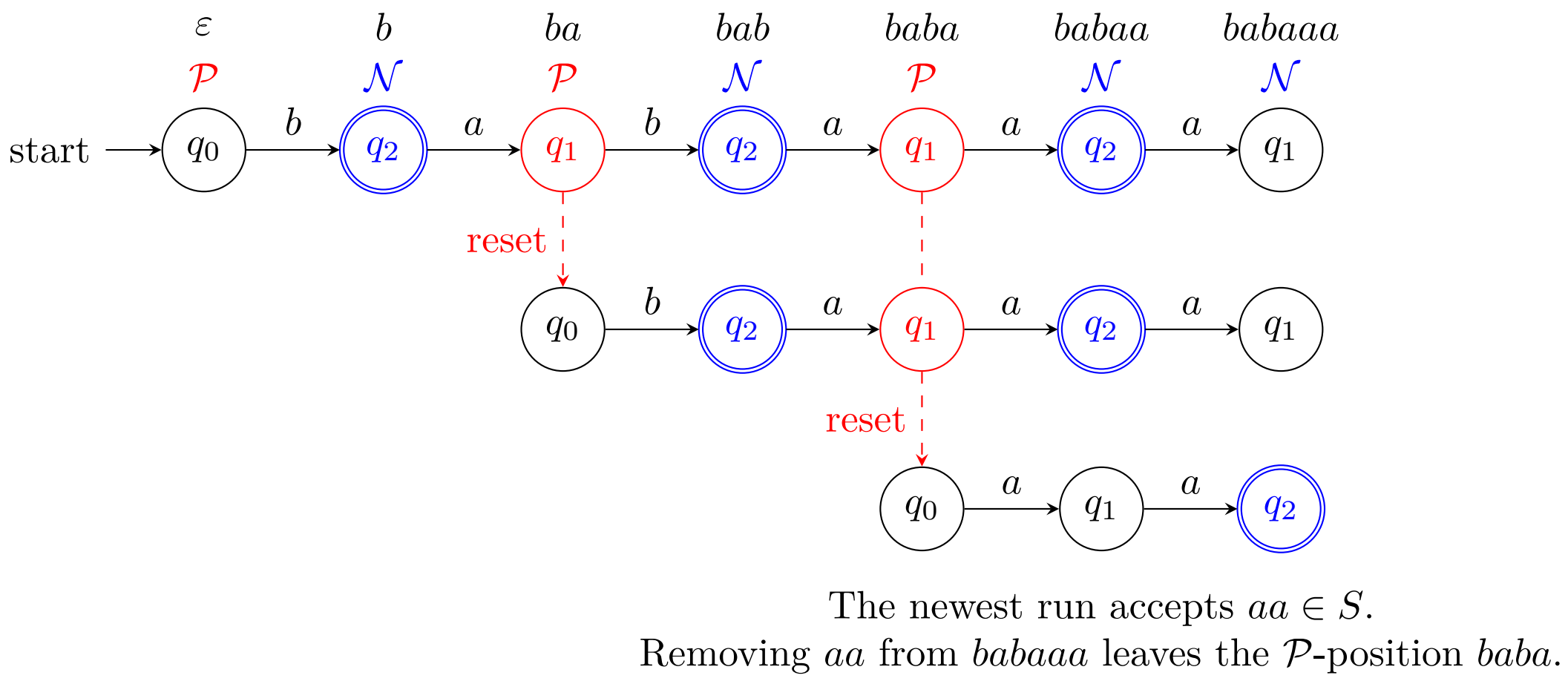
\begin{figure}[htbp]
\centering
\begin{tikzpicture}[
    >=stealth,
    every state/.style={minimum size=7mm},
    every node/.style={font=\small}
]
\node[state, initial]   (r0) at (0,0)   {$q_0$};
\node[state, accepting, color=blue] (r1) at (1.6,0) {$q_2$};
\node[state, color=red]            (r2) at (3.2,0) {$q_1$};
\node[state, accepting, color=blue] (r3) at (4.8,0) {$q_2$};
\node[state,, color=red]            (r4) at (6.4,0) {$q_1$};
\node[state, accepting, color=blue] (r5) at (8,0)   {$q_2$};
\node[state]            (r6) at (9.6,0) {$q_1$};

\path[->]
    (r0) edge node[above] {$b$} (r1)
    (r1) edge node[above] {$a$} (r2)
    (r2) edge node[above] {$b$} (r3)
    (r3) edge node[above] {$a$} (r4)
    (r4) edge node[above] {$a$} (r5)
    (r5) edge node[above] {$a$} (r6);

\node[state]            (s2) at (3.2,-1.6) {$q_0$};
\node[state, accepting, color=blue] (s3) at (4.8,-1.6) {$q_2$};
\node[state, color=red]            (s4) at (6.4,-1.6) {$q_1$};
\node[state, accepting, color=blue] (s5) at (8,-1.6)   {$q_2$};
\node[state]            (s6) at (9.6,-1.6) {$q_1$};

\path[->]
    (s2) edge node[above] {$b$} (s3)
    (s3) edge node[above] {$a$} (s4)
    (s4) edge node[above] {$a$} (s5)
    (s5) edge node[above] {$a$} (s6);

\node[state]            (t4) at (6.4,-3.2) {$q_0$};
\node[state]            (t5) at (8,-3.2)   {$q_1$};
\node[state, accepting, color=blue] (t6) at (9.6,-3.2) {$q_2$};

\path[->]
    (t4) edge node[above] {$a$} (t5)
    (t5) edge node[above] {$a$} (t6);

\draw[->, red, dashed]
    (r2) -- node[left] {reset} (s2);
\draw[->, red, dashed]
    (s4) -- node[left] {reset} (t4);
\draw[-, red, dashed]
    (r4) -- node[right] {} (s4);

\foreach \x/\w/\o in {
    0/\varepsilon/{\color{red}\Ppos},
    1.6/b/{\color{blue}\Npos},
    3.2/ba/{\color{red}\Ppos},
    4.8/bab/{\color{blue}\Npos},
    6.4/baba/{\color{red}\Ppos},
    8/babaa/{\color{blue}\Npos},
    9.6/babaaa/{\color{blue}\Npos}
}{
    \node at (\x,1.1) {$\w$};
    \node at (\x,0.65) {$\o$};
}

\node[align=center] at (8,-4.3) {
    The newest run accepts $aa\in S$.\\
    Removing $aa$ from $babaaa$ leaves the
    $\Ppos$-position $baba$.
};
\end{tikzpicture}
\caption{
Tracking runs on input $babaaa$ for $S=a\Sigma(\Sigma\Sigma)^*\cup b(\Sigma\Sigma)^*$.}
\label{fig3:tracked-runs}
\end{figure}

To implement this idea, take a DFA recognizing $S$ and read the input
from left to right.
At every prefix already identified as a $\Ppos$-position, start a new run of this DFA, while retaining all existing runs.
This includes the empty prefix, which is a $\Ppos$-position.
An accepting run then identifies a suffix in $S$ whose removal leaves a $\Ppos$-position.
Thus, if at least one run is accepting, the current word is an
$\Npos$-position.
If none is accepting, no move leads to a $\Ppos$-position, so the current word is a $\Ppos$-position.
Figure~\ref{fig3:tracked-runs} illustrates this process using the DFA from Section~\ref{subsec:reg-motivation}.

\vspace{8pt}
\vspace{8pt}

Each row follows one run, with accepting states shown
in blue.
After reading $ba$, the only existing run is rejecting, so we start a second run.
After reading $baba$, both existing runs are rejecting, so we start a third.
These resets are marked by red dashed lines.
At the end of $babaaa$, only the third run is accepting.
It has read $aa\in S$ from the prefix $baba\in L_\Ppos(S)$, so removing this suffix leaves a $\Ppos$-position.
Thus $babaaa$ is an $\Npos$-position.

We do not need to keep track of these runs separately:
it is enough to record their current states, just as in the subset construction.
Let $R\subseteq Q$ be the set of current states of all the runs started so far, where $Q$ is the state set of the DFA recognizing $S$.
For example, after reading $baba$ and starting the third run, we have $R=\{q_0,q_1\}$.
Whenever $R$ contains no accepting state, we add the initial state $q_0$ to $R$ and start a new run.
Otherwise, the current word is an $\Npos$-position, so no new run is started.

We first arrange that no transition of the original DFA returns to its initial state.
This will allow us to identify resets in the construction below.
\begin{lemma}[Non-Returning DFA]\label{lem3-no_return}
For every DFA $\mathbf{A}=(Q,\Sigma,\delta,q_0,F)$, there is an equivalent DFA $\mathbf{A}'=(Q',\Sigma,\delta',q_0',F')$ whose transitions
never return to the initial state.
More precisely, there is a DFA $\mathbf{A}'$ such that
\[
L(\mathbf{A}')=L(\mathbf{A})
\]
and
\[
\delta'(q,a)\neq q_0'
\]
for every $q\in Q'$ and $a\in \Sigma$.
\end{lemma}
\begin{proof}
    Let $Q'=Q\cup \{q_0'\}$ with a new distinct initial state $q_0'\notin Q$.
    In addition, let\[
    \delta'(q,a)=\begin{cases}
        \delta(q_0,a)&\text{if }q=q_0',\\
        \delta(q,a)&\text{otherwise}
    \end{cases}
    \]
    for each $q\in Q'$ and $a\in\Sigma$, and let\[
    F'=\begin{cases}
        F\cup\{q_0'\}&\text{if }q_0\in F,\\
        F&\text{otherwise}.
    \end{cases}
    \]
    No transition enters $q'_0$ because every value of $\delta'$ lies in $Q$.
By the definition of $F'$, the two automata
either both accept or both reject the empty word.
For a nonempty word $w=av$, where $a\in\Sigma$ and $v\in\Sigma^*$,
we have
\[
    \delta'^*(q'_0,av)
    =\delta'^*(\delta'(q'_0,a),v)
    =\delta^*(\delta(q_0,a),v)
    =\delta^*(q_0,av).
\]
Here, the middle equality holds because $\delta'(q,a)=\delta(q,a)$ for every $q\in Q$ and $a\in\Sigma$.
Because $F'\cap Q=F$, the two automata accept exactly the same nonempty words.
Therefore, $L(\mathbf{A}')=L(\mathbf{A})$.
\end{proof}
\noindent
In this note, we call a DFA with no transition returning to its initial state \emph{non-returning}.

We now turn to the main result of this section.
We will show that if $S$ is regular, then the $\Npos$-position language $L_{\Npos}(S)$ is also regular.
To prove this, we start with a DFA $\mathbf{A}$ recognizing $S$ and construct a new DFA $\mathbf{A}_{\Npos}$ recognizing $L_{\Npos}(S)$.
We first give the construction of $\mathbf{A}_{\Npos}$.

Let
\[
\mathbf{A}=(Q,\Sigma,\delta,q_0,F)
\]
be a DFA recognizing $S$.
By Lemma~\ref{lem3-no_return}, we can assume without loss of generality that $\mathbf{A}$ is non-returning.
Since $\varepsilon\notin S$, we have $q_0\notin F$.
We now define
\[
\mathbf{A}_{\Npos}=(Q_{\Npos},\Sigma,\delta_{\Npos},\{q_0\},F_{\Npos})
\]
as follows:
\begin{itemize}
    \item $Q_{\Npos}=2^Q$,
    \item for each $R\subseteq Q$ and $a\in\Sigma$,\[
    \delta_{\Npos}(R,a)=\begin{cases}
        \{\delta(q,a):q\in R\}\cup \{{\color{red}q_0}\}&{\color{red}\text{if }\delta(q,a)\notin F\text{ for all }q\in R},\\
        \{\delta(q,a):q\in R\}&\text{otherwise},
    \end{cases}
    \] 
    and
    \item $F_{\Npos}=\{R\subseteq Q:R\cap F\ne \emptyset\}$.
\end{itemize}
The construction is similar in spirit to the subset construction:
the states of $\mathbf{A}_{\Npos}$ keep track of the information needed to detect both $\Npos$- and $\Ppos$-positions simultaneously.
In particular, the part highlighted in red is crucial.
When the automaton detects a $\Ppos$-position, that is, a word for which none of the possible runs reaches an accepting state, we apply the reset rule by adding $q_0$ to the state set.

The following two lemmas describe the key properties of this construction.
The first lemma shows that an accepting run of $\mathbf{A}_\Npos$
can be traced through individual states of $\mathbf{A}$,
using original transitions and occasional resets.
\begin{lemma}\label{lem3-recovering_run}
    Let $\mathbf{A}=(Q,\Sigma,\delta,q_0,F)$ be a non-returning DFA, and let
$\mathbf{A}_{\Npos}=(Q_{\Npos},\Sigma,\delta_{\Npos},\{q_0\},F_{\Npos})$ denote the DFA obtained from $\mathbf{A}$ by the construction above.
If \[
\{q_0\}=R_0~\xrightarrow{a_1}~R_1~\xrightarrow{a_2}~\cdots ~\xrightarrow{a_n}~R_n
\]
is an accepting run of $\mathbf{A}_\Npos$ with 
$R_0,R_1,\ldots,R_n\in Q_\Npos$ and
$a_1,a_2,\ldots,a_n\in \Sigma$, then there is a sequence\[
q_0,q_1,\ldots,q_n\in Q
\]
such that\begin{itemize}
    \item $q_i\in R_i$ for each $0\le i\le n$,
    \item $q_i\in \delta_\Npos(\{q_{i-1}\},a_i)$ for each $1\le i\le n$, and
    \item $q_n\in F$.
\end{itemize}
\end{lemma}
\begin{proof}
Let us construct the sequence $q_0,q_1,\ldots,q_n$ backwards.
All the sets $R_i$ are nonempty because $R_0=\{q_0\}$ and the transition rule maps every nonempty set to a nonempty set.
In addition, $R_n$ is an accepting state, that is, $R_n\cap F\ne \emptyset$ holds.
Therefore, we can choose a state $q_n\in R_n\cap F$. 

Now suppose that we have chosen $q_j,\ldots,q_n$ satisfying the required conditions, and consider how to choose $q_{j-1}$.
If $q_j=q_0$, we can choose $q_{j-1}$ arbitrarily from $R_{j-1}$.
Indeed,
by the assumption that $\mathbf{A}$ is non-returning,
$q_0\in R_j$ implies that the reset rule was used in the transition from $R_{j-1}$ to $R_j$.
Thus, for every $q\in R_{j-1}$, we have $\delta(q,a_j)\notin F$ and $q_0\in\delta_{\Npos}(\{q\},a_j)=\{\delta(q,a_j)\}\cup\{q_0\}$.

On the other hand, suppose that $q_j\neq q_0$.
Because the reset rule adds only $q_0$, the state $q_j$ must come
from an original transition.
Thus, we can choose $q_{j-1}\in R_{j-1}$ such that
\[
    \delta(q_{j-1},a_j)=q_j.
\]
The definition of $\delta_\Npos$ then gives
\[
    q_j\in\delta_\Npos(\{q_{j-1}\},a_j).
\]
Repeating these steps backwards until we reach $\{q_0\}=R_0$ gives the desired sequence $q_0,q_1,\ldots,q_n\in Q$.
\end{proof}

The second lemma concerns the role of the initial state $q_0$.
Recall that the original automaton $\mathbf{A}$ is non-returning.
Thus, in the construction of $\mathbf{A}_{\Npos}$, the current state set contains $q_0$ only if the reset rule has been applied.
The following lemma shows that this happens precisely when the input word is rejected.
\begin{lemma}\label{lem3-reject_return}
Let $\mathbf{A}=(Q,\Sigma,\delta,q_0,F)$ be a non-returning DFA with $q_0\notin F$, and let
$\mathbf{A}_\Npos=(Q_\Npos,\Sigma,\delta_\Npos,\{q_0\},F_\Npos)$ denote the DFA obtained from $\mathbf{A}$ by the construction above.
Then, it holds that \[
w\in L(\mathbf{A}_\Npos)\quad\iff \quad q_0\notin{\delta_\Npos}^{*}(\{q_0\},w)
\]
for each $w\in\Sigma^*$.
\end{lemma}
\begin{proof}
    The case $w=\varepsilon$ is clear because ${\delta_\Npos}^{*}(\{q_0\},\varepsilon)=\{q_0\}$ and $\{q_0\}\notin F_\Npos$.
    Let $w=w'a$ with $w'\in\Sigma^*$ and $a\in \Sigma$, and let $R_{w'}={\delta_\Npos}^*(\{q_0\},w')$.

    Suppose that $q_0\in{\delta_\Npos}^{*}(\{q_0\},w)$.
    By the assumption that $\mathbf{A}$ is non-returning,
    the only way to have $q_0\in{\delta_\Npos}^*(\{q_0\},w)={\delta_\Npos}(R_{w'},a)$ is for the reset rule to have been applied.
    That is, $\delta(q,a)\notin F$ for all $q\in R_{w'}$.
Therefore,\[
    \delta_\Npos^*(\{q_0\},w)\cap F
    =
    \bigl(\{\delta(q,a):q\in R_{w'}\}\cup\{q_0\}\bigr)\cap F
    =\emptyset
\]
holds,
and thus, ${\delta_\Npos}^*(\{q_0\},w)\notin F_\Npos$. 
This implies that $w\notin L(\mathbf{A}_\Npos)$.

    Conversely, suppose that $q_0\notin{\delta_\Npos}^{*}(\{q_0\},w)={\delta_\Npos}(R_{w'},a)$.
    Then, there exists $q\in R_{w'}$ such that $\delta(q,a)\in F$ by the definition of $\delta_\Npos$.
    Therefore,
    \[
    \delta(q,a)\in {\delta_\Npos}^{*}(\{q_0\},w)\cap F
    \]
    holds, and thus, ${\delta_\Npos}^{*}(\{q_0\},w)\in F_\Npos$.
    This means that $w$ is accepted by $\mathbf{A}_\Npos$.
\end{proof}

We are now ready to prove the main theorem.

\begin{theorem}[Regularity for Regular Subtraction Sets]\label{th3-reg}
 Let $\Sigma$ be an alphabet and $S\subseteq \Sigma^+$ be a regular language.
    Then, the $\Npos$-position language $L_{\Npos}(S)$ of $\mathtt{WordSubtraction}(S)$ is regular.
\end{theorem}
\begin{proof}
    Let $\mathbf{A}=(Q,\Sigma,\delta,q_0,F)$ be a non-returning DFA recognizing $S$, and let $\mathbf{A}_{\Npos}$ be the DFA constructed above.
    We show that\[
    L(\mathbf{A}_\Npos)=L_\Npos(S)
    \]
    holds with $\mathbf{A}_\Npos=(Q_{\Npos},\Sigma,\delta_{\Npos},\{q_0\},F_{\Npos})$.
    
    We use induction on the length of $w$.
    For $w=\varepsilon$, the claim is immediate because $\varepsilon\notin L(\mathbf{A}_\Npos)$ and $\varepsilon$ is a $\Ppos$-position.
    Now let $w\neq\varepsilon$, and assume as the induction hypothesis that
    \[
    w'\in L(\mathbf{A}_\Npos)
\quad\Longleftrightarrow\quad
w'\in L_\Npos(S)
\]
for every proper prefix $w'$ of $w$.

If $w$ is an $\Npos$-position, then $w=w's$ for some
$\Ppos$-position $w'$ and some $s\in S$.
Because $s\neq\varepsilon$, the word $w'$ is a proper prefix of $w$.
By the induction hypothesis and Lemma~\ref{lem3-reject_return},
\[
    q_0\in\delta_\Npos^*(\{q_0\},w').
\]
The transition rule always retains the original transition:
if $q\in R$, then
\[
    \delta(q,a)\in\delta_\Npos(R,a).
\]
Applying this observation successively along $s$, we obtain
\[
    \delta^*(q_0,s)\in
    \delta_\Npos^*(\{q_0\},w's)
    =\delta_\Npos^*(\{q_0\},w).
\]
Because $s\in S=L(\mathbf{A})$, we have $\delta^*(q_0,s)\in F$.
Therefore, $w\in L(\mathbf{A}_\Npos)$.

Conversely, assume that $w=a_1a_2\cdots a_n\in L(\mathbf{A}_\Npos)$ and consider its run\[
R_0\xrightarrow{a_1}R_1\xrightarrow{a_2}\cdots \xrightarrow{a_n}R_n
\]
with $R_0,R_1,\ldots,R_n\subseteq Q$.
By Lemma~\ref{lem3-recovering_run}, there are states $q_0,q_1,\ldots,q_{n-1}\in Q$ and $q_n\in F$ such that $q_i\in R_i$ and $q_i\in \delta_\Npos(\{q_{i-1}\},a_i)$ for each $1\le i\le n$.
Let $j$ be the largest index such that $q_j=q_0$.
Since $q_n\in F$ and $q_0\notin F$, we have $j<n$.
Then, the sequence \[
q_0=q_j\xrightarrow{a_{j+1}}q_{j+1}\xrightarrow{a_{j+2}}\cdots \xrightarrow{a_{n}}q_n\in F
\]
is an accepting run of the original DFA $\mathbf{A}$.
Indeed, for each $j<i\leq n$, we have $q_i\neq q_0$ by the choice of $j$.
Because the only state that the reset rule can add is $q_0$, the relation $q_i\in\delta_\Npos(\{q_{i-1}\},a_i)$ implies
\[
    q_i=\delta(q_{i-1},a_i)
\]
for each $j<i\leq n$.
This implies that $a_{j+1}a_{j+2}\cdots a_n\in L(\mathbf{A})=S$.
In addition, $a_1a_2\cdots a_j\notin L(\mathbf{A}_\Npos)$ by $q_j=q_0\in R_j$ and Lemma~\ref{lem3-reject_return}. 
This means that $a_1a_2\cdots a_j$ is a $\Ppos$-position by the induction hypothesis.
Therefore, their concatenation $a_1a_2\cdots a_j\cdot a_{j+1}a_{j+2}\cdots a_n$ is indeed an $\Npos$-position.
Consequently, we can conclude that $L_\Npos(S)= L(\mathbf{A}_\Npos)$ and $L_\Npos(S)$ is regular.
\end{proof}
\begin{example}
    Let us illustrate the construction for $S=a\Sigma(\Sigma\Sigma)^*\cup b(\Sigma\Sigma)^*$.
That is, we construct $\mathbf{A}_{\Npos}$ from the
non-returning DFA $\mathbf{A}$ shown in Figure~\ref{fig3:DFA-A3}.
In principle, the subset construction gives $2^{\#(\{q_0,q_1,q_2\})}=8$ possible states.
However, several of these states are unreachable from the initial state. 
We therefore restrict our attention to the four states that actually occur:
\[
\{q_0\},\{q_2\},\{q_0,q_1\},\{q_1,q_2\}\subseteq Q.
\]

The initial state of $\mathbf{A}_{\Npos}$ is $\{q_0\}$.
Suppose that we read $b$ from $\{q_0\}$.
As in the usual subset construction, we move to $\{q_2\}$ because $\delta(q_0,b)=q_2$.
Moreover, $q_2$ is an accepting state of $\mathbf{A}$, so $b$ is detected as an $\Npos$-position.

Now consider reading $a$ from $\{q_0\}$.
In this case, we move to $\{q_0,q_1\}$.
The original transition leads to the rejecting state $q_1$, so the reset rule is applied and $q_0$ is added to the set of possible states.
In terms of the game, this means that $a$ is detected as a
$\Ppos$-position.

Similarly, the reset rule is applied when we read $a$ or $b$ from $\{q_2\}$. That is, $\delta_\Npos(\{q_2\},a)=\delta_\Npos(\{q_2\},b)=\{q_0,q_1\}$.
Thus, for example, both $ba$ and $bb$ are detected as $\Ppos$-positions.

For the remaining transitions, no reset is needed.
For example,
$\delta_{\Npos}(\{q_0,q_1\},a)=\{q_1,q_2\}$.
Indeed, although the transition from $q_0$ leads to the rejecting state $q_1$, the transition from $q_1$ leads to the accepting state $q_2$.
Hence there is a possible run that reaches an accepting state, and the word is detected as an $\Npos$-position.

Continuing in this way, we obtain the following DFA
$\mathbf{A}_{\Npos}$:
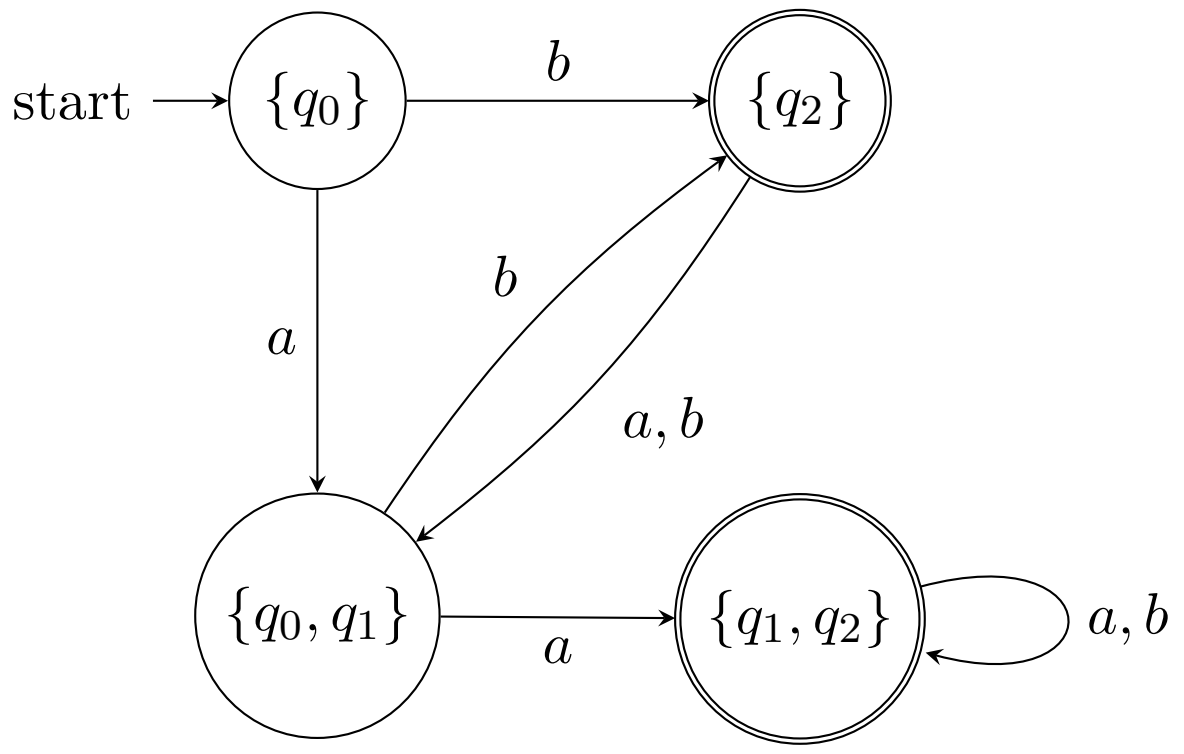
\begin{figure}[H]
\centering
\begin{tikzpicture}[
  ->,
  >=stealth,
  node distance=2cm
]
  \node[state, initial] (q0) {$\{q_0\}$};
  \node[state] (q01) [below=of q0] {$\{q_0,q_1\}$};
  \node[state,accepting] (q2) [right=of q0] {$\{q_2\}$};
  \node[state,accepting] (q12) [below=of q2] {$\{q_1,q_2\}$};

  \path
    (q0) edge node[left] {$a$} (q01)
    (q0) edge node[above] {$b$} (q2)
    (q01) edge[bend left=10] node[above left] {$b$} (q2)
    (q2) edge[bend left=10] node[below right] {$a,b$} (q01)
    (q01) edge node[below] {$a$} (q12)
    (q12) edge[loop right] node {$a,b$} ();
\end{tikzpicture}
\end{figure}\noindent
This automaton shows that
\[
    L_\Ppos(S)=(\varepsilon\cup a)(b\Sigma)^*,
\]
and $L_\Npos(S)$ is its complement.
In other words, a $\Ppos$-position consists of an optional initial
$a$, followed by zero or more two-letter blocks, each beginning
with $b$.\bqed
\end{example}

Taking complements gives the corresponding result for $\Ppos$-position languages.
\begin{corollary}
    Let $\Sigma$ be an alphabet and $S\subseteq \Sigma^+$ be a regular language.
    Then, the $\Ppos$-position language $L_{\Ppos}(S)$ of $\mathtt{WordSubtraction}(S)$ is regular.\bqed
\end{corollary}
This result also has a direct consequence for ordinary subtraction games.
If the subtraction set is ultimately periodic, then the sets of
$\Ppos$- and $\Npos$-positions are ultimately periodic as well.
Indeed, under the identification of $a^n$ with $n$, unary regular
languages correspond exactly to ultimately periodic subsets of
the nonnegative integers.
This includes, in particular, finite and cofinite subtraction sets.

For readers familiar with Grundy numbers, we briefly explain how the same idea extends to them.
This discussion is not needed for the later sections.
Let $G(w)$ denote the Grundy number of a position $w$.
For every fixed integer $g\geq 0$, we can construct a DFA recognizing
\[
    L_g(S)=\{w\in\Sigma^*:G(w)=g\}.
\]
Starting with the DFA $\mathbf{A}=(Q,\Sigma,\delta,q_0,F)$
recognizing $S$, we use tuples
\[
    (R_0,R_1,\ldots,R_g)\in(2^Q)^{g+1}
\]
as states.
The set $R_i$ tracks runs of $\mathbf{A}$ started at prefixes
with Grundy number $i$.
Because the empty word $\varepsilon$ has Grundy number $0$, the initial state is
\[
    (\{q_0\},\emptyset,\ldots,\emptyset).
\]

Suppose that the next input letter is $a$.
First, advance all tracked runs by setting
\[
    T_i=\{\delta(q,a):q\in R_i\}
\]
with $0\leq i\leq g$.
Then $T_i\cap F\neq\emptyset$ means that the current position has an option with Grundy number $i$.
By the mex rule, the Grundy number is the smallest nonnegative
integer that does not occur among the Grundy numbers of its options.
If there is an index $i\in\{0,\ldots,g\}$ with $T_i\cap F=\emptyset$,
let $h$ be the smallest such index; then the current Grundy number is $h$.
Otherwise, the current Grundy number exceeds $g$, and we set $h=g+1$.
The next state is $(R'_0,\ldots,R'_g)$, where
\[
    R'_i=
    \begin{cases}
        T_i\cup\{q_0\} & \text{if }i=h,\\
        T_i & \text{otherwise}.
    \end{cases}
\]
Thus, we start a new run in the component corresponding to the
current Grundy number, provided that this number is at most $g$.

A tuple $(R_0,\ldots,R_g)$ is declared accepting exactly when
\[
    R_g\cap F=\emptyset
    \quad\text{and}\quad
    R_i\cap F\neq\emptyset
\]
for every $0\leq i<g$.
Because $q_0\notin F$, adding $q_0$ does not change whether a component contains an accepting state.
Therefore, the acceptance condition expresses precisely that the mex is $g$.
The interpretation of each $R_i$ is preserved after every letter,
so induction on the input length shows that this DFA recognizes $L_g(S)$.
Values larger than $g$ need not be distinguished, and the state space $(2^Q)^{g+1}$ is finite.
Hence, $L_g(S)$ is regular.

We note that the regularity of $L_g(S)$ for each $g\geq 0$ was also independently established by Nanako Omiya, Ryo Yoshinaka, and Ayumi Shinohara of Tohoku University.

\subsection*{Exercises}
\begin{exercise}
    Let $\Sigma=\{a,b\}$.
    Construct an NFA with the specified number of states that recognizes each of the following languages:
    \begin{enumerate}
        \item $\Sigma^*abab\Sigma^*$ with five states
        \item $\{a,b,aa\}$ with three states
        \item $a^+b^*\cup a^*b^+$ with two states
        \item $\{\varepsilon\}$ with one state\diaqed
    \end{enumerate}
\end{exercise}
\begin{exercise}
    The class of regular languages is closed under union.
    This fact can be proved using NFAs.

    Let $L_1,L_2\subseteq\Sigma^*$ be regular languages. 
    For each $i\in\{1,2\}$, let $\mathbf{A}_i=(Q_i,\Sigma,\delta_i,q_{0i},F_i)$ be a DFA recognizing $L_i$, where $Q_1$ and $Q_2$ are disjoint.
    Construct an NFA $\mathbf{N}$ with an initial state $q_0\notin Q_1\cup Q_2$ as follows:\[
    Q=Q_1\cup Q_2\cup\{q_0\},
\]
and, for each $a\in\Sigma$,\[
\delta(q_0,a)=
\{\delta_1(q_{01},a),\delta_2(q_{02},a)\}.
\]
On states in $Q_1$ and $Q_2$, let $\mathbf{N}$ have the same transitions as $\mathbf A_1$ and $\mathbf A_2$, respectively. 
Make every state in $F_1\cup F_2$ accepting, and make $q_0$ accepting if and only if $q_{01}\in F_1$ or $q_{02}\in F_2$.
Write down the construction of $\mathbf{N}$ precisely, and prove that \[L(\mathbf{N})=L_1\cup L_2\]
holds.\diaqed
\end{exercise}
\begin{starexercise}
In Section~\ref{sec:intro}, we claimed that there is little difference between defining word subtraction games by removing suffixes and defining them by removing prefixes. 
One reason is that many language classes, including the class of regular languages, are closed under reversal.

For a regular language $L\subseteq\Sigma^*$,
let $L^{\mathsf{rev}}=\{w^{\mathsf{rev}}: w\in L\}$.
Prove that $L^{\mathsf{rev}}$ is also regular by the following steps:
Let $\mathbf{A}$ be a DFA recognizing $L$. Construct an automaton $\mathbf{N}$ by reversing the direction of every transition of $\mathbf{A}$, making the original initial state an accepting state, and making the original accepting states initial states.
Finally, show that
\[    L(\mathbf N)=L(\mathbf A)^{\mathsf{rev}}.
   \]
Strictly speaking, $\mathbf{N}$ need not be an NFA as defined above because its set of initial states need not contain exactly one state.
Adapt the construction in the previous exercise to obtain an NFA
with a single initial state that recognizes the same language.\diaqed
\end{starexercise}
\begin{exercise}
    Choose a regular subtraction set $S\subseteq\Sigma^+$.
Construct a DFA recognizing $L_\Npos(S)$ by following the
construction in Theorem~\ref{th3-reg}, and remove its unreachable states.
Check its answers on several short words directly from the game rules.\diaqed
\end{exercise}

\subsection*{Open Problems}
\begin{problem}[When is the $\Ppos$-position language a group language?]\label{prob3:group}
    For ordinary subtraction games, one can ask which subtraction sets $S$ give a purely periodic $\Ppos/\Npos$ sequence or Grundy sequence.
    Here, a sequence is called \emph{purely periodic} if it is periodic from the beginning, that is, if it has no preperiod.
    
    We omit the details here, but this question has a natural analogue for word subtraction games: for which subtraction sets $S$ is the $\Ppos$-position language a group language? Here, a regular language $L$ is called a \emph{group language}
    if its syntactic monoid is a group. 
    Equivalently, in the minimal DFA $(Q,\Sigma,\delta,q_0,F)$ for $L$,
the map
\[
    \delta(-,a):Q\to Q
\]
is a bijection for every $a\in\Sigma$.
    Over a unary alphabet, a regular language is a group language exactly when its membership sequence is purely periodic.\diaqed
\end{problem}

\newpage

\section{Beyond Regular Languages}\label{sec:cf}
\subsection{Motivation}
Theorem~\ref{th3-reg} shows that regular languages are, in a sense, closed under the operation of taking the $\Npos$-position language of a word subtraction game: starting from a regular subtraction set, we obtain a regular language again.
Therefore, if we want to go beyond regular languages, we must first allow the subtraction set itself to go beyond regular languages.

But how far should we go?
Although the definition of $\mathtt{WordSubtraction}(S)$ allows any language $S\subseteq\Sigma^+$ as a subtraction set,
we would like to study rules that have finite descriptions.
A natural next step is the class of \emph{context-free languages},
which extends the regular languages and can be described by finite grammars.

Context-free languages can express some conditions that finite
automata cannot.
For example, the language
\[
    \{a^n b a^n:n\geq 0\},
\]
which requires the two blocks of $a$'s to have the same length,
is context-free but not regular.
They also share some useful properties with regular languages.
In particular, Parikh's theorem states that the Parikh image
of every context-free language is semilinear \cite{DBLP:journals/jacm/Parikh66}.

Therefore, we can ask whether the regularity result from Section~\ref{sec:reg} extends to context-free languages:
if $S$ is context-free, must $L_\Ppos(S)$ also be context-free?
In contrast to the case that $S$ is regular, the answer is no.
In this section, we will study a context-free subtraction set
whose $\Ppos$-position language is not context-free.

The first concrete example of this phenomenon was discovered for the subtraction set
\[
S=a^+\cup b^+
\cup
\{w\in\{a,b\}^+: |w|_a=|w|_b\}.
\]
Thus, the game allows us to remove a nonempty suffix consisting only of $a$'s, a nonempty suffix consisting only of $b$'s, or a suffix containing the same number of $a$'s and $b$'s.
Yushin Kusajima and Shunya Oguchi, both at Nagoya University, discovered that the set of $\Ppos$-positions of this game is not context-free.
They later found that the same phenomenon also occurs when
\[
S=\mathsf{PAL}=\{w\in \{a,b\}^+:w=w^{\mathsf{rev}}\}.
\]

Of these two examples, we focus on the palindrome example, for which the proof is simpler.
This example also shows that the $\Npos$-position
language is not context-free.
At the end of the section, we give a simpler example in which only the $\Ppos$-position language is not context-free.
Before analyzing these games, we introduce context-free grammars and develop the tools needed for the proofs: a pumping lemma and several closure properties.

\subsection{Context-Free Grammars}
Let us begin by introducing context-free languages.
As regular languages can be described by finite
automata, context-free languages can be described
by \emph{context-free grammars} (\emph{CFGs}).
To see how context-free languages extend the class of regular languages, we begin with three special classes of CFGs:
\emph{regular grammars}, \emph{right-linear grammars},  and \emph{linear grammars}.

Regular grammars provide a third way of looking at regular languages,
alongside DFAs and NFAs.
In fact, their underlying structure is essentially the same as that of
NFAs.
Let us first give the definition.
\begin{definition}[Regular Grammars]
    A \emph{regular grammar} is a $4$-tuple $\mathbf{G}=(\Gamma,\Sigma,R,I)$
where
\begin{itemize}
\item $\Gamma$ is a finite set of \emph{nonterminal symbols}.
We usually write nonterminals using uppercase letters such as $A,B,C,\ldots$.
\item $\Sigma$ is an alphabet disjoint from $\Gamma$. Its elements are also called
\emph{terminal symbols}, or simply \emph{terminals}.
As before, we usually write terminals using lowercase letters such as $a,b,c,\ldots$.
\item $R$ is a finite set of \emph{production rules}, each of one of the following two forms:
\[
A\to aB
\qquad\text{or}\qquad
A\to \varepsilon,
\]
where $A,B\in\Gamma$ and $a\in\Sigma$.
\item $I\in\Gamma$ is the \emph{initial symbol}.
\end{itemize}
We now explain how a grammar generates words.
If $A\to\beta$ is a production rule, we can replace an occurrence of $A$ by $\beta$, leaving the surrounding symbols unchanged.
More precisely, for $\alpha_1,\alpha_2\in(\Sigma\cup\Gamma)^*$, we write
\[
    \alpha_1A\alpha_2\Rightarrow\alpha_1\beta\alpha_2
\]
and call this a \emph{one-step derivation}.
A sequence of zero or more one-step derivations
\[
\alpha_0\Rightarrow\alpha_1\Rightarrow\cdots\Rightarrow\alpha_n
\]
with $\alpha_0,\alpha_1,\ldots,\alpha_n\in (\Sigma\cup\Gamma)^*$ is called a \emph{derivation}. We sometimes also write $\alpha_0\Rightarrow^*\alpha_n$ when such a derivation exists.

A word $w\in\Sigma^*$ is said to be \emph{generated by $\mathbf{G}$}
if there is a derivation
\[
I\Rightarrow^*w.
\]
In other words, starting from the initial symbol $I$, we repeatedly apply production rules until we obtain the terminal word $w$.
The set\[
L(\mathbf{G})=\{w\in\Sigma^*:w\text{ is generated by $\mathbf{G}$}\}
\] is called the
\emph{language generated by $\mathbf{G}$}.\bqed
\end{definition}

\begin{example}\label{ex4-ends_with_ba}
    The regular grammar $\mathbf{G}=(\{A_0,A_1,A_2\},\{a,b\},R,A_0)$ with\[
    R=\{A_0\to aA_0,~A_0\to bA_0,~A_0\to bA_1,~A_1\to aA_2,~A_2\to \varepsilon\}
    \]
    generates the language\[
    L(\mathbf{G})=\Sigma^*ba.
    \]
    For example, \[{\color{red}A_0}\Rightarrow b{\color{red}A_1}\Rightarrow ba{\color{red}A_2}\Rightarrow ba\] and \[{\color{red}A_0}\Rightarrow b{\color{red}A_0}\Rightarrow ba{\color{red}A_0}\Rightarrow bab{\color{red}A_1}\Rightarrow baba{\color{red}A_2}\Rightarrow baba\] are derivations of this grammar.\bqed
\end{example}
As mentioned earlier, regular grammars are essentially the same as NFAs.
For example, in the grammar above, replace the nonterminals
$A_0,A_1,A_2$ by states $q_0,q_1,q_2$, respectively.
Then interpret a production such as
$A_0\to {\color{blue}b}A_1\in R$ as the transition $q_0\xrightarrow{{\color{blue}b}}q_1$.
Under this correspondence, the second derivation in Example~\ref{ex4-ends_with_ba}
corresponds directly to the run
\[
q_0\xrightarrow{b}q_0\xrightarrow{a}q_0\xrightarrow{b}q_1\xrightarrow{a}q_2.
\]
In addition, the production $A_2\to\varepsilon\in R$ means that a derivation is allowed to stop at $A_2$.
Therefore, we can declare the corresponding state $q_2$ to be accepting.
By this construction, we obtain exactly the same NFA as in Example~\ref{ex-NFA}.

More generally, for a given regular grammar $\mathbf{G}=(\Gamma,\Sigma,R,I)$, we can construct an equivalent NFA $\mathbf{A}=(\Gamma,\Sigma,\delta,I,F)$
by letting\[
\delta(A,a)=\{B\in \Gamma:A\to aB\in R\}
\]
for each $A\in \Gamma$ and $a\in \Sigma$, and\[
F=\{A\in \Gamma:A\to \varepsilon\in R\}.
\]
Conversely, every NFA can be translated into an equivalent regular grammar by reversing this construction.
Therefore, a language is generated by a regular grammar if and only if it is a regular language.

We can generalize the definition slightly by allowing production rules of the form
\[A\to uB\qquad\text{or}\qquad
A\to u,\]
where $A,B\in\Gamma$ and $u\in\Sigma^*$. 
Such a grammar is called a \emph{right-linear grammar}.
For example, the grammar $\mathbf{G}=(\{I\},\{a,b\},\{I\to aaI,I\to abI,I\to baI,I\to bbI,I\to \varepsilon\},I)$ has derivations such as\[
{\color{red}I}
\Rightarrow ab{\color{red}I}
\Rightarrow abaa{\color{red}I}
\Rightarrow abaa.
\]
This grammar generates the language $L(\mathbf{G})=(\Sigma\Sigma)^*$ consisting of all words of even length.

Despite this more general form of the production rules, right-linear grammars still generate only regular languages.
Intuitively, a rule that produces several letters at once can be replaced by a sequence of rules that produces them one letter at a time.
For example, a rule $I\to abI$ can be simulated by introducing a new nonterminal $I'$ and replacing
the original rule with
\[
    I\to aI',\qquad I'\to bI.
\]
Repeating this construction allows every right-linear grammar to be converted into an equivalent regular grammar.
Strictly speaking, one must also deal with rules of the form $A\to B$, which are called \emph{unit rules}.
This requires a small additional argument, but nothing fundamentally new is involved, so we will not go into the details here.

So far, regular grammars and right-linear grammars generate words by successively producing terminal symbols to the left of the remaining nonterminal.
Suppose that we also allow terminals to be produced on its right.
Then, we obtain a more general class of grammars that can generate nonregular languages.
\begin{definition}[Linear Grammars]
    A \emph{linear grammar} (or \emph{linear context-free grammar}) is a $4$-tuple $\mathbf{G}=(\Gamma,\Sigma,R,I)$
where
\begin{itemize}
\item $\Gamma$ is a finite set of \emph{nonterminal symbols}.
\item $\Sigma$ is an alphabet disjoint from $\Gamma$.
\item $R$ is a finite set of production rules, each of one of the following two forms:
\[
A\to uBv
\qquad\text{or}\qquad
A\to u,
\]
where $A,B\in\Gamma$ and $u,v\in\Sigma^*$.
\item $I\in\Gamma$ is the initial symbol.\bqed
\end{itemize}
\end{definition}
\begin{example}\label{ex4-a^nba^n}
    The linear grammar $\mathbf{G}=(\{I\},\{a,b\},R,I)$ with\[
    R=\{I\to aIa,I\to b\}
    \]
    generates the language\[
    L(\mathbf{G})=\{a^nba^n\in \{a,b\}^*:n\ge 0\}.
    \]
    For example, \[{\color{red}I}\Rightarrow b\] and\[
    {\color{red}I}\Rightarrow a{\color{red}I}a\Rightarrow aa{\color{red}I}aa\Rightarrow aabaa
    \] are derivations of this grammar.\bqed
\end{example}
\begin{example}
    The linear grammar $\mathbf{G}=(\{I\},\{a,b\},R,I)$ with\[
    R=\{I\to aIa,I\to bIb,I\to a,I\to b,I\to aa,I\to bb\}
    \]
    generates $\mathsf{PAL}=\{w\in\{a,b\}^+ : w=w^{\mathsf{rev}}\}$.
    The rules $I\to a$ and $I\to b$ give odd-length palindromes,
while $I\to aa$ and $I\to bb$ give even-length palindromes.
    For example, \[{\color{red}I}\Rightarrow b{\color{red}I}b\Rightarrow ba{\color{red}I}ab\Rightarrow babbab\] and\[
    {\color{red}I}\Rightarrow a{\color{red}I}a\Rightarrow aa{\color{red}I}aa\Rightarrow aab{\color{red}I}baa\Rightarrow aababaa
    \] are derivations of this grammar.\bqed
\end{example}
Recall from Propositions~\ref{prop2-pump_a^nba^n} and \ref{prop2-pump_pal} that these languages are not regular.
Therefore, the class of languages generated by linear grammars is strictly larger than the class of regular languages.

We now allow a production rule to create several nonterminals at once.
This gives a \emph{context-free grammar}, in which the right-hand side of a rule can be any word over $\Sigma\cup\Gamma$.
For example, a rule $A\to BC$ creates two nonterminals that can be rewritten independently.
\begin{definition}[Context-Free Grammars]
    A \emph{context-free grammar} (\emph{CFG} for short) is a $4$-tuple $\mathbf{G}=(\Gamma,\Sigma,R,I)$ where
\begin{itemize}
\item $\Gamma$ is a finite set of \emph{nonterminal symbols}.
\item $\Sigma$ is an alphabet disjoint from $\Gamma$.
\item $R$ is a finite set of production rules of the form $A\to \alpha$
where $A\in\Gamma$ and $\alpha\in(\Sigma\cup \Gamma)^*$.
\item $I\in\Gamma$ is the initial symbol.\bqed
\end{itemize}
\end{definition}
The name ``context-free'' means that a rule $A\to\alpha$ can be applied to any occurrence of $A$, regardless of the symbols surrounding it.\footnote{For readers familiar with disjunctive sums of games,
the independent rewriting of different nonterminals is similar to making a move in one component while leaving the others unchanged.}
\begin{example}\label{ex4-balance}
    The CFG $\mathbf{G}=(\{I\},\{a,b\},R,I)$ with\[
    R=\{I\to II,~I\to aIb,~I\to bIa,~I\to \varepsilon\}
    \]
    generates the language\[
    L(\mathbf{G})=\{w\in \{a,b\}^*:|w|_a=|w|_b\}.
    \]
    For example, \[
    {\color{red}I}\Rightarrow {\color{red}I}I\Rightarrow a{\color{red}I}bI\Rightarrow ab{\color{red}I}\Rightarrow abb{\color{red}I} a\Rightarrow abba
    \]
    and
    \[
    {\color{red}I}\Rightarrow {\color{red}I}I\Rightarrow^*  
    ab{\color{red}I}\Rightarrow
    ab{\color{red}I}I\Rightarrow^* abbbbaaa{\color{red}I} \Rightarrow^* abbbbaaaab
    \]
    are derivations of this grammar.
    In these examples, we always rewrite the leftmost
nonterminal, but this is not required.
For example, we can change the rewriting order
in the second derivation as follows:
\[
{\color{red}I}\Rightarrow I{\color{red}I}\Rightarrow II{\color{red}I}
\Rightarrow^* I{\color{red}I}ab
\Rightarrow^* {\color{red}I}bbbaaaab
\Rightarrow^* abbbbaaaab.
\]
The original derivation and this one are represented
by the same derivation tree, as we will see
in the next subsection.\bqed
\end{example}
\begin{definition}[Context-Free Languages]
A language $L$ is called a \emph{context-free language} (\emph{CFL} for short) if there exists a context-free grammar $\mathbf{G}$ such that $
L=L(\mathbf{G})$.\bqed
\end{definition}
It is known that the context-free language $\{w\in \{a,b\}^*:|w|_a=|w|_b\}$ cannot be generated by any linear grammar, although we omit the proof here. 
Thus, the class of context-free languages is strictly larger than the class of linear context-free languages.

Combining this observation with the results above, we obtain the following hierarchy:
\[ \text{regular languages} \subsetneq \text{linear context-free languages}\subsetneq \text{context-free languages}.\]

\subsection{Pumping Lemma for Context-Free Languages}
Our goal is to find an example in which the subtraction set $S$ is context-free, while the corresponding $\Ppos$- or $\Npos$-position language is not context-free.
To establish that a language is not context-free, we need a suitable tool for proving non-context-freeness.
Just as for regular languages, context-free languages admit a \emph{pumping lemma}.
To introduce the lemma, we first describe derivations using trees.

Earlier, we represented the derivation
\[
\rho:\quad I\Rightarrow II\Rightarrow^*  
    abI\Rightarrow
    abII\Rightarrow^* abbbbaaaI \Rightarrow^* abbbbaaaab
\]
as a sequence of rewriting steps (see Example~\ref{ex4-balance}).
This describes the process of rewriting one nonterminal at each step until a word consisting only of terminal symbols is obtained.
However, the derivation $\rho$ can also be viewed as a tree:
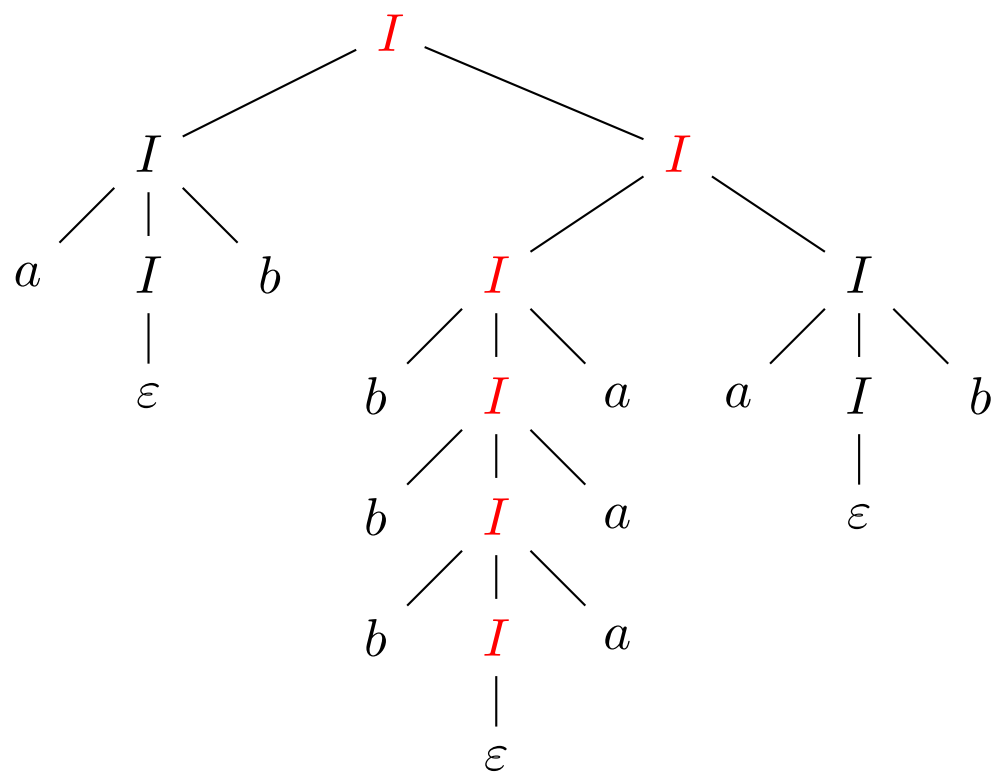
\begin{center}
\begin{tikzpicture}[
    level distance=8mm,
    sibling distance=8mm,
    every node/.style={font=\small}
]
\node[text=red] {$I$}
    child {
        node[xshift=-12mm] {$I$}
        child {
            node {$a$}
        }
        child {
            node {$I$}
            child {
                node {$\varepsilon$}
            }
        }
        child {
            node {$b$}
        }
    }
    child {
        node[xshift=15mm, text=red] {$I$}
        child {
            node[xshift=-8mm, text=red] {$I$}
            child {
                node {$b$}
            }
            child {
                node[text=red] {$I$}
                child {
                    node {$b$}
                }
                child {
                    node[text=red] {$I$}
                    child {
                        node {$b$}
                    }
                    child {
                        node[text=red] {$I$}
                        child {
                            node {$\varepsilon$}
                        }
                    }
                    child {
                        node {$a$}
                    }
                }
                child {
                    node {$a$}
                }
            }
            child {
                node {$a$}
            }
        }
        child {
        node[xshift=8mm] {$I$}
        child {
            node {$a$}
        }
        child {
            node {$I$}
            child {
                node {$\varepsilon$}
            }
        }
        child {
            node {$b$}
        }
    }
    };
\end{tikzpicture}
\end{center}
Each application of a production rule
$A\to X_1\cdots X_n$ is represented by a node labeled $A$ with children labeled $X_1,\ldots,X_n$, in that order.
%For a rule $A\to\varepsilon$, we draw a single child labeled $\varepsilon$.
Reading the terminal symbols at the leaves from left to right, with $\varepsilon$ contributing nothing, gives the generated word.
We call this a \emph{derivation tree}.

The part of the tree highlighted in red is responsible for making the tree deep.
For comparison, we can derive $(ab)^8$ by first using $I\to II$ to produce eight occurrences of $I$, and then deriving $ab$ from each one.
The resulting tree spreads out more widely, without requiring a long chain of nested nonterminals:
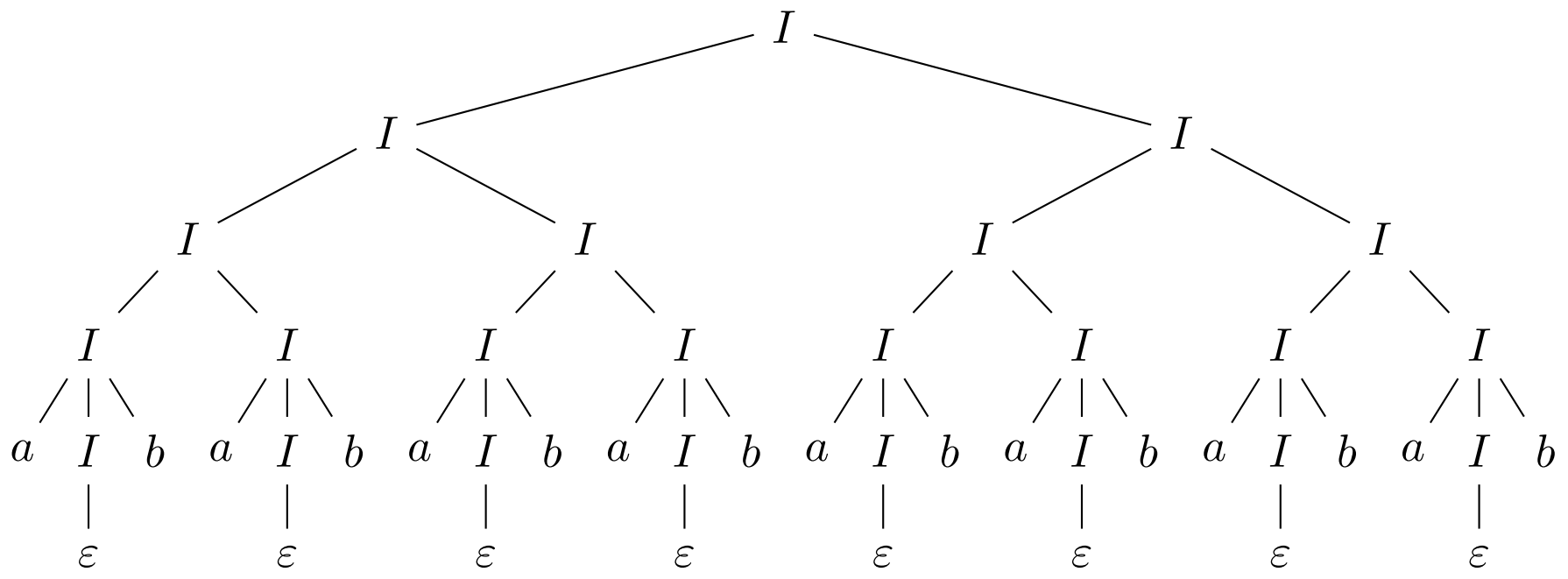
\begin{center}
\begin{tikzpicture}[
    level distance=8mm,
    level 1/.style={sibling distance=60mm},
    level 2/.style={sibling distance=30mm},
    level 3/.style={sibling distance=15mm},
    level 4/.style={sibling distance=5mm},
    every node/.style={font=\small}
]
\node {$I$}
    child {
        node {$I$}
        child {
            node {$I$}
            child {
                node {$I$}
                child { node {$a$} }
                child { node {$I$} child {node {$\varepsilon$}} }
                child { node {$b$} }
                }
            child {
                node {$I$}
                child { node {$a$} }
                child { node {$I$} child {node {$\varepsilon$}} }
                child { node {$b$} }
                }
        }
        child {
            node {$I$}
            child {
                node {$I$}
                child { node {$a$} }
                child { node {$I$} child {node {$\varepsilon$}} }
                child { node {$b$} }
                }
            child {
                node {$I$}
                child { node {$a$} }
                child { node {$I$} child {node {$\varepsilon$}} }
                child { node {$b$} }
                }
        }
    }
    child {
        node {$I$}
        child {
            node {$I$}
            child {
                node {$I$}
                child { node {$a$} }
                child { node {$I$} child {node {$\varepsilon$}} }
                child { node {$b$} }
                }
            child {
                node {$I$}
                child { node {$a$} }
                child { node {$I$} child {node {$\varepsilon$}} }
                child { node {$b$} }
                }
        }
        child {
            node {$I$}
            child {
                node {$I$}
                child { node {$a$} }
                child { node {$I$} child {node {$\varepsilon$}} }
                child { node {$b$} }
                }
            child {
                node {$I$}
                child { node {$a$} }
                child { node {$I$} child {node {$\varepsilon$}} }
                child { node {$b$} }
                }
        }
    };
\end{tikzpicture}
\end{center}
Here, we define the \emph{depth} of a derivation tree as the maximum number of nonterminal occurrences on a path from the root to a leaf.
Because the number of children of each node is bounded for this fixed grammar, a bound on the depth of the tree gives a bound on the number of leaves, and hence on the length of the generated word.
Therefore, for every positive integer $d$, there is a bound $\ell_d$ on the length of words generated by trees of depth at most $d$.

The pumping lemma for context-free languages, like the one for regular languages, ultimately comes from finiteness and the pigeonhole principle.
For regular languages, we use the finiteness of the set of states.
For context-free languages, we instead use the fact that the set $\Gamma$ of nonterminal symbols of the corresponding CFG is finite.
Let
\[
d=\#(\Gamma).
\]
As we have just seen, a derivation tree of depth $d$ can generate only words of length at most $\ell_d$.
In other words, any word longer than $\ell_d$ must have a derivation tree of depth at least $d+1$.
Thus, if we consider a derivation tree generating a word $w$ such that $|w|>\ell_d$, there must be a path on which at least $d+1=\#(\Gamma)+1$ nonterminals appear.
By the pigeonhole principle, some nonterminal $A\in\Gamma$ must appear at least twice on this path:
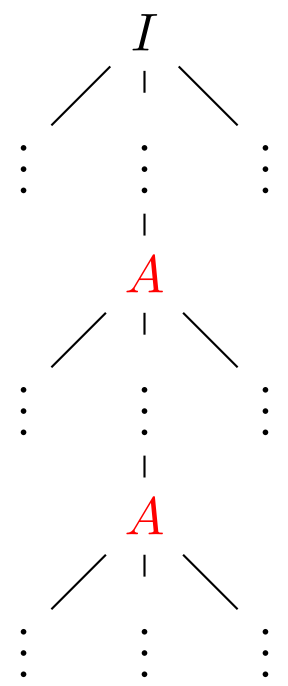
\begin{center}
\begin{tikzpicture}[
    every node/.style={font=\small},
    level distance=8mm,
    sibling distance=8mm
]
\node {$I$}
    child {node {$\vdots$}}
    child {
        node {$\vdots$}
        child {
            node[text=red] {$A$}
            child {node {$\vdots$}}
            child {
                node {$\vdots$}
                child {
                    node[text=red] {$A$}
                    child {node {$\vdots$}}
                    child {
                        node {$\vdots$}
                    }
                    child {node {$\vdots$}}
                }
            }
            child {node {$\vdots$}}
        }
    }
    child {node {$\vdots$}};
\end{tikzpicture}
\end{center}
Then, the part of the tree between the two occurrences of $A$ can be contracted to obtain a smaller tree:
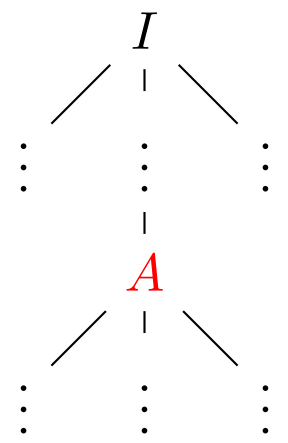
\begin{center}
\begin{tikzpicture}[
    every node/.style={font=\small},
    level distance=8mm,
    sibling distance=8mm
]
\node {$I$}
    child {node {$\vdots$}}
    child {
        node {$\vdots$}
        child {
                    node[text=red] {$A$}
                    child {node {$\vdots$}}
                    child {
                        node {$\vdots$}
                    }
                    child {node {$\vdots$}}
                }
    }
    child {node {$\vdots$}};
\end{tikzpicture}
\end{center}
Conversely, we can repeat this part any number of times and obtain another tree:
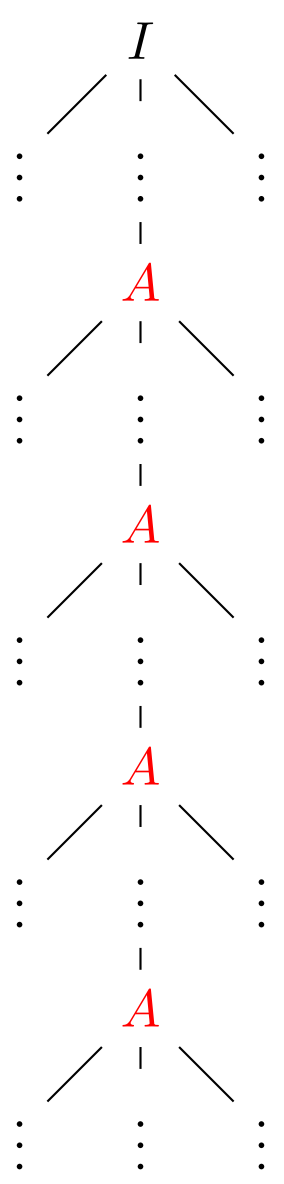
\begin{center}
\begin{tikzpicture}[
    every node/.style={font=\small},
    level distance=8mm,
    sibling distance=8mm
]
\node {$I$}
    child {node {$\vdots$}}
    child {
        node {$\vdots$}
        child {
            node[text=red] {$A$}
            child {node {$\vdots$}}
            child {
                node {$\vdots$}
                child {
            node[text=red] {$A$}
            child {node {$\vdots$}}
            child {
                node {$\vdots$}
                child {
            node[text=red] {$A$}
            child {node {$\vdots$}}
            child {
                node {$\vdots$}
                child {
                    node[text=red] {$A$}
                    child {node {$\vdots$}}
                    child {
                        node {$\vdots$}
                    }
                    child {node {$\vdots$}}
                }
            }
            child {node {$\vdots$}}
        }
            }
            child {node {$\vdots$}}
        }
            }
            child {node {$\vdots$}}
        }
    }
    child {node {$\vdots$}};
\end{tikzpicture}
\end{center}
Because the two subtrees have roots labeled by the same nonterminal $A$, the modified trees still follow the production rules of the grammar.
Therefore, their leaves give words generated by the same grammar.

Now, let $x_1$ be the word generated on the left side of the path from $I$ to the first occurrence of $A$, and let $x_2$ be the word generated on the right side.
Similarly, let $y_1$ and $y_2$ be the words generated on the left and right sides, respectively, between the two occurrences of $A$.
Finally, let $z$ be the word generated by the second occurrence of $A$.
Then, the original derivation tree corresponds to a derivation of the
form
\[
I\Rightarrow^* x_1 A x_2
\Rightarrow^* x_1 y_1 A y_2 x_2
\Rightarrow^* x_1 y_1 z y_2 x_2.
\]
Also, the derivations obtained by contracting and repeating this part can be written as
\[
I\Rightarrow^* x_1Ax_2\Rightarrow^* x_1zx_2
\]
and
\[
I\Rightarrow^* x_1y_1Ay_2x_2
\Rightarrow^* x_1y_1y_1Ay_2y_2x_2\Rightarrow^* x_1y_1y_1y_1Ay_2y_2y_2x_2\Rightarrow^* x_1y_1y_1y_1zy_2y_2y_2x_2,
\]
respectively.
Thus, for every $i\geq 0$, the word $x_1y_1^izy_2^ix_2$
is also generated by the grammar.
This is the basic intuition behind the pumping lemma for context-free languages.
\begin{theorem}[Pumping Lemma for Context-Free Languages]\label{th4-pump_cfl}
    Let $L\subseteq \Sigma^*$ be a context-free language. Then there exists a positive integer $N\ge 1$ such that, for every $w\in L$ such that $|w|\ge N$, there exist
$x_1,x_2,y_1,y_2,z\in\Sigma^*$ satisfying
\begin{itemize}
  \item $w=x_1y_1zy_2x_2$,
  \item $|y_1y_2|>0$, and
  \item $x_1y_1^izy_2^ix_2\in L$ for every $i\ge0$.
\end{itemize}
\begin{proof}[Proof sketch]
    Because $L$ is context-free, there is a CFG
$\mathbf{G}=(\Gamma,\Sigma,R,I)$ with $L=L(\mathbf{G})$.
Let $d=\#(\Gamma)$ and let $N=\ell_d+1$.

For a word $w\in L$ with $|w|\geq N$, choose a derivation tree with the fewest nodes among all derivation trees generating $w$.
The depth of the tree exceeds $d$, so some nonterminal occurs twice on a root-to-leaf path.
The construction above gives
\[
    w=x_1y_1zy_2x_2
\]
and shows that $x_1y_1^izy_2^ix_2\in L$ for every $i\geq 0$.

It remains to check that $|y_1y_2|>0$.
If both $y_1$ and $y_2$ were empty, replacing the subtree rooted at the upper occurrence by the subtree rooted at the lower occurrence would give a smaller derivation tree for the same word $w$.
This contradicts our choice of the tree.
Therefore, $|y_1y_2|>0$.
For a more detailed treatment, see, e.g., Chapter~2 of~\cite{sipser2013introduction}.
\end{proof}
\end{theorem}
Stronger versions of this pumping lemma give more control over the lengths and locations of the pumped factors.
The version above suffices for the next proposition.
We return to stronger versions in the exercises at the end of this section.
\begin{proposition}\label{prop4-pump_a^nba^nba^n}
    The language\[
    \{a^nba^nba^n\in \Sigma^*:n\ge 0\}
    \]
    is not context-free.
\end{proposition}
\begin{proof}
    Suppose, for contradiction, that
\[
L=\{a^n b a^n b a^n: n\geq 0\}
\]
is context-free. Let $N$ be the pumping length given by the pumping lemma, and consider
\[
w=a^Nba^Nba^N\in L.
\]
By the pumping lemma, we can write
\[
w=x_1y_1zy_2x_2
\]
such that $|y_1y_2|>0$
and
\[
x_1y_1^izy_2^ix_2\in L
\]
for every $i\ge 0$.

If either $y_1$ or $y_2$ contains some $b$, then pumping it twice introduces at least one additional $b$. Thus
\[
x_1y_1^2zy_2^2x_2
\]
contains at least three $b$'s and hence does not belong to $L$.

Therefore, both $y_1$ and $y_2$ consist only of $a$'s. By $|y_1y_2|>0$, at least one of them is nonempty.
Moreover, each of them lies entirely within one of the three blocks of $a$'s.
Thus, taking $i=2$ increases the lengths of at least one and at most two of the three blocks of $a$'s.
At least one block remains of length $N$, so the three lengths can no longer be equal.

For example, if $y_1$ lies in the first block of $a$'s and $y_2$ lies in the second block of $a$'s, then
\[
x_1y_1^2zy_2^2x_2
=
a^{N+|y_1|}ba^{N+|y_2|}ba^N,
\]
which is not in $L$.
The other possible cases are analogous. This contradicts the pumping lemma.
\end{proof}
Compare this with the context-free language
\[
\{a^n b a^n: n\geq 0\}
\]
from Example~\ref{ex4-a^nba^n}.
In this example, the two blocks of $a$'s can be kept in balance by the recursive rule
\[
I\to aIa.
\]
Each application of this rule adds one $a$ to each side, increasing the lengths of both blocks by one.
In contrast, the language
\[
    \{a^nba^nba^n:n\geq 0\}
\]
requires all three separated blocks of $a$'s to have the same length.
The pumping lemma above shows why this extra requirement cannot be maintained.

\subsection{Closure Properties of Context-Free Languages}
When working with context-free languages, it is useful to keep some of their closure properties in mind.
We begin with two basic properties: the class of context-free languages is closed under finite union and concatenation.
\begin{proposition}\label{prop4-closure_union_concat}
    Let $L_1,L_2\subseteq\Sigma^*$ be context-free languages. Then, the union $L_1\cup L_2$ and the concatenation $L_1L_2$ are context-free.
\end{proposition}
\begin{proof}
    Let $\mathbf{G}_1=(\Gamma_1,\Sigma,R_1,I_1)$ and
    $\mathbf{G}_2=(\Gamma_2,\Sigma,R_2,I_2)$ be CFGs such that $L(\mathbf{G}_1)=L_1$ and $L(\mathbf{G}_2)=L_2$.
    By renaming nonterminals if necessary, we can assume that $\Gamma_1\cap\Gamma_2=\emptyset$.
    Let $I$ be a new nonterminal symbol.
    Then, the CFG
    $\mathbf{G}=(\Gamma_1\cup\Gamma_2\cup\{I\},\Sigma,R,I)$
    with
    \[
    R=\{I\to I_1,\ I\to I_2\}\cup R_1\cup R_2
    \]
    generates $L_1\cup L_2$.
    Similarly, let $I'$ be a new nonterminal symbol.
    Then, the CFG $\mathbf{G}'=
    (\Gamma_1\cup\Gamma_2\cup\{I'\},\Sigma,R',I')$
    with
    \[
    R'=\{I'\to I_1I_2\}\cup R_1\cup R_2
    \]
    generates $L_1L_2$.
\end{proof}

On the other hand, the class of context-free languages is not closed under intersection.
\begin{proposition}\label{prop4-non_closure_intersection}
There exist context-free languages $L_1,L_2\subseteq\Sigma^*$ such that $L_1\cap L_2$ is not context-free.
\end{proposition}
\begin{proof}
    Let
\[
L_1=\{a^n b a^n:n\ge 0\}\cdot ba^*
\]
and
\[
L_2=a^*b\cdot\{a^nba^n:n\ge 0\},
\]
with $\Sigma=\{a,b\}$.
Note that the language $\{a^n b a^n:n\ge 0\}$ is context-free by Example~\ref{ex4-a^nba^n}.
Moreover, $ba^*$ and $a^*b$ are regular, and hence context-free.
Therefore, both $L_1$ and $L_2$ are context-free by Proposition~\ref{prop4-closure_union_concat}.
On the other hand,
\[
L_1\cap L_2
=
\{a^n b a^n b a^n:n\ge 0\},
\]
which is not context-free by Proposition~\ref{prop4-pump_a^nba^nba^n}.
\end{proof}
Although context-free languages are not closed under intersection, they are closed under intersection with regular languages.
Thus, restricting a context-free language to words accepted by a given DFA still gives a context-free language.
We omit the proof.
\begin{proposition}\label{prop4-closure_intersection_with_reg}
    Let $L_1\subseteq\Sigma^*$ be a context-free language and $L_2\subseteq\Sigma^*$ be a regular language.
    Then, $L_1\cap L_2$ is context-free.\bqed
\end{proposition}

Finally, the failure of closure under complement follows from the properties above.
Indeed, the class of context-free languages is closed under union.
If it were also closed under complement, then De Morgan's laws would imply that it is closed under intersection. This contradicts Proposition~\ref{prop4-non_closure_intersection}.
\begin{proposition}
There exists a context-free language whose complement is not context-free.\bqed
\end{proposition}
Let us illustrate this fact with an important example.
We have already seen in Proposition~\ref{prop4-pump_a^nba^nba^n} that
\[
\{a^n b a^n b a^n : n\ge 0\}
\]
is not context-free.
On the other hand, its complement is context-free:
\begin{proposition}\label{prop4-comp_of_a^nba^nba^n}
    The complement of the language\[
    L=\{a^nba^nba^n:n\ge0\}
    \]
    is context-free.
\end{proposition}
\begin{proof}
    Take complements relative to $\Sigma^*$, where $\Sigma=\{a,b\}$.
Let
\[
    K=\{a^mba^n:m,n\geq 0,\ m\neq n\}.
\]
This language is generated by the grammar with initial symbol $A$
and rules
\[
\begin{aligned}
    A\to aAa,~A\to aBb,~A\to bBa,~
    B\to aB,~B\to\varepsilon.
\end{aligned}
\]
Indeed, $A\to aBb$ makes the first block longer, while
$A\to bBa$ makes the second block longer.
The rule $A\to aAa$ adds one $a$ to each block.
Together, these rules generate exactly the words in $K$.

Now a word belongs to $L^c$ if it is not of the form
$a^{n_1}ba^{n_2}ba^{n_3}$, or if two adjacent blocks have
different lengths.
Therefore,
\[
    L^c
    =(a^*ba^*ba^*)^c
    \cup Kba^*
    \cup a^*bK.
\]
The first language is regular, and the other two are context-free by closure under concatenation.
Because context-free languages are closed under finite union,
$L^c$ is context-free.
(In fact, the same decomposition shows that $L^c$ is linear context-free.)
\end{proof}

\subsection{Back to Word Subtraction}\label{subsec:B2WS_cf}
We are now ready to return to word subtraction games.
We will show that, even when the subtraction set is linear
context-free, both position languages $L_\Ppos(S)$ and $L_\Npos(S)$ can fail to be
context-free.
Our example is
\[
    S=\mathsf{PAL}
    =\{w\in\{a,b\}^+:w=w^{\mathrm{rev}}\}.
\]
Readers who only want a simple example with a non-context-free
$\Ppos$-position language can skip ahead to
``A simpler example'' on page~\pageref{sec4-simple-example}.
\begin{theorem}\label{th4-PAL}
    The language $L_\Ppos(\mathsf{PAL})\subseteq \Sigma^*$ is not context-free.
\end{theorem}
\begin{proof}
    For integers $n_1,n_2,n_3\geq 5$, consider the word
\[
    w=aabba^{n_1}bbbba^{n_2}bbbba^{n_3}bbb.
\]
We first show that
\[
    w\in L_\Ppos(\mathrm{PAL})
    \quad\Longleftrightarrow\quad
    n_1=n_2=n_3.
\]
We will then use this characterization to obtain a contradiction to the pumping lemma.
To determine the status of $w$, we compute the outcomes of its prefixes from left to right.
    Let \[w'=aabba^{n_1}bbbba^{n_2},\] which is a prefix of $w$.
    The first part of the pattern is as follows:
    \begin{multline*}
        \mathsf{pat}(w')=\langle\Ppos ,a,\Npos,a,\Npos,b,\Ppos,b,\Npos,\underbrace{a,\Ppos,a,\Npos,a,\Npos,\ldots,a,\Npos}_{\text{the $a^{n_1}$ part}},  \\b,\Npos,b,\Ppos,b,\Npos,b,\Npos,\underbrace{a,\Ppos,a,\Npos,a,\Npos,\ldots,a,\Npos}_{\text{the $a^{n_2}$ part}}\rangle.
    \end{multline*}
    For example, $aabba$ is a $\Ppos$-position because the only possible options are \[aabb \quad\text{and}\quad a,\] and both are $\Npos$-positions:
    \[
    \mathsf{pat}(aabba)=\langle\Ppos ,a,{\color{blue}\Npos},a,\Npos,b,\Ppos,b,{\color{blue}\Npos},a,{\color{red}\Ppos}\rangle.
    \]
    Then, every word of the form $aabbaa^{j_1}$ with $1\le j_1< n_1$ is an $\Npos$-position because it has the option $aabba\in L_\Ppos(\mathsf{PAL})$.
    In addition, $aabba^{n_1}b$ is also an $\Npos$-position because it has the option $aab\in L_\Ppos(\mathsf{PAL})$.
    On the other hand, $aabba^{n_1}bb$ is a $\Ppos$-position because all of its possible options, \[aabba^{n_1}b, \quad aabba^{n_1}, \quad \text{and} \quad aa,
\]are $\Npos$-positions.
    In this way, we can compute the $\Ppos$/$\Npos$ pattern of $w'$ as described above.

    Next, we have that $w'b$ is a $\Ppos$-position if and only if $n_1\ne n_2$.
    This is because if $n_1=n_2$, $w'b=aabba^{n_1}bbbba^{n_2}b$ has the option $aab\in L_\Ppos(\mathsf{PAL})$.
    On the other hand, if $n_1\ne n_2$, then the only possible options are \[
    w'=aabba^{n_1}bbbba^{n_2} \quad\text{and} \quad aabba^{n_1}bbb,
    \]and both are $\Npos$-positions. 
    We next determine the status of $w'bbb$, which has $w'b$ as an option, by considering the same two cases.
    Specifically, $w'bbb$ is a $\Ppos$-position if and only if $n_1=n_2$. 
    On the other hand, regardless of the values of $n_1$ and $n_2$, both $w'bb$ and $w'bbbb$ are $\Npos$-positions. Furthermore, $w'bbbba^{j_3}$ with $1\leq j_3\leq n_3$ is a $\Ppos$-position if and only if $j_3=1$:
    \begin{multline*}
        \mathsf{pat}(aabba^{n_1}bbbba^{n_2}bbbba^{n_3})=\langle\Ppos ,a,\Npos,a,\Npos,b,\Ppos,b,\Npos,\\\underbrace{a,\Ppos,a,\Npos,a,\Npos,\ldots,a,\Npos}_{\text{the $a^{n_1}$ part}},  b,\Npos,b,\Ppos,b,\Npos,b,\Npos,\underbrace{a,\Ppos,a,\Npos,a,\Npos,\ldots,a,\Npos}_{\text{the $a^{n_2}$ part}},\\
        b,\theta_{n_1\ne n_2},b,\Npos,b,\theta _{n_1=n_2},b,\Npos,\underbrace{a,\Ppos,a,\Npos,a,\Npos,\ldots,a,\Npos}_{\text{the $a^{n_3}$ part}}\rangle
    \end{multline*}
    where\[
    \theta_\psi=\Ppos \quad\iff\quad\psi\text{ is true}.
    \]

    Now put
\[
    v=aabba^{n_1}bbbba^{n_2}bbbba^{n_3},
\]
so that $w=vbbb$.
It remains to determine the outcomes of $vb$, $vbb$, and $vbbb$,
in that order.
At each step, we list all options and use the outcomes
already determined.
    First, $vb$ is a $\Ppos$-position if and only if $n_1\ne n_2\land n_1\ne n_3$.
    This is because this word has options\begin{itemize}
        \item $v\in L_\Npos(\mathsf{PAL})$,
        \item $aabba^{n_1}bbbba^{n_2}bbb\in L_{\theta_{n_1= n_2}}(\mathsf{PAL})$,
        \item $aabba^{n_1}bbb\in L_\Npos(\mathsf{PAL})$ if $n_2=n_3$, and
        \item $aab\in L_\Ppos(\mathsf{PAL})$ if $n_1=n_3$.
    \end{itemize}
    Thus, the word is an $\Npos$-position if and only if either $n_1=n_2$, in which case the second option is a $\Ppos$-position,
or $n_1=n_3$, in which case the fourth option exists.

Then, $vbb$ is a $\Ppos$-position if and only if $(n_1=n_2\lor n_1=n_3)\land n_2\ne n_3$.
This is because this word has options\begin{itemize}
\item $vb\in L_{\theta_{n_1\ne  n_2\land n_1\ne n_3}}(\mathsf{PAL})$,
\item $v\in L_\Npos(\mathsf{PAL})$,
\item $aabba^{n_1}bbbba^{n_2}bb\in L_\Npos(\mathsf{PAL})$,
\item $aabba^{n_1}bb\in L_\Ppos(\mathsf{PAL})$ if $n_2=n_3$, and
\item $aa\in L_\Npos(\mathsf{PAL})$ if $n_1=n_3$.
\end{itemize}
Thus, the word is an $\Npos$-position if and only if either $n_1\ne  n_2\land n_1\ne n_3$, in which case the first option is a $\Ppos$-position,
or $n_2=n_3$, in which case the fourth option exists.

Finally, $w=vbbb$ is a $\Ppos$-position if and only if $n_1=n_2=n_3$.
This is because $w$ has options\begin{itemize}
    \item $vbb\in L_{\theta_{(n_1=n_2\lor n_1=n_3)\land n_2\ne n_3}}(\mathsf{PAL})$,
    \item $vb\in L_{\theta_{n_1\ne  n_2\land n_1\ne n_3}}(\mathsf{PAL})$,
    \item $v\in L_\Npos(\mathsf{PAL})$,
    \item $aabba^{n_1}bbbba^{n_2}b\in L_{\theta_{n_1\ne n_2}}(\mathsf{PAL})$, and
    \item $aabba^{n_1}b\in L_\Npos(\mathsf{PAL})$ if $n_2=n_3$.
\end{itemize}
Thus, if $n_1\ne n_2$, the fourth option is a $\Ppos$-position.
In addition, if $n_1=n_2$ but $n_2\ne n_3$, then the first option is a $\Ppos$-position.
Therefore, $\lnot (n_1=n_2=n_3)$ implies that $w$ is an $\Npos$-position.
Conversely, if $n_1=n_2=n_3$, then all options are $\Npos$-positions, and hence $w$ is a $\Ppos$-position.

Suppose, for contradiction, that $L_\Ppos(\mathrm{PAL})$
is context-free.
By Proposition~\ref{prop4-closure_intersection_with_reg}, its intersection
\[
    L=L_\Ppos(\mathrm{PAL})\cap aabba^*bbbba^*bbbba^*bbb
\]
with this regular language is also context-free.
(It is easily shown that $aabba^*bbbba^*bbbba^*bbb$ is regular.)
Let $N\ge 5$ be an integer greater than the pumping length given by the pumping lemma, and consider
\[
aabba^Nbbbba^Nbbbba^Nbbb\in L.
\]
Indeed, $aabba^Nbbbba^Nbbbba^Nbbb$ belongs to $L_{\Ppos}(\mathsf{PAL})$ because this word corresponds to $ w=vbbb$ with $n_1=n_2=n_3=N$.
By the pumping lemma, we can write
\[
w=x_1y_1zy_2x_2
\]
such that $|y_1y_2|>0$
and
\[
x_1y_1^izy_2^ix_2\in L
\]
for every $i\ge 0$.
Neither $y_1$ nor $y_2$ can contain the letter $b$ because taking $i=2$ would increase the number of $b$'s and take the word outside $aabba^*bbbba^*bbbba^*bbb$.
In addition, neither can contain a letter from the fixed initial block $aa$ because pumping that letter would change the required prefix $aabb$.
Thus, every nonempty $y_j$ lies entirely within one of the
three $a^N$-blocks.

Taking $i=2$ increases the lengths of at least one and at most two of these blocks.
Therefore, the three lengths are no longer all equal, and each is still at least $5$.
By the characterization proved above, the resulting word is not a $\Ppos$-position.
This contradicts $x_1y_1^2zy_2^2x_2\in L$.
\end{proof}
Note that, already at the stage where we obtained the condition \[
n_1\ne n_2 \land n_1\ne n_3
\]
characterizing when $vb$ is a $\Ppos$-position, one can show that the corresponding language is not context-free.
However, we used the condition $n_1=n_2=n_3$ to simplify the proof.
\begin{corollary}[Non-Context-Freeness for Context-Free Subtraction Sets]
Even when the subtraction set $S$ is context-free, the corresponding $\Ppos$-position language $L_\Ppos(S)$ need not be context-free.\bqed
\end{corollary}

Because context-free languages are not closed under complement,
Theorem~\ref{th4-PAL} does not settle whether $L_\Npos(\mathrm{PAL})$ is context-free.
Moreover, the same family of words does not give a contradiction: among words of the form
\[
    aabba^{n_1}bbbba^{n_2}bbbba^{n_3}bbb
\]
with $n_1,n_2,n_3\geq 5$,
the $\Npos$-positions are exactly those for which the three exponents are not all equal.
The language described by this condition is context-free, by an argument similar to Proposition~\ref{prop4-comp_of_a^nba^nba^n}.

To prove non-context-freeness for $\Npos$-positions, we instead use the words ending in $bb$, whose outcomes were also determined in the proof above.

\begin{theorem}\label{th4-Npos}
 The language $L_\Npos(\mathsf{PAL})\subseteq \Sigma^*$ is not context-free.
\end{theorem}
\begin{proof}[Proof sketch]
In the proof of Theorem~\ref{th4-PAL}, we showed that, for
$n_1,n_2,n_3\geq 5$, the word
\[
    aabba^{n_1}bbbba^{n_2}bbbba^{n_3}bb
\]
is an $\Npos$-position if and only if
\[
    (n_1\neq n_2\land n_1\neq n_3)\lor(n_2=n_3).
\]
Using this characterization, we can show that
\[
L_{\Npos}(\mathsf{PAL})\cap aabba^*bbbba^*bbbba^*bb
\]
is not context-free.

To prove this, we use a refined pumping lemma that gives more control over the positions of the pumped factors.
This allows us to analyze how pumping changes the comparisons between $n_1$, $n_2$, and $n_3$.
The details are developed in Exercises~\ref{ex4:ogden} and~\ref{ex4:ogden_Npos}.
\end{proof}

Because $\mathrm{PAL}$ is generated by a linear grammar,
the two non-context-freeness results already hold for linear context-free subtraction sets.
The following table summarizes the results obtained so far:
\begin{table}[H]
\centering
\begin{tabular}{c||c|c}
Subtraction set $S$ & $L_{\Ppos}(S)$ & $L_{\Npos}(S)$\\
\hline\hline
Finite & \multirow{2}{*}{Regular} & \multirow{2}{*}{Regular} \\
\cline{1-1}
Regular & & \\
\hline
Linear context-free &
\multirow{2}{*}{Not necessarily context-free} &
\multirow{2}{*}{Not necessarily context-free} \\
\cline{1-1}
Context-free & & \\
\hline
\end{tabular}
\caption{The relationship between subtraction sets and position languages.}
\end{table}

\paragraph{A simpler example.}\label{sec4-simple-example}
There is also a much simpler example in which the subtraction set is linear context-free but the $\Ppos$-position language is not context-free.
Let $\Sigma=\{a,b\}$ and set
\[
    L=\{a^nba^nba^n:n\geq 0\},
    \qquad
    S=\Sigma^+\setminus L.
\]
We claim that
\[
    L_{\Ppos}(S)=\{\varepsilon\}\cup L,
    \qquad
    L_{\Npos}(S)=S.
\]
The argument in
Proposition~\ref{prop4-comp_of_a^nba^nba^n} shows that $S$ is linear context-free.
Every word in $S$ is an $\Npos$-position because the whole word can be removed, leaving $\varepsilon$.
On the other hand, no word in $L$ is a proper prefix of another word in $L$.
Therefore, every legal move from a word in $L$ leaves a nonempty prefix in $S$, which is an $\Npos$-position.
This proves the claim.

Because $L$ is not context-free, neither is
$L_{\Ppos}(S)=\{\varepsilon\}\cup L$.
However, in this example, $L_{\Npos}(S)=S$ is context-free.
Thus, the palindrome example establishes a stronger conclusion: both the $\Ppos$- and $\Npos$-position languages can fail to be context-free, even when the subtraction set is linear context-free.

Interestingly, the reverse situation cannot occur when $S$ is context-free.
Indeed, the identity
\[
    L_{\Npos}(S)=L_{\Ppos}(S)S
\]
and closure under concatenation imply that, if both $S$ and $L_{\Ppos}(S)$ are context-free, then $L_{\Npos}(S)$ is context-free as well.

The same construction applies more generally to any prefix-free language $K\subseteq\Sigma^+$:
if $S=\Sigma^+\setminus K$, then
\[
    L_{\Ppos}(S)=\{\varepsilon\}\cup K,
    \qquad
    L_{\Npos}(S)=S.
\]
Here, \emph{prefix-free} means that no word in $K$ is a proper prefix of another word in $K$.
Hiyu Inoue of Hiroshima University also independently discovered an example of a context-free subtraction set $S$ for which $L_{\Ppos}(S)$ is not context-free but $L_{\Npos}(S)$ is context-free, together with the general observation above for prefix-free languages $K$.
This gives a simple way to construct further examples.

For instance, consider
\[
    K=\{a^kba^{k-1}b\cdots a^2ba:k\geq 2\},
    \qquad
    S=\Sigma^+\setminus K.
\]
The language $K$ is prefix-free because different initial blocks of $a$'s are followed by $b$.
Moreover, $S$ is context-free.
To see this, first consider words of the form
\[
    a^{n_1}ba^{n_2}b\cdots ba^{n_r}
\]
with $r\ge 2$ and $n_i\ge 1$.
Such a word belongs to $K$ exactly when its final block has length $1$ and each block is one letter shorter than the previous block.
A failure of the latter condition can be described by a context-free grammar that generates two adjacent blocks of lengths $m,n\geq 1$ with $m\neq n+1$, adding matching $a$'s on either side of the $b$ between them.
The blocks before and after this pair are unrestricted.
Nonempty words of the wrong form, or with a final block of length other than $1$, form regular languages.
Together, these observations show that $S$ is context-free.
Nevertheless, it is easily shown that the Parikh image
\[
    \mathsf{Pk}(L_{\Ppos}(S))=\{(0,0)\}\cup
    \left\{
        \left(\frac{k(k+1)}2,k-1\right):k\geq 2
    \right\}
\]
of
$L_{\Ppos}(S)=\{\varepsilon\}\cup K$ is not semilinear.
Thus, a context-free subtraction set can give rise to a $\Ppos$-position language whose Parikh image is not semilinear.

The position languages of several other context-free subtraction sets have also been studied, revealing different relationships between the language classes of subtraction sets and their position languages.
We leave these other examples for another occasion.

\subsection*{Exercises}
\begin{exercise}
    Let $\Sigma=\{a,b\}$. Give a context-free grammar generating each of the following languages:
    \begin{enumerate}
        \item $\{a^nb^n:n\ge 0\}$
        \item $\{a^mb^ma^nb^n:m,n\ge 0\}$
        \item $\{w\in\Sigma^*:|w|_a>|w|_b\}$
        \item $\mathsf{PAL}^c=\{\varepsilon\}\cup \{w\in \Sigma^*:w\ne w^\mathsf{rev}\}$\diaqed
    \end{enumerate}
\end{exercise}
\begin{exercise}
    Prove that the class of context-free languages is closed under reversal.
    In particular, given a CFG generating a language $L$, construct a CFG generating $L^{\mathsf{rev}}$.\diaqed
\end{exercise}
\begin{starexercise}\label{ex4:pump_cfl_bounded}
Show that Theorem~\ref{th4-pump_cfl} can be strengthened by
requiring the decomposition
\[
    w=x_1y_1zy_2x_2
\]
to satisfy
\[
    |y_1zy_2|\leq N
\]
in addition to the conditions stated in the theorem.\diaqed
\end{starexercise}
\begin{exercise}\label{ex4:crossing_copy}
Let $\Sigma=\{a,b\}$.
Using the strengthened pumping lemma from
    Exercise~\ref{ex4:pump_cfl_bounded}, prove that
    \begin{itemize}
        \item $\{a^mb^na^mb^n:m,n\geq 0\}$ and
        \item $\{ww:w\in\Sigma^*\}$
    \end{itemize}
    are not context-free.\diaqed
\end{exercise}
\begin{starexercise}\label{ex4:ogden}
    Exercise~\ref{ex4:pump_cfl_bounded} strengthens
Theorem~\ref{th4-pump_cfl} by bounding the length of $y_1zy_2$.
Another useful strengthening allows us to specify a sufficiently
long factor of the input word and guarantees that at least one
nonempty pumping part lies entirely within that factor:
    \begin{adjustwidth}{1.5em}{1.5em}\vspace{-15pt}
\begin{proposition}[Lemma~2 of~\cite{DBLP:journals/tcs/InoueHS23}]\label{prop4-ogden_pump}
Let $L\subseteq\Sigma^*$ be a context-free language.
Then there exists a positive integer $N\ge1$ such that, whenever $w=uw'v\in L$ for some $u,w',v\in\Sigma^*$ with $|w'|\ge N$, there exist
$x_1,x_2,y_1,y_2,z\in\Sigma^*$
satisfying:
\begin{itemize}
\item $w=x_1y_1zy_2x_2$,
\item either
\begin{itemize}
\item $1\le|y_1|< N$ and the occurrence of $y_1$ in $w$ is contained in the occurrence of $w'$ in $w$, or
\item $1\le|y_2|< N$ and the occurrence of $y_2$ in $w$ is contained in the occurrence of $w'$ in $w$, and
\end{itemize}
\item $x_1y_1^izy_2^ix_2\in L$ for every $i\ge0$.
\end{itemize}
Here, an occurrence is said to be contained in another occurrence if it lies entirely within the latter.\bqed
\end{proposition}
\end{adjustwidth}\vspace{5pt}
Intuitively, this lemma allows us to constrain the location of the pumping pair $(y_1,y_2)$ by specifying in advance a sufficiently long factor $w'$.
However, unlike in the case of regular languages, we can only guarantee that one of the occurrences of $y_1$ and $y_2$ is contained in the occurrence of $w'$.

Now, show that the language
\[
L=\{a^{n_1}ba^{n_2}ba^{n_3} :n_1,n_2,n_3\geq 0,~
n_1\ne n_2 \land n_1\ne n_3\}
\]
is not context-free.  The proof strategy is as follows: Let $N$ be the pumping length, and consider the word
\[
uw'v=a^Nba^{N+N!}ba^{N+N!}\in L
\]
with $u=\varepsilon$, $w'=a^N$, and $v=ba^{N+N!}ba^{N+N!}$.\diaqed
\end{starexercise}
\begin{starexercise}\label{ex4:ogden_Npos}
    Use the ideas from Exercise~\ref{ex4:ogden} to complete the proof of Theorem~\ref{th4-Npos}.\diaqed
\end{starexercise}

\subsection*{Open Problems}
\begin{problem}[Which subclasses of context-free languages preserve context-freeness?]
We have seen that a context-free subtraction set $S$ can give rise to a position language $L_{\Ppos}(S)$ that is not context-free.
Can we restrict $S$ to a natural subclass of context-free languages so that $L_{\Ppos}(S)$ is always context-free?
We are interested in classes that include nonregular languages, extending the result of Theorem~\ref{th3-reg}.

Two familiar candidates are linear context-free languages and \emph{deterministic context-free languages} (\emph{DCFLs}).
The former are generated by linear grammars, as discussed in the main text.
The latter are recognized by \emph{deterministic pushdown automata}: roughly speaking, finite automata that use a stack and operate without nondeterministic choices.

However, neither class has the desired property.
The language $\mathsf{PAL}$ is linear context-free, but $L_{\Ppos}(\mathsf{PAL})$ is not context-free, as shown above.
Similarly, as mentioned earlier, the subtraction set
\[
    S=a^+\cup b^+\cup
    \{w\in\{a,b\}^+:|w|_a=|w|_b\}
\]
is deterministic context-free, but $L_{\Ppos}(S)$
is not context-free.

Are there other natural candidates?
Yushin Kusajima and the first author have discussed this question, but so far have obtained positive results only under very restrictive assumptions.
\diaqed
\end{problem}

\begin{problem}[What is a natural upper bound?]
We have seen that, when $S$ is a context-free language, $L_{\Ppos}(S)$ need not be context-free.
Is there a natural class of languages that contains $L_{\Ppos}(S)$ for every context-free subtraction set $S$?

One natural candidate is the class of \emph{multiple context-free languages} (\emph{MCFLs}) \cite{DBLP:journals/tcs/SekiMFK91}, which properly contains the class of context-free languages.
Roughly speaking, a multiple context-free grammar generates tuples of words and uses rules that combine their components without copying them.
This allows dependencies between several parts of a word to be maintained during a derivation.
MCFLs retain many useful properties of context-free languages; in particular, their Parikh images are semilinear.

However, the example in Section~\ref{sec4-simple-example} shows that a context-free
subtraction set can have a $\Ppos$-position language whose Parikh image is not semilinear.
Thus, MCFLs are not sufficient.
More generally, any language class containing $L_{\Ppos}(S)$ for every context-free subtraction set $S$ must include languages with non-semilinear Parikh images.
What natural language class can provide such an upper bound?
\diaqed
\end{problem}

\newpage
\section{Variants}\label{sec:variant}
\subsection{Prefix and Suffix Removal}\label{subsec:two-sided}
So far, we have considered word subtraction games in which a move consists of removing a suffix from the current word. 
A natural variant is to allow moves from both ends of the word.
More precisely, let $S,T\subseteq\Sigma^+$ be languages. 
In the \emph{two-sided word subtraction game} $\mathtt{WordSubtraction}(S,T)$, a player can either remove a prefix belonging to $S$, or remove a suffix belonging to $T$. 
Thus, from a position $w$, a move consists of replacing $w$ by $w'$ whenever either
\[
w=sw',\quad s\in S,
\]
or
\[
w=w't,\quad t\in T.
\]
As before, the player who cannot make a move loses.
We expect that general results can also be obtained for the two-sided version.
However, we currently understand only a few special cases, which we discuss here.

As our first case study, consider $\Sigma=\{a,b\}$, $S=\Sigma^*a$, and $T=b\Sigma^*$. 
In other words, a player can remove either a prefix ending in $a$ or a suffix beginning with $b$.
Interestingly, this game is equivalent to a combinatorial game known as the \emph{Timber game} \cite{nowakowski2013game}. 
In the Timber game, a position is represented by a row of dominoes, each of which is tilted either to the left or to the right.
On each turn, a player chooses one domino and knocks it down. 
If the chosen domino is tilted to the left, then all dominoes to its left are also knocked down, regardless of their directions. 
Similarly, if it is tilted to the right, then all dominoes to its right are knocked down. 
The player who knocks down all the remaining dominoes wins.
Indeed, if we regard a domino tilted to the left as the letter $a$ and a domino tilted to the right as the letter $b$, then the Timber game is exactly the same as $\mathtt{WordSubtraction}(\Sigma^*a,b\Sigma^*)$.

The $\Ppos$-position language $L_\Ppos(\Sigma^*a,b\Sigma^*)$ of $\mathtt{WordSubtraction}(\Sigma^*a,b\Sigma^*)$ is the context-free language known as the \emph{Dyck language}, which is generated by the CFG $(\{I\},\{a,b\},R,I)$ with\[
R=\{I\to II,~I\to aIb,~I\to \varepsilon\}.
\]
For example,\begin{itemize}
    \item $I\Rightarrow aIb\Rightarrow ab$,
    \item $I\Rightarrow II\Rightarrow aIbI\Rightarrow abI\Rightarrow abaIb\Rightarrow abab$, and
    \item $I\Rightarrow^* III\Rightarrow^*abII\Rightarrow^*abaabbI\Rightarrow^*abaabbab$
\end{itemize}
are derivations of this grammar.
In fact, the Dyck language consists of all well-formed sequences of parentheses, where $a$ is regarded as an opening parenthesis and $b$ as a closing parenthesis. We call such a word a \emph{Dyck word}. (Dyck words can also be represented as lattice paths, called \emph{Dyck paths}.)
For example, the derived words $ab$, $abab$, and $abaabbab$ correspond to the parenthesis sequences $\mathtt{()}$, $\mathtt{()()}$, and $\mathtt{()(())()}$, respectively, and hence are Dyck words. On the other hand, $a$, $b$, $ba$, and $aaba$ are not Dyck words.

But why are Dyck words $\Ppos$-positions in $\mathtt{WordSubtraction}(\Sigma^*a,b\Sigma^*)$? 
To understand this, it is useful to consider a characterization of Dyck words in terms of their \emph{balance}.
For a word $w$, define its balance $\mathsf{bal}(w)\in\mathbb{Z}$ by
\[
\mathsf{bal}(w) =|w|_a-|w|_b.
\]
Then a word $w\in\{a,b\}^*$ is a Dyck word if and only if it satisfies the following two conditions:
\begin{itemize}
\item $\mathsf{bal}(w')\geq 0$ for every prefix $w'$ of $w$, and
\item $\mathsf{bal}(w')\leq 0$ for every suffix $w'$ of $w$.
\end{itemize}
Intuitively, when reading the word from left to right, we can never close more parentheses than we have opened. Similarly, when reading it from right to left, we can never open more parentheses than we have closed.
In particular, every Dyck word has balance $0$.

We can now see why the Dyck words are exactly the $\Ppos$-positions.
First, suppose that $w$ is a Dyck word, and consider a move that removes a suffix beginning with $b$, leaving a word $w'$. 
Then $w'b$ is a prefix of $w$, and hence
\[
\mathsf{bal}(w'b)\geq 0.
\]
Therefore, $\mathsf{bal}(w')>0$. Because a Dyck word has balance $0$, $w'$ cannot be a Dyck word. 
The same argument shows that removing a prefix ending in $a$ also leaves a word that is not a Dyck word. Thus, no move from a Dyck word leads to another Dyck word.

Conversely, suppose that $w$ is not a Dyck word.
By the characterization above, either some prefix of $w$
has negative balance or some suffix of $w$ has positive balance.
In the first case, let $w'b$ be the shortest prefix of $w$ having negative balance.
Then $\mathsf{bal}(w')=0$ by $\mathsf{bal}(w'b)=-1$.
In addition, for every decomposition $w'=ps$ of $w'$, $s$ has a non-positive balance because $\mathsf{bal}(s)=\mathsf{bal}(w')-\mathsf{bal}(p)=0-\mathsf{bal}(p)\le 0$.
This implies that $w'$ is a Dyck word.
Thus, by removing the suffix beginning with the $b$ immediately following $w'$, we can move to the Dyck word $w'$. 
The second case is analogous. 
Hence, from every non-Dyck word, there is a move to a Dyck word.
Summarizing, by the Partition Theorem, a word $w$ is a Dyck word if and only if $w$ is a $\Ppos$-position.

Both $S=\Sigma^*a$ and $T=b\Sigma^*$ in this example are regular languages. In contrast, the $\Ppos$-position language $L_\Ppos(S,T)$ of $\mathtt{WordSubtraction}(S,T)$, while context-free, is not regular. 
(This can be easily shown using the pumping lemma.) Thus, at least the closure property stated in Theorem~\ref{th3-reg} does not extend to the two-sided setting.

One might still expect that, if both $S$ and $T$ are regular,
then $L_{\Ppos}(S,T)$ is context-free.
The next example shows that this does not hold in general.

The example we consider next is obtained by reformulating a combinatorial game known as \emph{End-Nim} \cite{end-nim2001} as a word subtraction game.
In End-Nim, a position is a sequence of heaps of stones, such as $(3,2,5,4)$.
On each turn, a player chooses either the leftmost or the rightmost heap and removes any positive number of stones from it. 
The player who makes the last move wins.
The $\Ppos$-positions of End-Nim can be characterized as follows. 
For a position\[
A=(n_1,n_2,\ldots,n_k),
\]
let $l(A)$ be the largest integer $i$ such that
\[
n_1=\cdots=n_{i-1}\le n_i,
\]
and let $r(A)$ be the largest integer $j$ such that
\[
n_{k-j+1}\ge n_{k-j+2}=\cdots=n_k.
\]
When $n_1=\cdots=n_k$, we define $r(A)=0$.
\begin{proposition}[Theorem~2.1 of \cite{endnim}]\label{prop5-end_nim}
    For a position $A=(n_1,n_2,\ldots,n_k)$ of End-Nim such that $n_1\le n_k$, $A$ is a $\Ppos$-position if and only if either
\begin{itemize}
\item $n_1=n_k$ and $l(A)+r(A)$ is even, or
\item $n_k=n_1+1$, $l(A)$ is odd, and $r(A)$ is even.\bqed
\end{itemize}
\end{proposition}
\noindent
For example, $A=(3,2,5,4)$ is a $\Ppos$-position because $l(A)=1$ and $r(A)=2$.

Now, End-Nim behaves in the same way as $\mathtt{WordSubtraction}(b^*a^+b^*,b^*a^+b^*)$.
To see this correspondence, represent each heap by a nonempty block of $a$'s, with one or more $b$'s separating consecutive heaps. For example, the position
$(3,2,5,4)$ can be represented by a word such as
\[
b a^3 b^2 a^2 b a^5 b^3 a^4 b^2.
\]
The precise number of $b$'s used as separators is irrelevant; they merely serve to separate consecutive heaps. 
Removing a prefix or suffix in $b^*a^+b^*$ corresponds to removing a positive number of stones from the leftmost or rightmost heap.
If the whole heap is removed, some or all of the adjacent $b$'s can also be removed.
Thus, words with different numbers of $b$'s between the blocks of $a$'s or at either end can represent the same End-Nim position.
Therefore, for a word
\[
    w\in b^*a^{n_1}b^+a^{n_2}b^+\cdots b^+a^{n_k}b^*
\]
with $n_1,\ldots,n_k\geq 1$, let $A=(n_1,\ldots,n_k)$ be the corresponding End-Nim position.
By Proposition~\ref{prop5-end_nim} and symmetry, $w$ is a $\Ppos$-position
if and only if one of the following holds:
\begin{itemize}
\item $n_1=n_k$ and $l(A)+r(A)$ is even.
\item $n_k=n_1+1$, $l(A)$ is odd, and $r(A)$ is even.
\item $n_1=n_k+1$, $l(A)$ is even, and $r(A)$ is odd.
\end{itemize}
\begin{theorem}
    The $\Ppos$-position language $L_\Ppos(S,T)$ with $S=T=b^*a^+b^*\subseteq\{a,b\}^+$ is not context-free.
\end{theorem}
\begin{proof}[Proof sketch]
    We prove that $L_\Ppos(S,T)\cap a^+ba^+ba^+ba^+$ is not context-free by using the pumping lemma from Proposition~\ref{prop4-ogden_pump}.
Suppose that this language is context-free,
and let $N\geq 2$ be a pumping length.
Consider
\[
    w=a^Nba^{N-1}ba^{N+2}ba^{N+1}.
\]
The corresponding End-Nim position is
$A=(N,N-1,N+2,N+1)$.
Because $l(A)=1$ and $r(A)=2$, this is a $\Ppos$-position.

Apply the lemma with the first block $a^N$ as the specified
factor $w'$.
We obtain a decomposition
\[
    w=x_1y_1zy_2x_2
\]
in which at least one nonempty pumping factor lies entirely within the first block.
Now consider
\[
    \widehat{w}=x_1y_1^4zy_2^4x_2.
\]
The first heap of $\widehat{w}$ has size at least $N+3$.
Unless $y_2$ contains an $a$ from the last block, the last heap still has size $N+1$. In that case, the two end heaps differ by at least $2$, so $\widehat{w}$ cannot be a $\Ppos$-position.
Moreover, if $y_2$ contains the letter $b$, then $\widehat{w}$ contains more than three $b$'s and is not of the form $a^+ba^+ba^+ba^+$.
Therefore, $y_2=a^s$ for some $s\geq 1$, and its occurrence lies entirely within the last block.

It remains to consider the case where $y_1=a^t$ lies in the first block and $y_2=a^s$ lies in the last block, with $t,s\geq 1$.
The end heaps have sizes $N+3t$ and $N+1+3s$.
For their difference to be at most $1$, we must have $t=s$.
But the resulting position is then
\[
    \widehat{A}=(N+3t,N-1,N+2,N+1+3t),
\]
for which $l(\widehat{A})=r(\widehat{A})=1$.
Because the last heap is one larger than the first,
the characterization requires $r(\widehat{A})$ to be even.
Thus, $\widehat{w}$ is not a $\Ppos$-position,
contradicting the pumping lemma.
\end{proof}
\begin{corollary}
    There exist regular languages $S,T\subseteq\Sigma^+$ such that the $\Ppos$-position language $L_\Ppos(S,T)\subseteq\Sigma^*$ of $\mathtt{WordSubtraction}(S,T)$ is not context-free.\bqed
\end{corollary}

\subsection{Infix Removal and Merging}\label{subsec:infix}
Another variant is to remove an arbitrary infix, rather than a prefix or suffix, and merge the remaining parts into a single word. 
More precisely, if a position $w$ can be written as $w=usv$ for some $u,v\in\Sigma^*$ and $s\in S$, then one can replace $w$ by $w'=uv$.
Here we give two examples showing that simple subtraction sets can lead to $\Ppos$-position languages that are not context-free.

A familiar example is \emph{Wythoff's game}~\cite{Wythoff1907}.
Let
\[
    S=a^+\cup b^+\cup\{a^nb^n:n\geq1\},
\]
and consider the infix-removal game on words in $a^*b^*$.
A word $a^mb^n$ represents two heaps of sizes $m$ and $n$.
A move either reduces one heap by a positive amount or reduces both heaps by the same positive amount.
Thus, this restricted game is exactly Wythoff's game.

The $\Ppos$-positions of Wythoff's game are described by Beatty sequences involving the golden ratio (see, e.g., Theorem~3.4 in Chapter~IV of~\cite{CGT}).
The resulting set of pairs $(m,n)$ is not semilinear.
By Parikh's theorem, the language of $\Ppos$-positions within $a^*b^*$ is therefore not context-free.
Because context-free languages are closed under intersection with regular languages, the full $\Ppos$-position language is not context-free either.

Even restricting the subtraction set to a regular language does not guarantee a context-free $\Ppos$-position language.
For example, take $\Sigma=\{a,b,c\}$ and
\[
    S=a^+\cup b^+\cup c^+.
\]
This infix-removal game is known as \emph{StrNim},
and its $\Ppos$-position language is not context-free~\cite{mizuno2025strnim}.
Thus, the regularity result of Theorem~\ref{th3-reg} does not extend to infix removal.

We leave a more general study of infix-removal games for future work.

\subsection*{Exercises}
\begin{exercise}
    Let $S=a^+$ and $T=b^+$ be languages over $\Sigma=\{a,b\}$.
    Show that the $\Ppos$-position language $L_\Ppos(S,T)$ of $\mathtt{WordSubtraction}(S,T)$ is not regular.
    (\textbf{Hint}: If $L_\Ppos(S,T)$ is regular, then $L_\Ppos(S,T)\cap a^*b^*$ is also regular.)\diaqed
\end{exercise}
\begin{starexercise}
Let $S,T\subseteq\Sigma^+$ be finite subtraction sets, and let $\$\notin\Sigma$.
Consider the two-sided game $\mathtt{WordSubtraction}(S,T)$ over the alphabet $\Sigma\cup\{\$\}$.
Show that
\[
    L_{\Ppos}(S,T)\cap\Sigma^*\$\Sigma^*
\]
is regular. (\textbf{Hint}: No move can remove the symbol $\$$.
Interpret a position $u\$v$ as a disjunctive sum and use Grundy numbers.)\diaqed
\end{starexercise}
\begin{starexercise}\label{ex5:generalized_timber}
    As a slight generalization of the Timber game, consider $\mathtt{WordSubtraction}(\Sigma^*\Sigma_1,\Sigma_2\Sigma^*)$, where $\Sigma_1,\Sigma_2\subseteq\Sigma$. 
    Thus, in each move, a player can remove either a prefix ending in a letter from $\Sigma_1$ or a suffix beginning with a letter from $\Sigma_2$.
   Characterize the $\Ppos$-positions in each of the following cases:
    \begin{enumerate}
        \item $\Sigma=\Sigma_1\cup\Sigma_2$ and $\Sigma_1\cap\Sigma_2=\emptyset$,
        \item $\Sigma_1\cap\Sigma_2=\emptyset$,
        \item the general case.\diaqed
    \end{enumerate}
\end{starexercise}
\begin{starexercise}
    Let $\varphi=\frac{1+\sqrt{5}}{2}$ be the golden ratio.
    It is well known that, in Wythoff's game, a position $(i,j)\in\mathbb{N}\times\mathbb{N}$ with $i\le j$ is a $\Ppos$-position if and only if\[
    i=\lfloor k\cdot  \varphi\rfloor\quad \text{and}\quad j=\lfloor k\cdot  \varphi^2\rfloor\quad
    \]
    for some integer $k\ge 0$.
\begin{enumerate}
\item Prove that the language\[
\{a^{\lfloor k\cdot  \varphi\rfloor }b^{\lfloor k\cdot  \varphi^2\rfloor}:k\ge 0\}\cup \{a^{\lfloor k\cdot  \varphi^2\rfloor}b^{\lfloor k\cdot  \varphi\rfloor}:k\ge 0\}
\]
over $\Sigma=\{a,b\}$ is not context-free.
You can use the fact that a Beatty sequence
\[
\{\lfloor \alpha k\rfloor\mid k\ge 0\},
\]
where $\alpha>1$ is irrational, does not contain an infinite arithmetic progression.
\item Prove that the set of $\Ppos$-positions of the infix-removal word subtraction game with subtraction set
\[
S=a^+\cup b^+\cup \{a^n b^n : n\geq 1\}\subseteq\Sigma^+
\]
is not context-free.\diaqed
\end{enumerate}
\end{starexercise}

\subsection*{Open Problems}
\begin{problem}[Can the Dyck-word characterization be generalized?]
Exercise~\ref{ex5:generalized_timber} considers a generalization of Timber in which
moves are determined by individual letters.
More generally, let $U,V\subseteq\Sigma^+$ be languages and consider
\[
    \mathtt{WordSubtraction}(\Sigma^*U,V\Sigma^*).
\]
Can the characterization by Dyck words be extended to this setting?
In particular, how should such a characterization account for overlapping occurrences of words in $U\cup V$?
\diaqed
\end{problem}

\begin{problem}[Can Beatty sequences arise in suffix-removal games?]
In the infix-removal version, a context-free subtraction set can realize a $\Ppos$-position language whose restriction to $a^*b^*$ has a Parikh image described by irrational Beatty sequences, as in Wythoff's game.

This suggests a related question for the ordinary suffix-removal version:
can a context-free subtraction set $S$ yield a $\Ppos$-position language $L_{\Ppos}(S)$ whose Parikh image is described by irrational Beatty sequences?

The example in Section~\ref{sec4-simple-example}
shows that, even in the suffix-removal version, a context-free subtraction set can realize a
$\Ppos$-position language with a non-semilinear Parikh image.
This suggests that context-free subtraction sets can give rise to a broad range of structures in their $\Ppos$-positions.
The question here is whether the particular structure of irrational Beatty sequences can also arise.\diaqed
\end{problem}

\newpage
\section{Advanced Topics}\label{sec:advanced}
\subsection{Misère Play}\label{subsec:misere}
So far, we have considered combinatorial games under the normal-play convention, in which the player who makes the last move wins. Another important convention is misère play, in which the player who makes the last move loses. Equivalently, a player who has no legal move wins.

For a subtraction set $S \subseteq \Sigma^+$, we write $\mathtt{WordSubtraction}^-(S)$ for the word subtraction game played under the misère convention. The legal moves are exactly the same as those in
$\mathtt{WordSubtraction}(S)$; only the winning condition is changed. 
We denote its $\Ppos$- and $\Npos$-position languages by $L_\Ppos^-(S)$ and $L_\Npos^-(S)$, respectively.

Although this change may appear minor, it changes the starting point of the recursive classification: Under normal play, every terminal position is a $\Ppos$-position, whereas under mis\`ere play, every terminal position is an $\Npos$-position. For nonterminal positions, the recursive characterization remains unchanged.
\begin{theorem}[Partition Theorem for Misère Play]\label{th6-misere_partition}
For an impartial short mis\`ere-play game,\begin{itemize}
        \item a position is a $\Ppos$-position if and only if it has at least one option and all of its options are $\Npos$-positions, and
        \item a position is an $\Npos$-position if and only if it has no option or it has an option that is a $\Ppos$-position.
    \end{itemize}
    In particular, the set of positions is partitioned into $\Ppos$-positions and $\Npos$-positions.\bqed
\end{theorem}
Based on this characterization, we will show that even in the misère-play setting, the $\Npos$-position language $L_\Npos^-(S)$ is regular if $S$ is regular.
Let us begin with the following lemma:
\begin{lemma}\label{lem6-nonterminal_positions}
    If the subtraction set $S\subseteq \Sigma^+$ is regular, then the set of nonterminal positions $\Sigma^*S$ is also regular.
\end{lemma}
\begin{proof}
    Let $\mathbf{A}_S=(Q_S,\Sigma,\delta_S,q_S,F_S)$ be a DFA recognizing $S$.
    Note that $q_S\notin F_S$ holds by $\varepsilon\notin S$.
    We can construct an NFA $\mathbf{N}_{\mathrm{NT}}$ recognizing $\Sigma^*S$ in the following way:
    For each $a\in \Sigma$, add a transition
\[
q_S\xrightarrow{a}q_S
\]
to $\mathbf{A}_S$. More precisely, let $\mathbf{N}_{\mathrm{NT}}=(Q_S,\Sigma,\delta'_S,q_S,F_S)$
where
\[
\delta'_S(q,a)=
\begin{cases}
\{\delta_S(q,a)\}\cup\{q_S\} & \text{if }q=q_S,\\
\{\delta_S(q,a)\} & \text{otherwise}.
\end{cases}
\]
Indeed, $w\in \Sigma^*S$ if and only if there exists a run of the form\[
q_S\xrightarrow{a_1}q_S\xrightarrow{a_2}\cdots q_S\xrightarrow{a_m}q_S\xrightarrow{b_1}q_1\xrightarrow{b_2}\cdots \xrightarrow{b_n}q_n\in F_S
\]
on $\mathbf{N}_{\mathrm{NT}}$ where $w=a_1a_2\cdots a_mb_1b_2\cdots b_n$ with $a_1a_2\cdots a_m\in \Sigma^*$ and $b_1b_2\cdots b_n\in S$.
Thus, $L(\mathbf{N}_{\mathrm{NT}})=\Sigma^*S$ holds and $\Sigma^*S$ is regular.
\end{proof}
More generally, regular languages are closed under concatenation; that is, if $L_1$ and $L_2$ are regular, then $L_1L_2$ is also regular.

We now construct a DFA recognizing the $\Npos$-position language $L_\Npos^-(S)$ of $\mathtt{WordSubtraction}^{-}(S)$.
The strategy is very similar to that used for the normal-play setting in Theorem~\ref{th3-reg}.
We first identify a set of base $\Npos$-positions.
We then identify the nonterminal positions that can move only to these base $\Npos$-positions as $\Ppos$-positions.
At each such $\Ppos$-position, we apply the reset rule.
From each reset point, positions for which the suffix read from that point belongs to $S$ are then identified as $\Npos$-positions.
We repeat this process to obtain a DFA recognizing $L_\Npos^-(S)$.

The only difference is how the base $\Npos$-positions are obtained.
In the normal-play case, the base $\Npos$-positions are precisely the elements of the subtraction set $S$.
On the other hand, in the misère-play case, the terminal positions are the base $\Npos$-positions.
Thus, we begin by identifying the terminal positions
and then identify the nonterminal positions that can move only to terminal positions as $\Ppos$-positions.
At each such $\Ppos$-position, we apply the reset rule and proceed in the same way as in the normal-play case.
\begin{theorem}\label{th6-misere_reg}
    Let $\Sigma$ be an alphabet and $S\subseteq \Sigma^+$ be a regular language.
    Then, the $\Npos$-position language $L_{\Npos}^-(S)$ of $\mathtt{WordSubtraction}^-(S)$ is regular.
\end{theorem}
\begin{proof}
    Let $\mathbf{A}_S=(Q_S,\Sigma,\delta_S,q_S,F_S)$ be a non-returning DFA recognizing the subtraction set $S$.
    Also, let $\mathbf{A}_T=(Q_T,\Sigma,\delta_T,q_T,F_T)$ be a DFA recognizing the set of terminal positions $(\Sigma^*S)^c$.
    Such a DFA $\mathbf{A}_T$ indeed exists because $\Sigma^*S$ is regular by Lemma~\ref{lem6-nonterminal_positions}, and hence its complement $(\Sigma^*S)^c$ is also regular.
    Note that $q_T\in F_T$ because $\varepsilon$ is a terminal position.
    By renaming states if necessary, assume that
$Q_S\cap Q_T=\emptyset$.
    Let $\mathbf{A}=(Q,\Sigma,\delta,q_T,F)$ be the DFA defined as\begin{itemize}
        \item $Q=Q_S\cup Q_T$,
        \item for each $q\in Q$ and $a\in \Sigma$,\[\delta(q,a)=\begin{cases}
            \delta_S(q,a)&\text{if }q\in Q_S,\\
            \delta_T(q,a)&\text{if }q\in Q_T,
        \end{cases}\] and
        \item $F=F_S\cup F_T$.
    \end{itemize}
    Finally, let $\mathbf{A}_\Npos=(Q_\Npos,\Sigma,\delta_\Npos,\{q_T\},F_\Npos)$ be the DFA defined as
    \begin{itemize}
    \item $Q_{\Npos}=2^Q$,
    \item for each $R\subseteq Q$ and $a\in\Sigma$,\[
    \delta_{\Npos}(R,a)=\begin{cases}
        \{\delta(q,a):q\in R\}\cup \{{\color{red}q_S}\}&{\color{red}\text{if }\delta(q,a)\notin F\text{ for all }q\in R},\\
        \{\delta(q,a):q\in R\}&\text{otherwise},
    \end{cases}
    \] 
    and
    \item $F_{\Npos}=\{R\subseteq Q:R\cap F\ne \emptyset\}$.
\end{itemize}
The construction always keeps track of the run of $\mathbf{A}_T$ from $q_T$.
In addition, each reset starts a new run of $\mathbf{A}_S$ from $q_S$.

We prove correctness by induction on the length of the input,
showing that a word is accepted exactly when it is an $\Npos$-position, and that a reset occurs exactly at each nonempty $\Ppos$-position.

For the empty word, the initial state $\{q_T\}$ is accepting,
as required because $\varepsilon$ is a terminal $\Npos$-position.
Now, let $w\neq\varepsilon$, and assume the claim for every proper prefix of $w$.
Suppose that, after reading $w$, the current set of states contains an accepting state $q_f\in F$.
There are two cases:
\begin{itemize}
    \item If $q_f\in F_T$, then $\mathbf{A}_T$ accepts $w$, so $w$ is a terminal position.
    \item If $q_f\in F_S$, then a run of $\mathbf{A}_S$
    started by a reset at some proper prefix $w'$
    accepts the remaining suffix $s\in S$, where $w=w's$.
    By the induction hypothesis, $w'$ is a $\Ppos$-position.
\end{itemize}
In either case, $w$ is an $\Npos$-position by
Theorem~\ref{th6-misere_partition}.

Conversely, suppose that $w$ is an $\Npos$-position.
By Theorem~\ref{th6-misere_partition}, either $w$ is terminal or $w=w's$ for some $\Ppos$-position $w'$ and $s\in S$.
In the first case, the run of $\mathbf{A}_T$ reaches an accepting state.
In the second case, the induction hypothesis ensures that a reset occurs at $w'$.
The run of $\mathbf{A}_S$ started there accepts $s$, so the current set again contains an accepting state.
Thus, $w$ is accepted if and only if it is an $\Npos$-position.
\end{proof}

As an example, let us consider $\mathtt{WordSubtraction}^{-}(S)$ for the subtraction set $S=a\Sigma(\Sigma\Sigma)^*\cup b(\Sigma\Sigma)^*$, which we considered in Section~\ref{sec:reg}. 
It is clear that the set of terminal positions is $(\Sigma^*S)^c=(ba)^*\cup a(ba)^*$. 
Thus, for example, $\mathbf{A}_T$ is given by the following automaton:
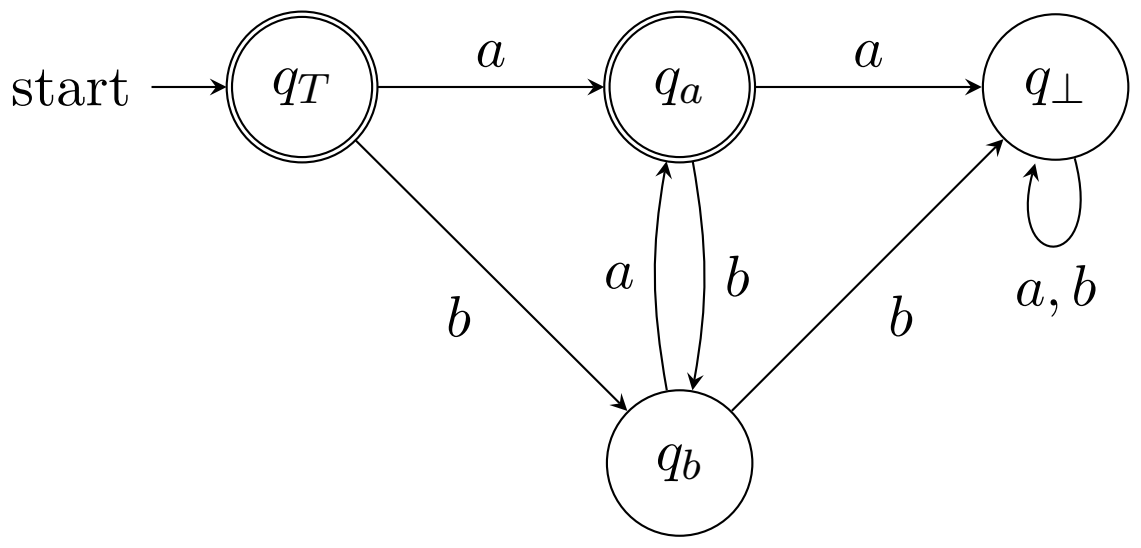
\begin{figure}[H]
\centering
\begin{tikzpicture}[
  ->,
  >=stealth,
  node distance=1.5cm
]
\node[state, initial,accepting] (qT) {$q_T$};
\node[state,accepting] (qa) [right=of qT] {$q_a$};
\node[state] (qb) [below=of qa] {$q_b$};
  \node[state] (qbot) [right=of qa]  {$q_\bot$};
  
  \path
    (qT) edge node[above] {$a$} (qa)
    (qT) edge node[below left] {$b$} (qb)
    (qa) edge node[above] {$a$} (qbot)
    (qb) edge node[below right] {$b$} (qbot)
    (qa) edge[bend left=10] node[right] {$b$} (qb)
    (qb) edge[bend left=10] node[left] {$a$} (qa)
    (qbot) edge[loop below] node {$a,b$} ();
\end{tikzpicture}
\end{figure}

Combining this with the DFA $\mathbf{A}_S$ recognizing $S$ given in Example~\ref{ex-DFAs}, we obtain the following automaton $\mathbf{A}$.
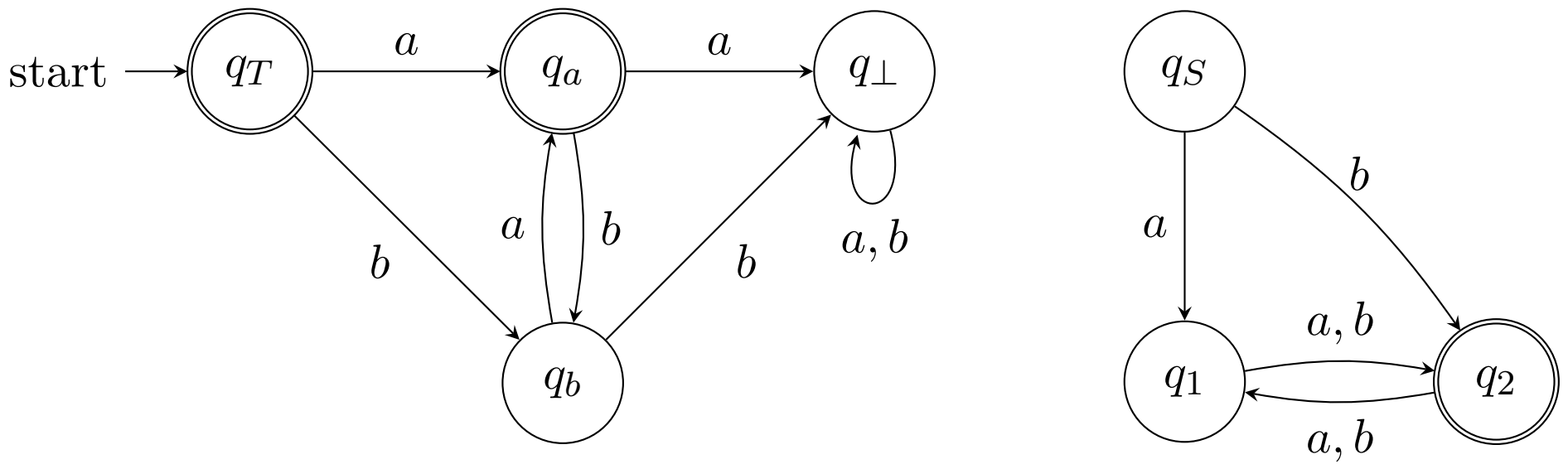
\begin{figure}[H]
\centering
\begin{tikzpicture}[
  ->,
  >=stealth,
  node distance=1.5cm
]
    \node[state, initial,accepting] (qT) {$q_T$};
\node[state,accepting] (qa) [right=of qT] {$q_a$};
\node[state] (qb) [below=of qa] {$q_b$};
  \node[state] (qbot) [right=of qa]  {$q_\bot$};
  \node[state] (q0) [right=of qbot] {$q_S$};
  \node[state] (q1) [below=of q0] {$q_1$};
  \node[state,accepting] (q2) [right=of q1] {$q_2$};

  \path
    (q0) edge node[left] {$a$} (q1)
    (qT) edge node[above] {$a$} (qa)
    (qT) edge node[below left] {$b$} (qb)
    (qa) edge node[above] {$a$} (qbot)
    (qb) edge node[below right] {$b$} (qbot)
    (qa) edge[bend left=10] node[right] {$b$} (qb)
    (qb) edge[bend left=10] node[left] {$a$} (qa)
    (qbot) edge[loop below] node {$a,b$} ()
    (q0) edge[bend left=10] node[above] {$b$} (q2)
    (q1) edge[bend left=10] node[above] {$a,b$} (q2)
    (q2) edge[bend left=10] node[below] {$a,b$} (q1);
\end{tikzpicture}
\end{figure}

Although the two DFAs appear to be disconnected in the figure, they together form a single DFA $\mathbf{A}$.
Now, applying the construction of $\mathbf{A}_\Npos$ to this DFA $\mathbf{A}$, we obtain a DFA recognizing $L_\Npos^-(S)$. 
However, writing down all of its states would be inconvenient. 
Instead, let us illustrate the construction by considering the run of the word $baab$:\[
\{q_T\}\xrightarrow{b}\{q_b,q_S\}\xrightarrow{a}\{q_a,q_1\}\xrightarrow{a}\{q_\bot,q_2\}\xrightarrow{b}\{q_\bot,q_S,q_1\}\notin F_\Npos.
\]
We start from the initial state $\{q_T\}$. 
Following the transition of $\mathbf{A}_T$, we move to the state $\{q_b\}$. 
However, because $q_b$ is not an accepting state of $\mathbf{A}$, the reset rule is applied, and the actual state in $\mathbf{A}_N$ is $\{q_b,q_S\}$.
Next, the respective transitions from $q_b$ and $q_S$ on $a$ lead to $q_a$ and $q_1$. 
Because $q_a$ is an accepting state, the transition in $\mathbf{A}_N$ leads to $\{q_a,q_1\}$.
Indeed, $ba$ is a terminal position and hence an $\Npos$-position.
On reading the next $a$, the respective transitions lead to $q_\bot$ and $q_2$. 
Because $q_2$ is an accepting state, the resulting state in $\mathbf{A}_N$ is $\{q_\bot,q_2\}$.
In fact, $baa$ is not a terminal position, but it is an $\Npos$-position: the first player can remove $aa$, leaving $b$ to the second player, who is then forced to remove it.
Finally, on reading $b$, the respective transitions lead to the rejecting states $q_\bot$ and $q_1$. 
Thus, the reset rule is applied, and the resulting state is $\{q_\bot,q_S,q_1\}$.
Therefore, $baab$ is a $\Ppos$-position.

An important special case is when $\Sigma\subseteq S$, that is, when the only terminal position is $\varepsilon$. 
In this case, we have the following result.
\begin{theorem}\label{th6-misere_epsilon_is_only_terminal}
Let $S\subseteq\Sigma^+$ be a regular subtraction set such that $\Sigma\subseteq S$. Then, $L_\Npos^-(S)=\Sigma\cdot L_\Npos(S)\cup \{\varepsilon\}$ holds.
In other words, an $\Npos$-position in misère play is precisely either the empty word $\varepsilon$ or a word obtained by adding a single letter to the left end of a normal-play $\Npos$-position.
\end{theorem}
\begin{proof}[Proof sketch]
    When the only terminal position is $\varepsilon$, $\mathbf{A}_T$ is given by the following two-state DFA:
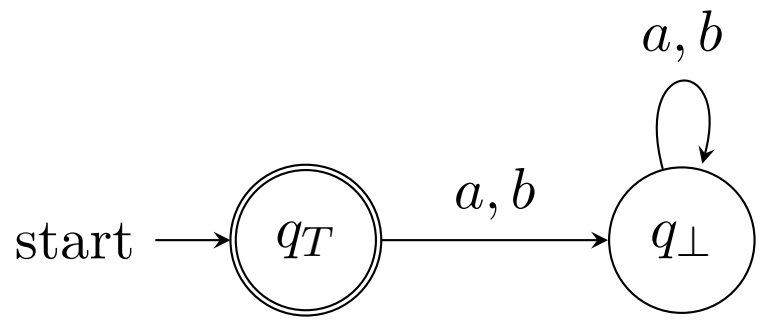
\begin{figure}[H]
\centering
\begin{tikzpicture}[
  ->,
  >=stealth,
  node distance=1.5cm
]
\node[state, initial,accepting] (qT) {$q_T$};
\node[state] (qbot) [right=of qT] {$q_\bot$};
  
  \path
    (qT) edge node[above] {$a,b$} (qbot)
    (qbot) edge[loop above] node {$a,b$} ();
\end{tikzpicture}
\end{figure}
Then, the DFA $\mathbf{A}_\Npos$ constructed in the proof of Theorem~\ref{th6-misere_reg} is isomorphic to the corresponding DFA for the normal-play version with the following transitions added at the beginning:
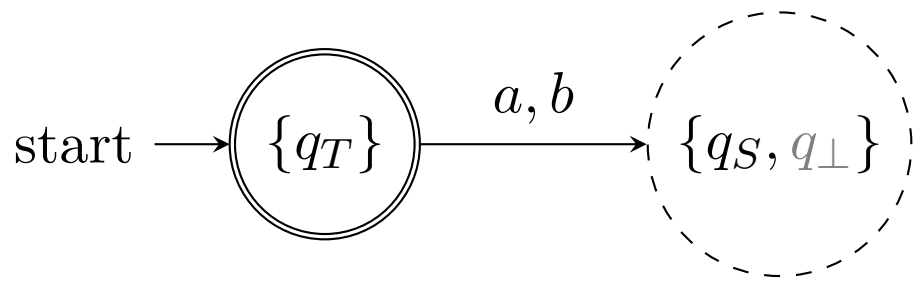
\begin{figure}[H]
\centering
\begin{tikzpicture}[
  ->,
  >=stealth,
  node distance=1.5cm
]
\node[state, initial,accepting] (qT) {$\{q_T\}$};
\node[state, dashed] (qbot) [right=of qT] {$\{q_S,{\color{gray} q_\bot}\}$};
  \path
    (qT) edge node[above] {$a,b$} (qbot);
\end{tikzpicture}
\end{figure}
The only difference is that each state additionally contains $q_\bot$. 
However, this does not affect the behavior of the automaton because no accepting state can be reached from $q_\bot$.
\end{proof}\noindent
We discuss a generalization of this fact in Exercise~\ref{ex6:prefix-closed}.

\subsection{Reflexivity}\label{subsec:reflexivity}
In ordinary subtraction games, a subtraction set $S\subseteq\mathbb{N}_+$ is called \emph{reflexive} if the $\Ppos$-position set of the game with subtraction set $L_\Ppos(S)\setminus\{0\}\subseteq\mathbb{N}$ coincides with $S$ after removing $0$; that is,
\[
L_\Ppos\bigl(L_\Ppos(S)\setminus\{0\}\bigr)\setminus\{0\}=S.
\]
This property can be used to study other properties of subtraction sets.
Takuto Maki of Kagoshima University observed that reflexivity and related properties should have natural analogues for word subtraction games, and suggested that they should be investigated in this setting. 
The results presented in this section were developed with his advice.

We first introduce a notion of reflexivity for word subtraction games.
Several results on reflexivity for ordinary subtraction games were proved in~\cite{DBLP:journals/tcs/Larsson12}.
To follow the same line of argument for word subtraction games, we use both prefix and suffix removal for technical reasons.
Unlike addition of integers, concatenation of words is not commutative, so we must distinguish between the two kinds of removal.
Therefore, we write $\WSsuf(S)$ and $\WSpre(S)$ for the suffix-removal and prefix-removal versions, respectively.
We use the corresponding notation $\LPsuf(S)$, $\LNsuf(S)$, $\LPpre(S)$, and $\LNpre(S)$ for their $\Ppos$-position and $\Npos$-position languages, respectively.

A subtraction set $S\subseteq\Sigma^+$ is said to be \emph{reflexive} if\[
S=\LPpre(\LPsuf(S)\setminus\{\varepsilon\})\setminus\{\varepsilon\}.
\]
In other words, consider the word subtraction game $\WSpre(\LPsuf(S)\setminus\{\varepsilon\})$, whose subtraction set consists of the nonempty $\Ppos$-positions of $\WSsuf(S)$. 
Then the set of nonempty $\Ppos$-positions in this new game coincides with the original subtraction set $S$.
For example, the subtraction set $S=a^+b^*\cup a^*b^+$ with $\Sigma=\{a,b\}$ is reflexive.
This is because
$L_{\Ppos}^{\mathsf{suf}}(S)=\Sigma^*ba\cup\{\varepsilon\}$,
as discussed in Exercise~\ref{ex1:a*b*}.
The $\Npos$-positions in
$\mathtt{WordSubtraction}^{\mathsf{pre}}(\Sigma^*ba)$
are exactly the words containing $ba$.
Indeed, from any such word, the first player can remove the prefix ending at the last occurrence of $ba$, immediately reaching a terminal position.
Thus, the set of nonempty $\Ppos$-positions is precisely\[
(\Sigma^*ba\Sigma^*)^c\setminus\{\varepsilon\}=a^+b^*\cup a^*b^+=S.
\]

We next give a characterization of reflexivity. 
In \cite{DBLP:journals/tcs/Larsson12}, it was shown that, for an ordinary subtraction set $S\subseteq \mathbb{N}_+$, $S$ is reflexive if and only if every element of $\mathcal{D}(S)$ is an $\Npos$-position, where $\mathcal{D}(S)$ is defined as\[
\mathcal{D}(S)=\{x-x'\in \mathbb{N}_+:x,x'\in S\}.
\]
To obtain a language-theoretic counterpart of this characterization, we introduce the notion of derivatives of languages.
For a language $L\subseteq\Sigma^*$ and a word $v\in\Sigma^*$, the \emph{right derivative} (or \emph{right quotient}) of $L$ with respect to $v$ is the language
\[
Lv^{-1}=\{w\in\Sigma^* : wv\in L\}.
\]
Thus, $Lv^{-1}$ consists of the prefixes that remain after removing the suffix $v$ from words in $L$.
More generally, for languages $L,V\subseteq\Sigma^*$, we define the \emph{right derivative} of $L$ by $V$ as
\[
LV^{-1}=\bigcup_{v\in V}Lv^{-1}
=\{w\in\Sigma^* : wv\in L\text{ for some }v\in V\}.
\]
For a subtraction set $S\subseteq\Sigma^+$, define
\[
    \mathcal{D}(S)=(SS^{-1})\setminus\{\varepsilon\}
    =\{u\in\Sigma^+:us'\in S
      \text{ for some }s'\in S\}.
\]
The following can be proved using the same approach as Theorem~2 in \cite{DBLP:journals/tcs/Larsson12}.
\begin{theorem}\label{th6-reflexivity}
    A subtraction set $S\subseteq\Sigma^+$ is reflexive if and only if $\mathcal{D}(S)\subseteq \LNsuf(S)$.
\end{theorem}
\begin{proof}
    Let $S^\mathsf{ref}=\LPpre(\LPsuf(S)\setminus\{\varepsilon\})\setminus\{\varepsilon\}$.
    We first prove the only-if direction by showing its contrapositive.
    If $\mathcal{D}(S)\not\subseteq \LNsuf(S)$, there exists a nonempty $\Ppos$-position $p\in\Sigma^+$ in $\WSsuf(S)$ such that $ps'=s$ with some $s,s'\in S$.
    Thus, the move from $s$ to $s'$ is legal in $\WSpre(\LPsuf(S)\setminus\{\varepsilon\})$, and therefore, $s'\in \LPpre(\LPsuf(S)\setminus\{\varepsilon\})$ implies $s\notin \LPpre(\LPsuf(S)\setminus\{\varepsilon\})$.
    Thus, $s$ and $s'$ cannot both belong to $S^{\mathsf{ref}}$.
Because both belong to $S$, we conclude that
$S\neq S^{\mathsf{ref}}$.

    We next prove the if direction.
   Suppose that $S$ is not reflexive. 
   Then there exists $w\in\Sigma^+$ such that
\[
w\in S\setminus S^\mathsf{ref}
\quad\text{or}\quad
w\in S^\mathsf{ref}\setminus S.
\]
Choose such a word $w$ that is minimal with respect to the suffix order; that is, no proper nonempty suffix of $w$ satisfies either of the two conditions above.
We claim that in fact
\[
w\in S\setminus S^\mathsf{ref}.
\]
Indeed, suppose that $w\in S^{\mathsf{ref}}\setminus S$.
If $w\in L_{\Ppos}^{\mathsf{suf}}(S)$, then $w$ itself belongs to the subtraction set of the new prefix-removal game.
Removing the whole word would make $w$ an $\Npos$-position in that game, contradicting $w\in S^{\mathsf{ref}}$.
Thus,
\[
    w\in L_{\Npos}^{\mathsf{suf}}(S).
\]
By $w\notin S$, there must be a move from $w$ to a nonempty $\Ppos$-position $p$ in $\WSsuf(S)$ obtained by removing some suffix $s\in S$. 
By the minimality of $w$, we have $s\in S^\mathsf{ref}$ because $s\in S\setminus S^\mathsf{ref}$ does not hold. 
Here, $p$ belongs to the subtraction set $\LPsuf(S)\setminus\{\varepsilon\}$ and $s\in S^\mathsf{ref}$ is a $\Ppos$-position in $\WSpre(\LPsuf(S)\setminus\{\varepsilon\})$. 
This implies that $w=ps$ is an $\Npos$-position in $\WSpre(\LPsuf(S)\setminus\{\varepsilon\})$, and this contradicts $w\in S^\mathsf{ref}$.
Consequently, we have $w\in S\setminus S^\mathsf{ref}$.

Now, $w\notin S^\mathsf{ref}$ implies $w\in \LNpre(\LPsuf(S)\setminus\{\varepsilon\})$.
Therefore, there exists $p\in \LPsuf(S)\setminus\{\varepsilon\}$ and $w'\in \LPpre(\LPsuf(S)\setminus\{\varepsilon\})$ such that $w=pw'$.
In addition, $w\ne p$, that is, $w'\ne\varepsilon$ because $w\in S$ implies $w\notin \LPsuf(S)$. 
Thus, we have $w'\in \LPpre(\LPsuf(S)\setminus\{\varepsilon\})\setminus\{\varepsilon\} =S^\mathsf{ref}$ and $w'\in S$ by the minimality of $w$.
Finally,\[
p=w\, (w')^{-1}\in  SS^{-1}=\mathcal{D}(S)\cup\{\varepsilon\}
\]
with $p\in \LPsuf(S)$, and hence $\mathcal{D}(S)\not\subseteq \LNsuf(S)$.
\end{proof}
For example, $\mathcal{D}(a^+b^*\cup a^*b^+)=a^+b^*\cup a^*b^+\subseteq\LNsuf(a^+b^*\cup a^*b^+)$ holds, and indeed $a^+b^*\cup a^*b^+$ is reflexive.

In this subsection, we have defined reflexivity using both suffix and prefix removal.
As noted above, we used this setting for technical reasons in order to follow the argument in~\cite{DBLP:journals/tcs/Larsson12}.
Hiyu Inoue of Hiroshima University has pointed out that a useful characterization can also be obtained when reflexivity is defined using suffix removal alone, that is, by
\[
    S=\LPsuf\bigl(\LPsuf(S)\setminus\{\varepsilon\}\bigr)
      \setminus\{\varepsilon\}.
\]
We leave this discussion for another occasion.

\subsection{Realizability}\label{subsec:realizability}
In Section~\ref{sec:reg}, we showed that if the subtraction set $S\subseteq \Sigma^+$ is regular, then the $\Ppos$-position language $L_\Ppos(S)$ is also regular. 
This naturally raises the converse question: can every regular language $W\subseteq \Sigma^*$ be realized as the $\Ppos$-position language of some $\mathtt{WordSubtraction}(S)$ with a regular subtraction set $S\subseteq\Sigma^+$?

The answer is no. 
In fact, a simple counterexample can already be found in the setting of ordinary subtraction games. Consider\[
W=\{\varepsilon,aa,aaa\}
\]over the alphabet $\Sigma=\{a\}$. 
In this case, $W$ cannot be the $\Ppos$-position language $L_\Ppos(S)$ for any subtraction set $S\subseteq\Sigma^+$, even without requiring $S$ to be regular.
To see this, suppose that $L_\Ppos(S)=W$. 
Because $a\notin W$, the position $a$ is an $\Npos$-position. 
This implies $a\in S$. 
On the other hand, $aaa$ must be an $\Npos$-position because $aa$ is a $\Ppos$-position and $aaa$ can be moved to $aa$ by removing $a\in S$. This contradicts $aaa\in L_\Ppos(S)=W$.

Takuto Maki previously established necessary and sufficient conditions for a set $W$ to be realizable as the set of $\Ppos$-positions of an ordinary subtraction game.
He also suggested that these results can be extended to word subtraction games.
This subsection summarizes these results and their extensions, and was completed with his advice and collaboration.

For a language $W\subseteq\Sigma^*$, let\[
\mathscr{S}(W)=\{S\subseteq\Sigma^+:L_\Ppos(S)=W\}.
\]
In other words, $\mathscr{S}(W)$ is the collection of all subtraction sets that realize $W$.
In the previous subsection, we considered right derivatives of languages. Left derivatives are defined similarly.
For languages $L,U\subseteq\Sigma^*$, we define the \emph{left derivative} of $L$ by $U$ as
\[
U^{-1}L=\{w\in\Sigma^* : uw\in L\text{ for some }u\in U\}.
\]
In particular, when $L=U=W$, we have\[
W^{-1}W=\{s\in \Sigma^*:w=w's \text{ with some }w,w'\in W\}.
\]
If $W$ is the $\Ppos$-position language,
no legal move can take a word $w\in W$ to a shorter word $w'\in W$; such a move would make $w$ an
$\Npos$-position because $w$ has an option $w'\in W$ that is a $\Ppos$-position.
Thus, every word in $W^{-1}W$ must be excluded from the subtraction set.
The following theorem shows that, whenever $W$ is realizable,
all remaining nonempty words can be allowed as members of a subtraction set.
\begin{theorem}\label{th6-maximum_subtraction_set}
    Let $W\subseteq\Sigma^*$ be a language.
    If $\mathscr{S}(W)\ne \emptyset$, then the language $S_{\max} \subseteq \Sigma^+$ defined as
    \[
    S_{\max}=(W^{-1}W)^c
    \]
    is the maximum element of $\mathscr{S}(W)$ with respect to set inclusion.
\end{theorem}
\begin{proof}
Assume that $\mathscr{S}(W)$ is nonempty, and let $S\in\mathscr{S}(W)$ be a subtraction set realizing $W$.
Because no legal move in $\mathtt{WordSubtraction}(S)$ can take a word in $W$ to another word in $W$, we have $S\cap W^{-1}W=\emptyset$, that is, $S\subseteq (W^{-1}W)^c=S_{\max}$.
Therefore, it suffices to show that $L_\Ppos(S)=L_\Ppos(S_{\max})$.

Let us show that for every $w\in\Sigma^*$, $w\in L_\Ppos(S)\iff w\in L_\Ppos(S_{\max})$ by induction on $|w|$.
    First, it is clear that $w\in L_\Ppos(S)\cap L_\Ppos(S_{\max})$ if $w=\varepsilon$.
    Next let $w\ne \varepsilon$, and assume as the induction hypothesis that\[
    w'\in L_\Ppos(S)\quad \iff\quad w'\in L_\Ppos(S_{\max})
    \]
    for every proper prefix $w'$ of $w$.

    Suppose that $w\in L_\Ppos(S)$ but $w\notin L_\Ppos(S_{\max})$, and derive a contradiction.
    By $w\in L_\Npos(S_{\max})$, there are a $\Ppos$-position $w'\in L_\Ppos(S_{\max})$ and a word $s\in S_{\max}$ such that $w=w's$.
    Note that $w'$ is also a $\Ppos$-position in $\mathtt{WordSubtraction}(S)$ by the induction hypothesis.
    Now, $w\in L_\Ppos(S)=W$ and $w'\in L_\Ppos(S)=W$ imply that $s=(w')^{-1}w\in W^{-1} W$.
    This contradicts $s\in S_{\max}=(W^{-1}W)^c$.
Therefore, $w\in L_\Ppos(S_{\max})$.
    
    Conversely, suppose that $w\in L_\Ppos(S_{\max})$.
    Then, every option $w'$ of $w$ is an $\Npos$-position in $\mathtt{WordSubtraction}(S_{\max})$. 
    Because $S\subseteq S_{\max}$ by the first paragraph, every legal move
in $\mathtt{WordSubtraction}(S)$ is also legal
in $\mathtt{WordSubtraction}(S_{\max})$.
    Thus, every option of $w$ in $\mathtt{WordSubtraction}(S)$ must also belong to $L_\Npos(S_{\max})$. 
    By the induction hypothesis, every option of $w$ belongs to $L_\Npos(S)$, and hence $w\in L_\Ppos(S)$.
    This completes the induction and proves that
$L_\Ppos(S)=L_\Ppos(S_{\max})$.
\end{proof}
Applying this result, we obtain a characterization of when a language $W$ is realizable.
\begin{theorem}
Let $W\subseteq\Sigma^*$ be a language.
Then $\mathscr{S}(W)\neq\emptyset$ if and only if
\[
    W(W^{-1}W)^c=W^c.
\]
\end{theorem}
\begin{proof}
Suppose first that $\mathscr{S}(W)\neq\emptyset$.
By Theorem~\ref{th6-maximum_subtraction_set}, the subtraction set
$S_{\max}=(W^{-1}W)^c$ realizes $W$.
Therefore,
\[
    W^c=L_{\Npos}(S_{\max})
       =L_{\Ppos}(S_{\max})S_{\max}
       =W(W^{-1}W)^c.
\]

Conversely, suppose that $W(W^{-1}W)^c=W^c$.
It suffices to show that $W=L_{\Ppos}(S_{\max})$.
By the Partition Theorem, we need to show that every option of a word in $W$ lies outside $W$, and that every word outside $W$ has an option in $W$.
First, suppose that $w\in W$.
If a legal move from $w$ led to another word $w'\in W$,
we would have $w=w's$ for some $s\in S_{\max}$.
But $w',w\in W$ would imply $s\in W^{-1}W$, contradicting the definition of $S_{\max}$.
Therefore, every option of a word in $W$ lies outside $W$.
Next, suppose that $w\notin W$.
By the assumed equality $W(W^{-1}W)^c=W^c$, we can write $w=w's$ with $w'\in W$ and $s\in S_{\max}$.
Removing the suffix $s$ is therefore a legal move from $w$ to a word in $W$.
Hence $W=L_{\Ppos}(S_{\max})$ and $\mathscr{S}(W)\neq\emptyset$.
\end{proof}
\noindent
Thus, realizability can be characterized using only concatenation, left derivatives, and complement.
For a regular language $W$, these operations preserve regularity.
\begin{corollary}\label{cor-smax_is_regular}
    Let $W\subseteq\Sigma^*$ be a regular language.
    If $\mathscr{S}(W)\ne \emptyset$, then there exists a regular subtraction set $S\subseteq \Sigma^+$ such that $W=L_\Ppos(S)$.
\end{corollary}
\begin{proof}
    Let $S=(W^{-1}W)^c$. Note that $S$ is regular because the class of regular languages is closed under taking derivatives and complements.
    By Theorem~\ref{th6-maximum_subtraction_set}, $\mathscr{S}(W)\ne \emptyset$ implies that the regular language $S$ belongs to $\mathscr{S}(W)$.
\end{proof}\noindent
Exercise~\ref{ex6:recognition_of_smax} strengthens this result:
if $W$ is recognized by a monoid $M$ and is realizable,
then there exists a subtraction set recognized by $M$ that realizes $W$.

We have focused on the maximum subtraction set $S_{\max}$ realizing $W$.
It is natural to ask whether there is also a minimum subtraction set with respect to set inclusion.
The situation is more complicated: even when $W$ is realizable, such a minimum element need not exist.
In such cases, we can instead ask how to find minimal subtraction sets realizing $W$.
For ordinary subtraction games, Takuto Maki has established a method for finding such minimal sets using reflexivity, the notion discussed in the previous subsection.
Maki and the first author expect that this method extends naturally to word subtraction games, but this remains to be investigated.

\subsection{Decidability}
So far, we have studied the languages of winning and losing positions.
We now ask whether properties of these languages can be decided algorithmically.

When $S$ is given by a DFA or a CFG, the outcome of any particular position can be computed by examining its finitely many possible moves recursively.
But deciding a property of all positions at once is a different question.
For example, can we decide whether every nonterminal position is an $\Npos$-position?
Checking positions one by one does not by itself give an algorithm that always terminates.

In this subsection, we consider the following two decision problems: 
\begin{itemize}
    \item Are all positions in $\mathtt{WordSubtraction}(S)$ $\Ppos$-positions?
    \item Are all nonterminal positions in $\mathtt{WordSubtraction}(S)$ $\Npos$-positions?
\end{itemize}
The first problem is, in fact, rather simple: it asks only whether $S$ is empty. 
When $S$ is given by a DFA or a CFG, there is an algorithm for determining whether $S=\emptyset$. 
In other words, the problem is \emph{decidable}.
The second problem leads to a more interesting phenomenon. 
When $S$ is a regular language given by a DFA, the problem is also decidable. In contrast, when $S$ is a context-free language given by a CFG, the problem is \emph{undecidable}. 
This means that no algorithm can always terminate with
the correct answer for every valid input CFG.

We first recall two basic facts from formal language theory.
\begin{proposition}\label{prop6-emptiness_reg}
    Given a regular language $L\subseteq\Sigma^*$, for example by a DFA, it is decidable whether $L=\emptyset$.
\end{proposition}
\begin{proof}
    Let $\mathbf{A}=(Q,\Sigma,\delta,q_0,F)$ be a DFA. 
    It is clear that $L(\mathbf{A})\ne\emptyset$ if and only if there is an accepting state that is reachable from the initial state $q_0$. 
    Here, we say that a state $q'\in Q$ is \emph{reachable} from a state $q\in Q$ if there is a word $w\in\Sigma^*$ such that $\delta^*(q,w)=q'$.

    We can decide whether $L(\mathbf{A})\ne\emptyset$ by computing the set $X\subseteq Q$ of states that can reach an accepting state.
    Initially, set $X:=F$.
    Indeed, every accepting state can reach itself by reading the empty word $\varepsilon$. 
    We then repeatedly apply the following rule: for each $q\in Q$ and $a\in\Sigma$, if
\[
\delta(q,a)\in X,
\]
then add $q$ to $X$. We repeat this process until no new state is added. 
Because $X$ contains at most $\#(Q)$ states, this process must terminate.
After the process terminates, we have
\[
q_0\in X \quad\iff\quad L(A)\ne\emptyset.
\]
Thus, the problem is decidable.
\end{proof}
\begin{proposition}\label{prop6-emptiness_cf}
Given a context-free language $L\subseteq\Sigma^*$, for example by a CFG, it is decidable whether $L=\emptyset$.
\end{proposition}
\begin{proof}
Let $\mathbf{G}=(\Gamma,\Sigma,R,I)$ be a CFG. 
We can decide whether $L(\mathbf{G})=\emptyset$ by a procedure similar to the one used in Proposition~\ref{prop6-emptiness_reg}. 
We compute the set $X\subseteq\Gamma$ of productive nonterminals. 
Here, a nonterminal $A\in\Gamma$ is called \emph{productive} if $A$ can derive a word consisting entirely of terminal symbols; that is, if there exists some $w\in\Sigma^*$ such that $A\Rightarrow^* w$.

We start with $X:=\emptyset$.
For each terminal rule $A\to u\in R$ with $u\in\Sigma^*$, we add $A$ to $X$. 
We then repeatedly apply the following rule until $X$ is no longer updated: for each nonterminal rule
\[
A\to u_0A_1u_1\cdots A_nu_n\in R,
\]
where $u_0,u_1,\ldots,u_n\in\Sigma^*$ and $A_1,\ldots,A_n\in\Gamma$, if
\[
A_1,\ldots,A_n\in X,
\]
then we add $A$ to $X$.

Because $X$ contains at most $\#(\Gamma)$ nonterminals, this procedure must terminate. 
After the process terminates, $X$ consists exactly of the productive nonterminals. Therefore, we have
\[
I\in X \quad\iff\quad L(\mathbf{G})\ne\emptyset,
\]
and the problem is decidable.
\end{proof}

It follows that the first problem is decidable:
\begin{theorem}
Given a DFA or a CFG representing $S\subseteq\Sigma^+$, it is decidable whether every position in $\mathtt{WordSubtraction}(S)$ is a $\Ppos$-position.
\end{theorem}
\begin{proof}
Every position in $\mathtt{WordSubtraction}(S)$ is a $\Ppos$-position if and only if $S=\emptyset$. 
Indeed, if $S=\emptyset$, then no move is possible, so every position is a $\Ppos$-position. 
Conversely, if there exists some $s\in S$, then $s$ itself is an $\Npos$-position because the first player can remove $s$ and reach the empty word. 
Therefore, by Propositions~\ref{prop6-emptiness_reg} and~\ref{prop6-emptiness_cf}, the problem is decidable.
\end{proof}

We next consider the second problem: whether every nonterminal position is an $\Npos$-position. This problem is slightly less trivial than the first one. For example, consider the case $S=a\Sigma^*$ with $\Sigma=\{a,b\}$. 
The terminal positions are precisely the words in $b^*$, while every nonterminal position contains at least one occurrence of $a$. Therefore, from any nonterminal position, the first player can win by removing the suffix beginning at the leftmost occurrence of $a$. Hence, every nonterminal position is an $\Npos$-position.

First, we show that the problem is decidable when $S$ is regular.
\begin{theorem}\label{th6-decide_every_ntposition_is_n}
Given a DFA representing $S\subseteq\Sigma^+$, it is decidable whether every nonterminal position in $\mathtt{WordSubtraction}(S)$ is an $\Npos$-position.
\end{theorem}
\begin{proof}
The set of nonterminal positions is $\Sigma^*S$. Thus, it suffices to decide whether
\[
L_\Ppos(S)\cap\Sigma^*S=\emptyset.
\]
For this purpose, we construct a DFA recognizing $L_\Ppos(S)\cap\Sigma^*S$ and apply Proposition~\ref{prop6-emptiness_reg}.

First, given a DFA $\mathbf{A}_S=(Q_S,\Sigma,\delta_S,q_S,F_S)$ such that $L(\mathbf{A}_S)=S$, we can construct a DFA recognizing $L_\Npos(S)$ using the method described in Section~\ref{sec:reg}. 
By interchanging its accepting and rejecting states, we obtain a DFA $\mathbf{A}_\Ppos$ recognizing $L_\Ppos(S)$.
Moreover, we can also construct a DFA  $\mathbf{A}_{\mathrm{NT}}$ recognizing $\Sigma^*S$ by applying the subset construction to $\mathbf{N}_{\mathrm{NT}}$ constructed in Lemma~\ref{lem6-nonterminal_positions}.

We now have a DFA $\mathbf{A}_\Ppos$ recognizing the $\Ppos$-position language $L_\Ppos(S)$ and a DFA $\mathbf{A}_{\mathrm{NT}}$ recognizing the set of nonterminal positions $\Sigma^*S$. 
We can construct a DFA recognizing $L_\Ppos(S)\cap\Sigma^*S$ using a standard construction called the \emph{product automaton}.
For two DFAs
\[
\mathbf{A}_1=(Q_1,\Sigma,\delta_1,q_{01},F_1)
\quad\text{and}\quad
\mathbf{A}_2=(Q_2,\Sigma,\delta_2,q_{02},F_2),
\]
their product automaton is the DFA
\[
\mathbf{A}_1\times\mathbf{A}_2
=(Q_1\times Q_2,\Sigma,\delta,(q_{01},q_{02}),F_1\times F_2),
\]
where
\[
\delta((q_1,q_2),a)=(\delta_1(q_1,a),\delta_2(q_2,a))
\]
for each $(q_1,q_2)\in Q_1\times Q_2$ and $a\in\Sigma$. 
The two components simulate $\mathbf{A}_1$ and
$\mathbf{A}_2$ on the same input.
The product accepts exactly when both components accept,
so
\[
    L(\mathbf{A}_1\times\mathbf{A}_2)
    =L(\mathbf{A}_1)\cap L(\mathbf{A}_2).
\]
We apply this construction to $\mathbf{A}_{\Ppos}$
and $\mathbf{A}_{\mathrm{NT}}$, and then use
Proposition~\ref{prop6-emptiness_reg} to test emptiness.
This determines whether
\[
    L_{\Ppos}(S)\cap\Sigma^*S=\emptyset
\]
and gives the desired answer.
\end{proof}

For context-free subtraction sets, the same automata construction is no longer available.
However, the failure of this method alone does not imply undecidability.
The next theorem shows that no algorithm can solve the problem for all context-free subtraction sets.
\begin{theorem}\label{th6-undecidability}
Given a CFG representing $S\subseteq \Sigma^+$, it is undecidable whether every nonterminal position in $\mathtt{WordSubtraction}(S)$ is an $\Npos$-position.
\end{theorem}
\begin{proof}
    We give a reduction from the \emph{universality problem} for CFGs, which is undecidable. That is, given a CFG $\mathbf{G}$ over an alphabet $\Delta$, it is undecidable whether $L(\mathbf{G})=\Delta^*$.
    (See, e.g., Theorem 5.13 of~\cite{sipser2013introduction} for details.)

    Let $L=L(\mathbf{G})\subseteq\Delta^*$.
Choose a fresh symbol $\$\notin\Delta$, let $\Sigma=\Delta\cup\{\$\}$, and define
\[
    S=(L\$)^+\cup\{\$\}\cup\Sigma^*\Delta.
\]
A CFG generating $S$ can be effectively constructed from $\mathbf{G}$ using the standard constructions for union, concatenation, and positive closure.
Moreover, $S\subseteq\Sigma^+$.

Because $\Sigma\subseteq S$, the only terminal position is $\varepsilon$.
We show that every nonterminal position is an $\Npos$-position if and only if $L=\Delta^*$.

Suppose first that $L=\Delta^*$.
Every nonempty word ending in a letter of $\Delta$ belongs to $\Sigma^*\Delta$.
Also, every word ending in $\$$ can be written as
\[
    u_1\$u_2\$\cdots u_k\$
\]
with $u_i\in\Delta^*,\ k\geq1$,
and hence belongs to $(L\$)^+$.
Thus $S=\Sigma^+$, so every nonempty position can move directly to $\varepsilon$ and is an $\Npos$-position.

Conversely, suppose that $L\neq\Delta^*$, and choose $w\in\Delta^*\setminus L$.
If $w=\varepsilon$, then $\$\$\notin S$.
Its only option is $\$$, which is an $\Npos$-position.
Thus $\$\$$ is a nonterminal $\Ppos$-position.
If $w\neq\varepsilon$, consider $w\$$.
Because it contains exactly one $\$$ and $w\notin L$, we have $w\$\notin S$.
Thus no move removes the whole word.
Every option of $w\$$ is a nonempty prefix of $w$ and therefore belongs to $\Delta^+\subseteq S$.
Every such option is an $\Npos$-position because it can move directly to $\varepsilon$.
Hence, $w\$$ is a $\Ppos$-position, which is nonterminal. 

Therefore, the required property holds exactly when $L(\mathbf{G})=\Delta^*$.
An algorithm deciding this property from a CFG for $S$ would decide universality for CFGs, which is impossible.
\end{proof}
These examples show that the answer depends on how the subtraction set is specified:
the same question is decidable for regular subtraction sets but undecidable for context-free subtraction sets.
Many other decision problems remain to be explored.

\subsection*{Exercises}
\begin{exercise}
    Consider $\mathtt{WordSubtraction}^{-}(S)$ for\[
    S=a\Sigma^*
    \]
    with $\Sigma=\{a,b\}$.
    Let us compute its $\Npos$-position language.
\begin{enumerate}
\item Construct a DFA $\mathbf{A}_S$ recognizing $S$.
\item Determine the set of terminal positions and construct a DFA $\mathbf{A}_T$ recognizing it.
\item Following the construction in Theorem~\ref{th6-misere_reg}, construct a DFA $\mathbf{A}_{\Npos}$ recognizing $L_{\Npos}^{-}(S)$.
\item Verify that
\[
L_{\Npos}^{-}(S)=(b^*a\cdot L_\Npos(S))\cup b^*
\]
holds.\diaqed
\end{enumerate}

\end{exercise}
\begin{starexercise}\label{ex6:prefix-closed}
    The result of Theorem~\ref{th6-misere_epsilon_is_only_terminal} can be extended to regular subtraction sets such that $(\Sigma^*S)^c$ is prefix-closed. Here, we say that a language $L\subseteq\Sigma^*$ is \emph{prefix-closed} if every prefix of every word in $L$ also belongs to $L$. 
    For example, $b^*\subseteq\{a,b\}^*$ is prefix-closed.
\begin{enumerate}
\item Let $L$ be a regular language and $\mathbf{A}$ be a DFA recognizing $L$ in which every state is reachable from the initial state.
Prove that $L$ is prefix-closed if and only if no rejecting state can reach an accepting state in $\mathbf{A}$.
\item Prove that for a regular language $L$, if $L$ is prefix-closed and nonempty, then\[
L^c=((L\Sigma)\setminus L)\Sigma^*
\]
holds.
\item Let $S\subseteq\Sigma^+$ be a regular subtraction set, and let $L=(\Sigma^*S)^c$ be the set of terminal positions. 
Suppose that $L$ is prefix-closed. 
Following the proof of Theorem~\ref{th6-misere_epsilon_is_only_terminal}, show that
\[
L_{\Npos}^{-}(S)
=
\bigl((L\Sigma\setminus L)\cdot L_{\Npos}(S)\bigr)\cup L
\]
holds. \diaqed
\end{enumerate}
\end{starexercise}
\begin{exercise}
    Show that the subtraction set $S=a\Sigma(\Sigma\Sigma)^*\cup b(\Sigma\Sigma)^*$ with $\Sigma=\{a,b\}$ is not reflexive.\diaqed
\end{exercise}
\begin{exercise}
    Let $\Sigma=\{a,b\}$ and $k\ge 1$. 
    Let $W\subseteq\Sigma^*$ be the language defined as\[
    W=\{w\in \Sigma^*:|w|_a\text{ is a multiple of }k\}.
    \]
    Show that $\mathscr{S}(W)\ne \emptyset$.\diaqed
\end{exercise}
\begin{starexercise}\label{ex6:recognition_of_smax}
A language $L\subseteq \Sigma^*$ is said to be \emph{recognized} by a monoid $M$ if there are a monoid homomorphism $\zeta:\Sigma^*\to M$ and a subset $P\subseteq M$ such that $L=\zeta^{-1}(P)$.
For example, the language $(\Sigma\Sigma)^*\subseteq \Sigma^*$ of even length words is recognized by the cyclic group $\mathbb{Z}/2\mathbb{Z}$ of order $2$.
Indeed, $(\Sigma\Sigma)^*=\zeta^{-1}(\{0\})$ holds with the homomorphism $\zeta:\Sigma^*\to \mathbb{Z}/2\mathbb{Z}=\{0,1\}$ defined as $\zeta(a)=1$ for each $a\in\Sigma$.
\begin{enumerate}
    \item Show that if a language $L\subseteq\Sigma^*$ is recognized by a monoid $M$, then the complement $L^c\subseteq\Sigma^*$ is also recognized by $M$.
    \item Show that if a language $L\subseteq\Sigma^*$ is recognized by a monoid $M$, then $L^{-1}L\subseteq\Sigma^*$ is also recognized by $M$. (\textbf{Hint}: Put $P_0=\zeta(L)$ and consider $P_0^{-1}P_0
    =\{m\in M:pm\in P_0
      \text{ for some }p\in P_0\}.$)
    \item Based on Theorem~\ref{th6-maximum_subtraction_set}, show that if a language $W\subseteq\Sigma^*$ is recognized by a monoid $M$ and $\mathscr{S}(W)\ne \emptyset$, then there is a language $S\in \mathscr{S}(W)$ such that $S$ is also recognized by $M$.
\end{enumerate}
It is well known that a language is regular if and only if
it is recognized by some finite monoid.
By item~3, if $W$ is recognized by a finite monoid $M$
and $\mathscr{S}(W)\neq\emptyset$, then some
$S\in\mathscr{S}(W)$ is recognized by the same monoid $M$.
This gives another proof of Corollary~\ref{cor-smax_is_regular}.

Studying languages through recognition by monoids is a central approach in \emph{algebraic language theory}.
For an introduction, see~\cite{DBLP:books/sp/necs2005/Lawson05};
for a comprehensive reference, see~\cite{Pin2025MFAT}.
\diaqed
\end{starexercise}

\begin{starexercise}
For an ordinary subtraction game, the set of nonnegative multiples of any integer $k\geq 1$ can be realized as the
set of $\Ppos$-positions by taking its complement as the subtraction set.
We consider a natural generalization to word subtraction games.

A regular language $W\subseteq\Sigma^*$ is called an \emph{identity group language} if there exist a finite
group $G$ and a monoid homomorphism
$\zeta:\Sigma^*\to G$ such that
\[
    W=\zeta^{-1}(\{1_G\}),
\]
where $1_G$ denotes the identity element of $G$.

Let $W$ be an identity group language.
\begin{enumerate}
    \item Show that $\varepsilon\in W$ and
    \[
        W^{-1}W=W.
    \]

    \item Put $S=W^c$.
    Show that no legal move can take a word in $W$ to another word in $W$, while every word outside $W$ has a legal move to $\varepsilon$.
    Use the Partition Theorem to conclude that
    \[
        L_{\Ppos}(S)=W.
    \]

    \item Apply Theorem~\ref{th6-maximum_subtraction_set}
    to show that $W^c$ is the maximum element of
    $\mathscr{S}(W)$ with respect to set inclusion.
\end{enumerate}
Thus, every identity group language can be realized
as the $\Ppos$-position language of a word subtraction game,
with its complement as the maximum subtraction set.
\diaqed
\end{starexercise}
\begin{starexercise}
Let $W\subseteq\Sigma^*$ be a regular language given by a DFA $\mathbf{A}=(Q,\Sigma,\delta,q_0,F)$.
\begin{enumerate}
    \item Let $F_{\mathsf{reach}}\subseteq F$ be the set of accepting states reachable from $q_0$.
For each $q\in F_{\mathsf{reach}}$, let $\mathbf{A}_q=(Q,\Sigma,\delta,q,F)$
be the DFA obtained from $\mathbf{A}$ by changing
only the initial state to $q$.
Show that
\[
    W^{-1}W
    =\bigcup_{q\in F_{\mathsf{reach}}}L(\mathbf{A}_q).
\]
    \item Using item~1 and closure under complement and concatenation, describe how to construct a DFA recognizing $W(W^{-1}W)^c$.
    \item Show that it is decidable whether there exists
    a subtraction set $S\subseteq\Sigma^+$ such that $W=L_{\Ppos}(S)$.\diaqed
\end{enumerate}
\end{starexercise}
\begin{exercise}
    Let $S\subseteq\Sigma^+$ and $L\subseteq\Sigma^*$
be regular languages given by DFAs.
    For each of the following problems, describe an algorithm that decides whether the answer is yes or no.
    \begin{enumerate}
        \item Does $L$ contain at least one word that is an $\Npos$-position in $\mathtt{WordSubtraction}(S)$?
        \item Is every word in $L$ an $\Npos$-position in $\mathtt{WordSubtraction}(S)$?\diaqed
    \end{enumerate}
\end{exercise}

\subsection*{Open Problems}
\begin{problem}[Can the problem in Theorem~\ref{th6-decide_every_ntposition_is_n} be solved in polynomial time?]
    As stated in Theorem~\ref{th6-decide_every_ntposition_is_n}, 
    given a DFA representing a subtraction set $S$, it is decidable whether every nonterminal position is an $\Npos$-position. 
    However, the decision procedure constructs a DFA $\mathbf A_{\Ppos}$ recognizing $L_\Ppos(S)$. 
    As discussed in Section~\ref{sec:reg}, this construction is essentially an instance of the subset construction, and the resulting product\[
    \mathbf{A}_{\Ppos}\times \mathbf{A}_{\mathrm{NT}}
    \]
    can have exponentially many states.

    Can this property be decided in time polynomial in the size of the input DFA, without explicitly constructing a DFA
for $L_{\Ppos}(S)$?
Even if recognizing the full position language requires many states, deciding this particular property could be easier.\diaqed
\end{problem}

\newpage
\section{Research Background and Further Directions}
This project began when Yuji Komatsu of Nagoya University, who introduced the first author to combinatorial game theory, encouraged him to explore connections between combinatorial games and formal language theory. Around the same time, Professor Ryuya Hora of ZEN University, who has contributed extensively to both formal language theory and combinatorial game theory, pointed out several connections between the two fields and helped bridge the communities working in them. In particular, he drew attention within these communities to the analogy between game quotients and syntactic monoids, which we discuss below.

The idea of word subtraction games emerged from a more general framework of \emph{derivation games}.
Around the end of 2025, the first author noticed that the computation of Grundy numbers using mex and nim-sum in normal-play games has a structural resemblance to weight computation in weighted context-free grammars. 
The first author formulated this analogy in terms of derivation games in \cite{DBLP:conf/wia/InoueKIK26}.
In our work, word subtraction games first appeared there as a concrete example of this framework.
In fact, the prefix-removal version of Example~\ref{ex1-ws_a_b_aa} in this note already appeared in \cite{DBLP:conf/wia/InoueKIK26}.
Its interesting behavior was investigated by Hiyu Inoue and Professor Shun-ichi Kimura, together with other members of the research group at Hiroshima University, including the second author, Hiroki Inazu,
Takahiro Yamashita, and Kosaku Watanabe.

It is worth noting that word subtraction games are not an entirely new idea.
The suffix-removal version considered here already appears in Hetyei's work \cite{DBLP:journals/jct/Hetyei10} under the name \emph{truncation game},
in connection with Bernoulli games and enumeration problems.
As briefly noted in the footnote on p.~\pageref{fn:prefix-closed},  results from that paper also apply to certain word subtraction games considered here.
However, our main emphasis is different: we focus on the relationship between the language class of the subtraction set and those of the corresponding position languages.

Over the following months, several interesting results on word subtraction emerged through discussions in different communities. 
In particular, Professor Hiroyuki Seki of Mukogawa Women's University and Professor Yoshiaki Takata of Kochi University of Technology studied derivation games restricted to right-linear grammars in detail \cite{takata2026constructiondfacomputinggrundy}. 
This special class is closely related to word subtraction. 
Their work was the first to give an explicit proof of the fact stated in Theorem~\ref{th2-fin} although the techniques are quite different from those used here. Moreover, many of the ideas underlying Section~\ref{sec:reg} are due to Yoshiaki Takata.

Around the same time, Yushin Kusajima and Shunya Oguchi, students in the first author's research group, investigated the case where the subtraction set is context-free, which is the subject of Section~\ref{sec:cf}. 
Their work showed that the general case is considerably more complicated, and raised several questions about where the boundary lies between the regular and context-free cases.

Since the summer of 2026, the first author has had several opportunities to present these results to researchers in combinatorial game theory. 
The discussions on these occasions have led to further observations and questions.

In particular, with the help of the second author, we obtained several interesting results concerning misère play, which are discussed in Section~\ref{subsec:misere}. 
He had already been in contact with the first author for some time and has a strong understanding of both formal language theory and combinatorial game theory. 
This naturally led to his joining the project.

We also benefited greatly from the advice and collaboration of Takuto Maki of Kagoshima University, who has extensive knowledge of reflexivity and realizability for ordinary subtraction sets. 
His insights and suggestions directly contributed to the material presented in Sections~\ref{subsec:reflexivity} and~\ref{subsec:realizability}. 
These discussions also provided an answer to a question that had been of particular interest to the first author: which languages can arise as $\Ppos$-position languages of word subtraction games?

Several specific open problems have already been stated throughout the note.
We conclude with some broader topics for further study.

\vspace{7pt}
\noindent\textbf{Grundy functions.} 
This note has focused primarily on $\Ppos$- and $\Npos$-positions, but the full Grundy function
\[
    G:\Sigma^*\to\mathbb{N}
\]
contains more information.
Section~\ref{sec:reg} shows that, for a regular subtraction set and each fixed $g\geq 0$,
the words with Grundy number $g$ form a regular language.
This provides a starting point for further study.
For example, when is the Grundy function bounded, and how efficiently can its values be computed?
Understanding these functions is also important when several word subtraction games are combined by disjunctive sum.

\vspace{7pt}
\noindent\textbf{Games with heap splitting.}
Subtraction games can be generalized to games in which a heap can be split as part of a move, giving rise to the well-known classes of \emph{octal games} and \emph{code games}; see, for example, \cite{WW, CGT}. 
We would like to define corresponding games on words and find methods for analyzing them.
For example, after removing an infix $s$ from a word $w=usv$, we could keep $u$ and $v$ as two independent components, rather than concatenate them as in Section~\ref{subsec:infix}.
Positions would then consist of finite collections of words,
and Grundy numbers would describe how the components combine.

A related construction appears in the derivation game
introduced in \cite{DBLP:conf/wia/InoueKIK26}, which was formulated as a special case of a word version of a code game.
Its main purposes were to explore the connection between
Grundy computation and context-free grammars and to prove
an undecidability result.
It was not introduced as a general framework for studying word versions of code games.

\vspace{7pt}
\noindent\textbf{Partizan games.}
Throughout this note, both players have had the same legal moves.
A natural extension is to allow Left and Right to use different subtraction sets.
One can then ask how the language classes of these sets affect the languages of positions in the four outcome classes $\mathcal{L}$, $\mathcal{R}$, $\mathcal{N}$, and $\mathcal{P}$.
Another possibility is to allow Left to remove prefixes and Right to remove suffixes.
With the subtraction sets used for End-Nim in Section~\ref{subsec:two-sided},
this gives a word version of partizan End-Nim, which has already been studied (see Section~2.6 of~\cite{Albert2019}).

\vspace{7pt}
\noindent\textbf{Computational complexity.}
Although computational complexity has been briefly mentioned in several open problems, it has not been a main focus of this note.
The complexity of various problems concerning word subtraction games is another interesting direction for further study.
For example, consider a word $w$ of length $n$ in the two-sided game of Section~\ref{subsec:two-sided}.
Once membership in $S$ and $T$ has been recorded for every infix of $w$, its outcome can be computed in $O(n^3)$ time by dynamic programming over intervals:
there are $O(n^2)$ intervals, and each has $O(n)$ possible moves.
This resembles the interval-based computation used in the CYK algorithm.
More generally, computational complexity questions for word subtraction games could provide an interesting meeting point for combinatorial game theory, formal language theory, and algorithms.

\vspace{7pt}
\noindent\textbf{Concatenation sum and game quotients.}
In combinatorial game theory, one can compare positions by asking whether they have the same outcome when combined with other positions.
In the theory of misère quotients, for example, one considers a set of positions closed under disjunctive sum and taking options, and identifies two positions when they have the
same outcome after adding any other position in the set \cite{miserequotient}.
As Ryuya Hora pointed out in discussions between the two communities, this idea is closely related to the construction of syntactic monoids in formal language theory.

The second author recently proposed considering concatenation as a sum operation on word positions.
For simplicity, fix a subtraction set $S$ and consider the suffix-removal version.
For two positions $u,v\in\Sigma^*$, define their \emph{concatenation sum} by
\[
    u\odot v=uv.
\]
The resulting word is played as a single position, so a move can cross the boundary between $u$ and $v$.
For example, if $u=u_1u_2$ and $u_2v\in S$, then
\[
    u\odot v=u_1u_2v\longrightarrow u_1
\]
is a legal move.
Thus, the two original words do not remain independent components during play.

This operation gives a concrete connection between game outcomes and familiar constructions in language theory.
Write $W=L_{\Ppos}(S)$.
We compare two words $u$ and $v$ by appending the same word $x$ to both.
The resulting positions $ux$ and $vx$ have the same outcome for every $x\in\Sigma^*$ if and only if
\[
    ux\in W\iff vx\in W
    \qquad\text{for every }x\in\Sigma^*.
\]
This equivalence relation is the \emph{Myhill--Nerode right congruence} of $W$.
When $W$ is regular, its equivalence classes form the state set of the minimal DFA recognizing $W$.
However, these classes do not in general form a monoid under concatenation: the proposed product
$[u][v]=[uv]$ can depend on the representatives chosen.
The right congruence guarantees only that appending a fixed word gives a well-defined transition between classes.

To obtain a monoid quotient, we instead compare positions by placing words on both sides:
\[
    xuy\in W\iff xvy\in W
    \qquad\text{for every }x,y\in\Sigma^*.
\]
This gives the \emph{syntactic congruence} of $W$.
Because this relation is preserved by concatenation on both sides, its equivalence classes form a monoid, called the \emph{syntactic monoid}.
When $W$ is regular, this monoid is isomorphic to the transition monoid of its minimal DFA.

Concatenation differs from the disjunctive sum used in misère quotients, but the underlying idea is shared:
two objects are identified when no allowed context can distinguish their behavior.
Here, membership in the position language is exactly the outcome of the game.
Thus, the analogy between game quotients and syntactic monoids becomes a direct correspondence for word positions under concatenation.

This viewpoint also suggests a connection with algebraic language theory, which studies regular
languages through finite monoids.
Exercise~\ref{ex6:recognition_of_smax} gives an example of how monoid recognition can be used in the study of word subtraction games.
We hope that these connections will encourage further exchanges between combinatorial game theory and formal language theory.

\newpage
\bibliographystyle{plain}
\bibliography{bibtex}

\end{document}